\documentclass{aa}  

\usepackage{graphicx}
\usepackage{txfonts}
\usepackage{lipsum}
\usepackage{subcaption}         
\usepackage{lscape}             
\usepackage{placeins}           

\usepackage[hyperindex=true,colorlinks=true,citecolor=blue,linkcolor=blue,breaklinks=true]{hyperref}

\newcommand{\vip}{\texttt{VIP}}

\makeatletter
\@ifundefined{nolinenumbers}{}{ \nolinenumbers }
\@ifundefined{linenumbers}{}{ \let\linenumbers\relax }
\makeatother

\begin{document}

        \title{High-contrast imaging of Galactic Cepheids with VLT/SPHERE}
 \titlerunning{High-contrast imaging of Galactic Cepheids}
 
 \subtitle{}
 \author{A.~Gallenne\inst{1}\corrauth{agallenne@academicos.uta.cl}        
        \and P.~Kervella\inst{2,3}
        \and N.~R.~Evans\inst{4}
        \and J.~Milli\inst{5}
        \and E.~Sivkova\inst{2,3}
        \and W.~Gieren\inst{6}
        \and G.~Pietrzy\'nski\inst{7}
        \and G.~Bras\inst{2}
        \and V.~Hocd\'e\inst{8}
        \and W.~Kiviaho\inst{2,3}
        \and N.~Nardetto\inst{8}
        \and B.~Pilecki\inst{7}
        \and B.~Zgirski\inst{6}
}
 
 \authorrunning{A. Gallenne et al.}
 
 \institute{Instituto de Alta Investigaci\'on, Universidad de Tarapac\'a, Casilla 7D, Arica, Chile
        \and LIRA, Observatoire de Paris, Universit\'e PSL, Sorbonne Universit\'e, Universit\'e Paris Cit\'e, CY Cergy Paris Universit\'e, CNRS, 5 place Jules Janssen, 92195 Meudon, France
        \and French-Chilean Laboratory for Astronomy, IRL 3386, CNRS and U. de Chile, Casilla 36-D, Santiago, Chile
        \and Smithsonian Astrophysical Observatory, MS 4, 60 Garden Street, Cambridge, MA 02138, USA
        \and Univ. Grenoble Alpes, CNRS, IPAG, 38000 Grenoble, France
    \and Universidad de Concepci\'on, Departamento de Astronom\'ia, Casilla 160-C, Concepci\'on, Chile
    \and Centrum Astronomiczne im. Miko\l{}aja Kopernika, PAN, Bartycka 18, 00-716 Warsaw, Poland
    \and Universit\'e C\^ote d'Azur, Observatoire de la C\^ote d'Azur, CNRS, Laboratoire Lagrange, Nice, France
 }

 \abstract
 {Classical Cepheids are key distance indicators and benchmarks for stellar evolution, yet most of them are members of binary or multiple systems. While spectroscopic surveys and \textit{Gaia} proper-motion anomalies reveal a high binary fraction, the population of resolved companions remains poorly characterised.}
 {We aim to search for and characterise visual companions to bright Galactic Cepheids using high-contrast imaging and to derive quantitative limits on undetected companions to constrain the architecture of Cepheid multiple systems.}
 {We observed 47 Galactic Cepheids with VLT/SPHERE using the ZIMPOL instrument in classical imaging mode and the $V, R^\prime$, and $I^\prime$ filters. The data were obtained in pupil-stabilised mode and analysed using PCA-based imaging technique. For detected companions, we injected negative fake companions in a Monte Carlo approach  to measure the relative astrometry. For non-detections, synthetic companions were injected to compute $5\sigma$ contrast curves as a function of separation.}
 {We detected companions with a signal-to-noise ratio of S/N $> 5$ for eight Cepheids ($\eta$~Aql, AX~Cir, S~Nor, AP~Pup, W~Sgr, T~Vel, TX~Del, and V659~Cen), corresponding to about 17\,\% of the sample. Our SPHERE imaging confirms previously known visual companions with improved astrometry and reveals new wide components for AP~Pup, T~Vel, and TX~Del) at projected separations of $\sim 0.16-0.9\arcsec$. For the remaining Cepheids, we derived typical maximum contrasts of $\sim 10, 11$, and 12\,mag at 0.25\arcsec, 0.5\arcsec, and $> 1\arcsec$, respectively. For a sub-set of targets, these limits ruled out main sequence companions more massive than late-K dwarfs beyond 0.5\arcsec.}
 {Our SPHERE survey provides the first homogeneous set of high-contrast optical constraints on wide companions of Galactic Cepheids. The low detection rate of visual companions compared to the high overall binary fraction implies that most companions inferred from radial velocities and \textit{Gaia} astrometry are either closer than $\sim 20$\,mas or significantly fainter than the limits reached here.}
 
 
 \maketitle
 
 %

 \section{Introduction}
 
 Classical Cepheids are pulsating evolved stars that are widely used as standard candles to estimate distances in nearby galaxies. Beyond their utility in terms of the cosmic distance scale with the period-luminosity relation (PLR), they are also powerful astrophysical laboratories, providing fundamental clues for studying the pulsation and evolution of intermediate-mass stars. The physics of classical Cepheids has been studied for decades, but there are still some issues that are still ripe for investigation, such as the mass loss mechanism \citep[e.g.][]{Neilson_2008_09_0,Neilson_2016_06_0}, the mass discrepancy between pulsation and evolution models \citep[e.g.][]{Keller_2008_04_0,Neilson_2011_05_0}, or the multiplicity fraction \citep[e.g.][noting that here we  use 'binary"'as shorthand for 'binary or multiple']{Evans_1995_05_0,Evans_2005_08_0}. Binary Cepheids offer the possibility to probe several aspects, such as the Cepheid evolution or the determination of the Cepheid mass. 
 
 More than 50\,\% of intermediate mass main sequence stars, which are the progenitors of Cepheids, are in binary systems \citep{Duchene_2013_08_4}. Therefore, we expect a similar probability for Cepheids; however, the binary fraction of Cepheid seems lower than expected. However, population synthesis models suggest  that the binary fraction of Cepheids should be significantly less than that of their progenitors due to binary interaction in the red giant branch phase \citep{Neilson_2015_02_0}. In addition, \citet{Evans_2005_08_0} noted that about 35\,\% of the Galactic Cepheids have a spectroscopic companion \citep[see also][]{Shetye_2024_10_0} and 44\,\% of those being in a multiple system. This would correspond to a spectroscopic binary fraction of those intermediate mass main sequence progenitors of $\sim40-45$\,\% \citep{Neilson_2015_02_0}. However, this low binary fraction is due to incompleteness. Recent work of \citet{Kervella_2019_03_0,Kervella_2019_03_1,Kervella_2022_01_0} combined the \textsc{Hipparcos} and \textit{Gaia} DR2 positions to determine the mean proper motion of nearby classical Cepheids and derived a binary fraction of $> 80$\,\%. This is significantly higher than the predicted \citep{Neilson_2015_02_0} and the observed values: $32-52$\,\% from \citealt{Anderson_2016_10_0}, $40-70\,\%$ for their Milky Way progenitors from \citealt{Chini_2012_08_3} and $\sim50$\,\% for their progenitors in the cluster Scorpius OB2 from \citealt{Kouwenhoven_2005_01_9}. We note that eight of the nine nearest classical Cepheids within 500 pc are members of binary systems and that they host, on average, more than one companion \citep{Kervella_2019_03_1}. The only object not yet established as a binary is $\zeta$~Gem, although it does have a visual companion.
 
 Binary statistics are fraught with selection effect due to the observing techniques used to detect the companions, either because they cannot reach a high dynamic range, the angular resolution is limited, or they cannot detect small orbital motions from radial velocities. Detecting the companions is a challenging task because of the brightness of the Cepheid and close components are hardly detectable from imaging. Most of the binary Cepheids were detected from radial velocities and are single-line spectroscopic binaries \citep[see e.g.][]{Lloyd-Evans_1982_06_0,Welch_1987_07_0,Evans_1988_03_0,Szabados_1991_01_0,Evans_1994_11_0,Szabados_2012_11_1,Evans_2013_10_0,Szabados_2013_09_0,Evans_2015_07_0,Pilecki_2015_06_0,Anderson_2016_10_0,Pilecki_2018_07_0,Gallenne_2019_02_0,Shetye_2024_10_0,Hocde_2024_09_0}. From spectroscopic observations, we can detect the orbital motion of the Cepheid if the companion is close and massive enough to produce a detectable effect in the radial velocities. Detection of the companion's spectral lines is difficult because most of the known companions are hot main sequence stars with broad and blended lines. Ultraviolet (UV) spectroscopy has proven to be efficient in detecting the hottest companions and characterising their spectral types and temperatures from low- and high-resolution spectra obtained with the International Ultraviolet Explorer (IUE) and Hubble Space Telescope (HST) satellites \citep[see e.g.][]{Evans_1992_01_0,Massa_2008_01_0}. Combined with a mass-temperature relation, the companion mass can be inferred. High-resolution UV spectra also make possible the detection of spectral lines for a few companions and the measurements of their radial velocities. When taken at maximum and minimum orbital velocities, the velocity amplitude of the companion combined with the orbital velocity amplitude of the Cepheid determined from the ground and the inferred mass of the companion provides an estimate of the Cepheid mass \citep[see e.g.][]{Bohm-Vitense_1997_03_0,Evans_1998_03_0,Evans_2011_09_0,Evans_2018_10_0}. Radial velocity measurements are the easiest and most frequently applied method to identify and characterise binary Cepheids \citep{Evans_2015_07_0}.
 
 Dynamical mass determinations (and, hence, model-independent determinations) requires astrometric measurements of the companion's orbit, in addition to the spectroscopic orbit and an assumed distance to the Cepheid. However, this is a challenging task due to small separations ($\lesssim50$\,mas) and contrast ($\gtrsim 3$\,mag in $H$). The first dynamical mass was measured by \citet{Evans_2008_09_0} who spatially resolved the spectroscopic companion of \object{Polaris} located at $\sim170$\,mas using the High Resolution Channel (HRC) of the Advanced Camera for Surveys (ACS) on board HST. Follow-up observations provided a mass precise to $\sim 20$\,\% \citep{Evans_2018_08_1}. New observing techniques such as near-infrared (NIR) long-baseline interferometry have strong potential for spatially resolving binary Cepheids \citep[see e.g.][]{Gallenne_2013_02_0}. \citet{Gallenne_2018_11_0} combined astrometric measurements from interferometry with ground-based radial velocities of the Cepheid and space-based radial velocities (HST/STIS) of the companion. They measured the most precise mass for a Galactic Cepheid (\object{V1334~Cyg}, 3\,\%) and its companion (1\,\%), together with the most precise distance for a Cepheid (1\,\%). Other binary Cepheid systems are also promising for precisely measuring their masses and distances \citep{Gallenne_2019_02_0,Gallenne_2015_07_0,Gallenne_2014_01_0}. 
 
 Wider separations were explored for 70 bright Cepheids using the HST and its Wide Field Camera 3 \citep[WFC3;][]{Evans_2020_12_0,Evans_2016_05_0,Evans_2013_10_0}. Several companions were detected in the range 0.5\arcsec-5\arcsec\ for 13 Cepheids. Their colour-magnitude diagram showed that the companion masses are evenly distributed, although a bias is expected at small separations as low-mass stars are more difficult to detect. In addition, the HST point-spread-function (PSF) within 2\arcsec\ is complicated and required sophisticated image processing. Interestingly, all detected companions within 2\arcsec\ are themselves spectroscopic binaries. \citet{Evans_2020_12_0} suggested that the wider companion is likely formed first, followed by the formation of the inner spectroscopic binary through disk fragmentation. \citet{Gallenne_2014_07_0} also investigated the presence of visual companions around five Cepheids using adaptive optics (AO) imaging with the VLT/NACO instrument, detecting a companion only for $\eta$~Aql.
 
 In this paper, we want to explore the Cepheid duplicity using high-contrast imaging to detect faint and nearby orbiting companions. We probe the spatial scale 0.02\arcsec-3.5\arcsec\ in three different photometric filters with the SPHERE instrument. This paper is organised as follows. In Sect.~\ref{section__observations_and_data_reduction}, we present the observations and data reduction. We then describe our data analysis in Sect.~\ref{section__data_analysis}, including the method we used to search for companions, the position and flux estimates, and the detection limits for undetected components. We discuss the results in Sect.~\ref{section__discussion} and the conclusions in Sect.~\ref{section__conclusion}.

 \section{Observations and data reduction}
 \label{section__observations_and_data_reduction}
 
 We observed 47 Galactic Cepheids with the extreme adaptive optics instrument SPHERE \citep[Spectro-Polarimetric High-contrast Exoplanet REsearch;][]{Beuzit_2019_11_0}, corresponding too all observable Cepheids with $V < 8$\,mag to have the best AO performances. Data were acquired in 2018 and 2020 with the ZIMPOL instrument \citep{Schmid_2017_06_0,Schmid_2018_11_0} in classical imaging mode (i.e. without coronagraph) in the optical. Observations were taken in pupil-stabilised mode to obtain the best PSF stability. ZIMPOL is a two-arm imager with two CCD detectors, which enabled us to acquire quasi-simultaneous observations in two different filters. The field of view of each detector is $3.5\arcsec\times3.5\arcsec$. As the identified  companion of Cepheids are mostly hot main sequence stars \citep{Evans_2022_10_0}, our observing strategy was to obtain images in the $V$ band in one arm and in two red filters ($R'$ and $I'$) on the second arm. This combination of filters should allow us to determine the spectral type of the detected companions. The full log of our observations is listed in Table~\ref{table__log}. Each observation include three dithering positions to remove artefacts such as bad pixels, cosmic rays, or fixed pattern noise.
 
  \begin{figure*}[!h]
        \centering
        \resizebox{\hsize}{!}{\includegraphics{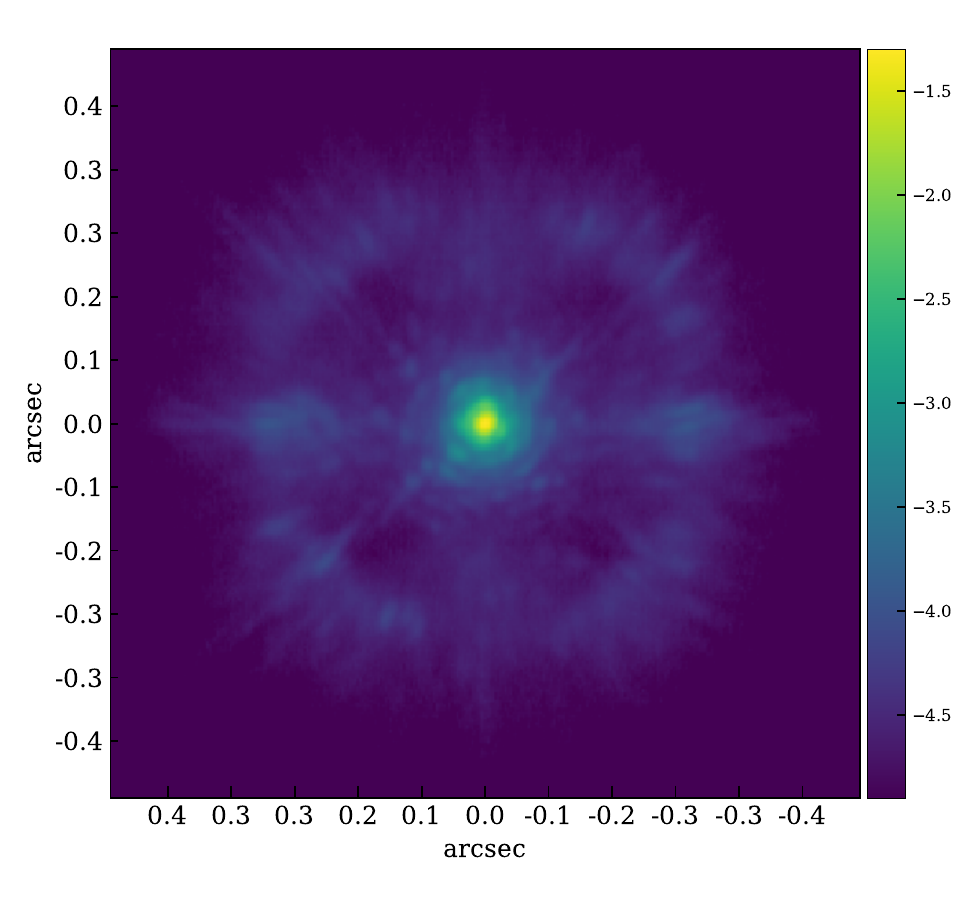}
                \includegraphics{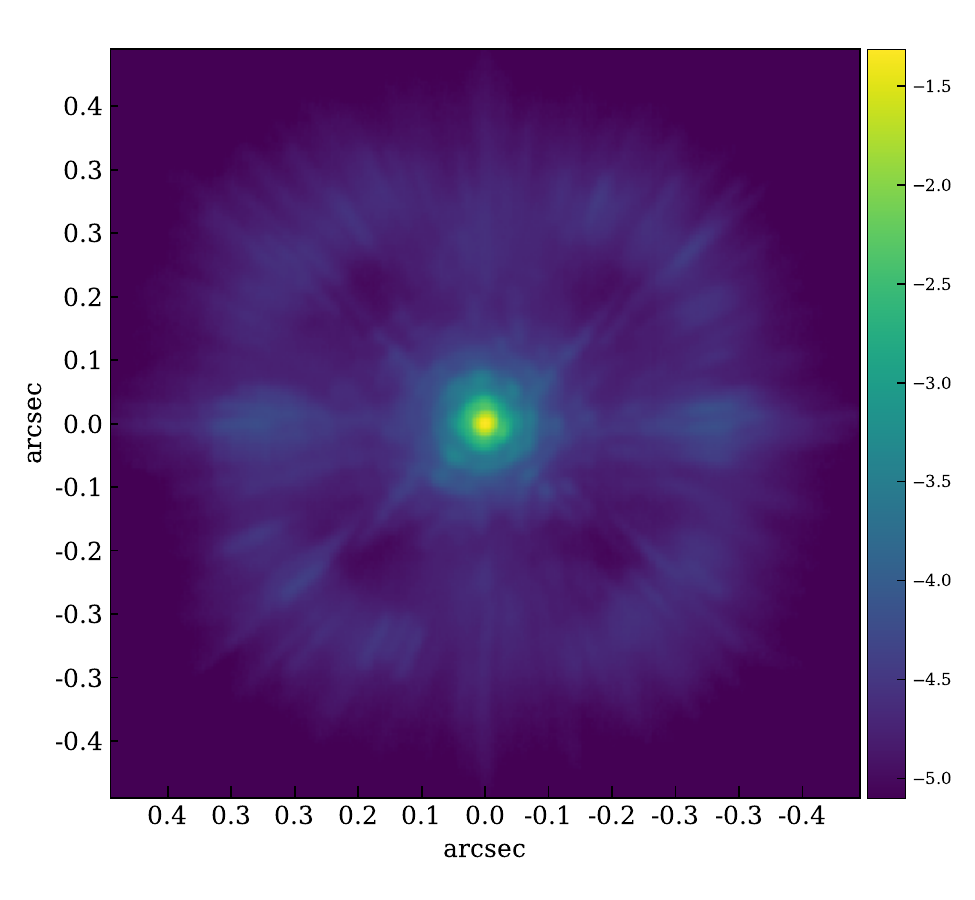}
                \includegraphics{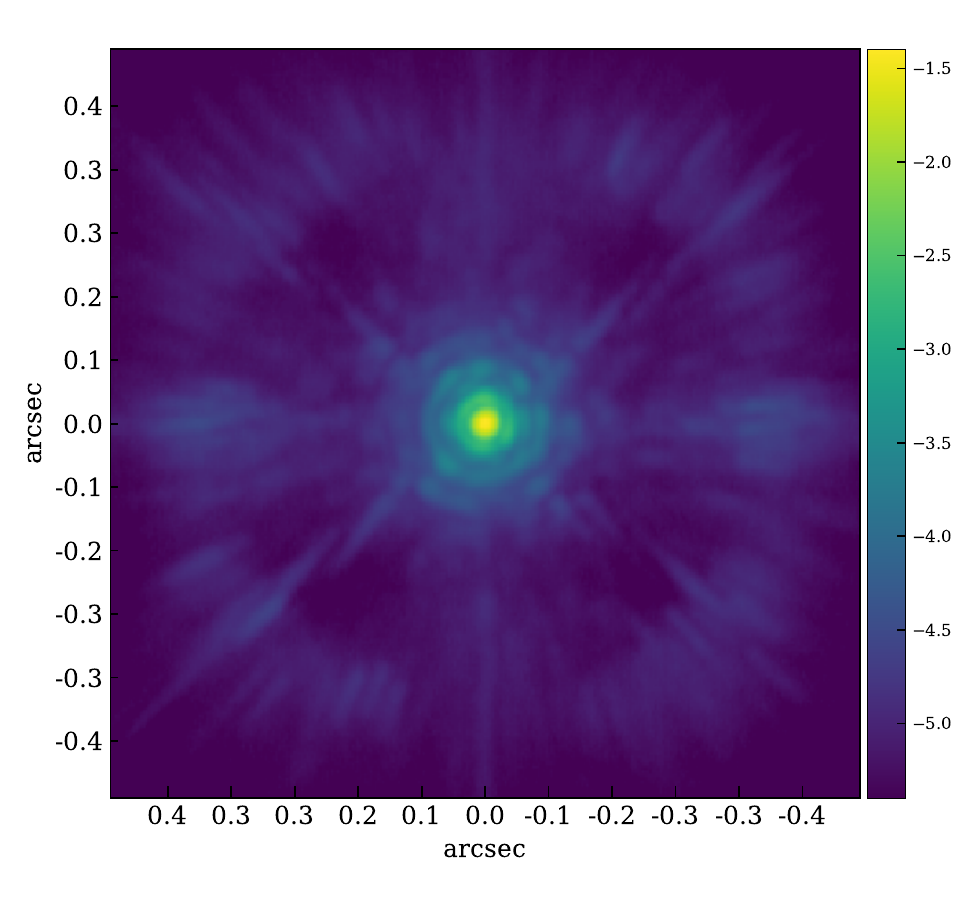}}
        \caption{Average images of FF~Aql in $V$ (left), $R'$ (middle) and $I'$ band. The scale is logarithmic and the images have been cropped to $1\arcsec\times1\arcsec$ to better see the PSF quality.}
        \label{fig__images_ffaql}
 \end{figure*}
 
 To save observing time, no photometric calibrators were observed. In addition, the brightest Cepheids saturated the detectors so that a neutral density filter had to be used. Therefore, in this work, we can only estimate $V, R'$, and $I'$ flux ratios between the detected companion and the Cepheid, $f = f_\mathrm{comp}/f_\mathrm{ceph}$.
 
 The data reduction of ZIMPOL images was carried out by the High-Contrast Data Center \citep{Delorme_2017_12_0,Beuzit_2019_11_0}\footnote{\url{https://sphere.osug.fr/spip.php?rubrique16}} using an implementation of the SPHERE-ZIMPOL software package described in \citet{Schmid_2017_06_0}. It includes standard procedures such as dark, flat, and bad pixel correction. Individual frames were also de-dithered and re-aligned so that the final output of the data reduction consists in a single non de-rotated cube for each filter. Most of the observations were obtained under very good seeing conditions (average seeing of 0.73\arcsec; see Table~\ref{table__log}), so that each frame in a cube has a good point spread function (PSF). The worst conditions were for the Cepheid V659~Cen, with an average seeing of $\sim 1.5$. To present the quality of the PSFs, we display as examples in Fig.~\ref{fig__images_ffaql} the cube-averaged images of FF~Aql in $V$, $R'$, and $I'$ bands, where several Airy rings are visible, even in $V$.

 \section{Data analysis}
 \label{section__data_analysis}
 
 \subsection{Searching for companions}
 
 To search for companions around our Cepheid sample, we modelled and subtracted the PSF using a principal component analysis \citep[PCA;][]{Amara_2012_12_0} implemented in the Vortex Image Processing library, which is a \texttt{python} package dedicated to high-contrast imaging \citep[VIP;][]{Gomez-Gonzalez_2017_07_0}\footnote{\url{https://github.com/vortex- exoplanet/VIP}}. We chose the PCA approach instead of other post-processing algorithms (e.g.  locally optimised combination of images, LOCI, \citealt{Lafreniere_2007_05_2},  non-negative matrix factorisation, NMF, \citealt{Lee_1999_10_5}, or local low-rank plus sparse plus Gaussian-noise decomposition, LLSG, \citealt{Gomez-Gonzalez_2016_05_1}) because of its good performance and processing speed. A PCA uses all images in a given data cube to model the variation of the PSF in time by identifying the principal linear components (PCs). The number of PCs determines how well the PSF is fitted, with the first components containing the most common structures. An image is built from these first components and then subtracted from the images to remove the speckle pattern. This image processing method is commonly used for exoplanet direct detection and is usually combined with additional post-processing techniques, such as angular differential imaging \citep[ADI;][]{Marois_2006_04_9} or reference star differential imaging \citep[RDI;][]{Lafreniere_2009_04_6,Soummer_2011_11_2,Gerard_2016_07_2}.
 
 \begin{figure}[!h]
        \centering
        \resizebox{\hsize}{!}{\includegraphics{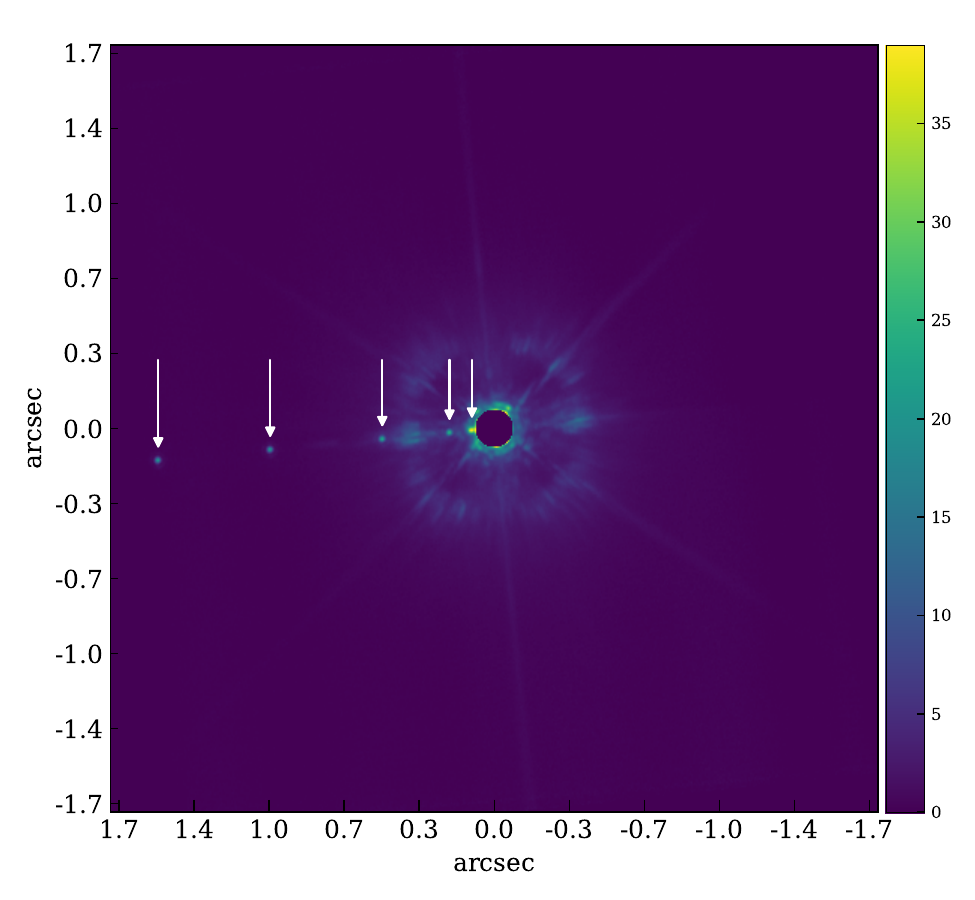}}
        \caption{De-rotated average image of FF~Aql in the $I'$ band, where we injected five fake companions (indicated with arrows). A fake central mask is inserted for the display purposes.}
        \label{fig__cube_injected_companions}
 \end{figure}
 
 \begin{figure*}[!t]
        \centering
        \resizebox{\hsize}{!}{\includegraphics{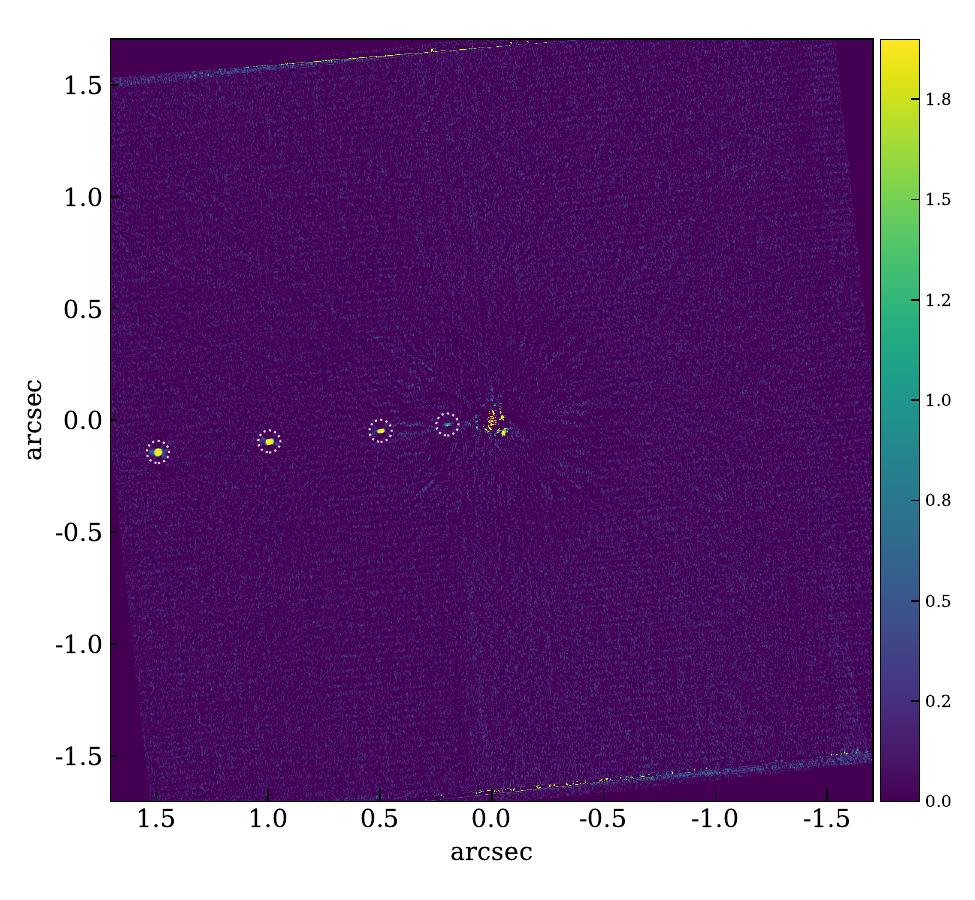}\includegraphics{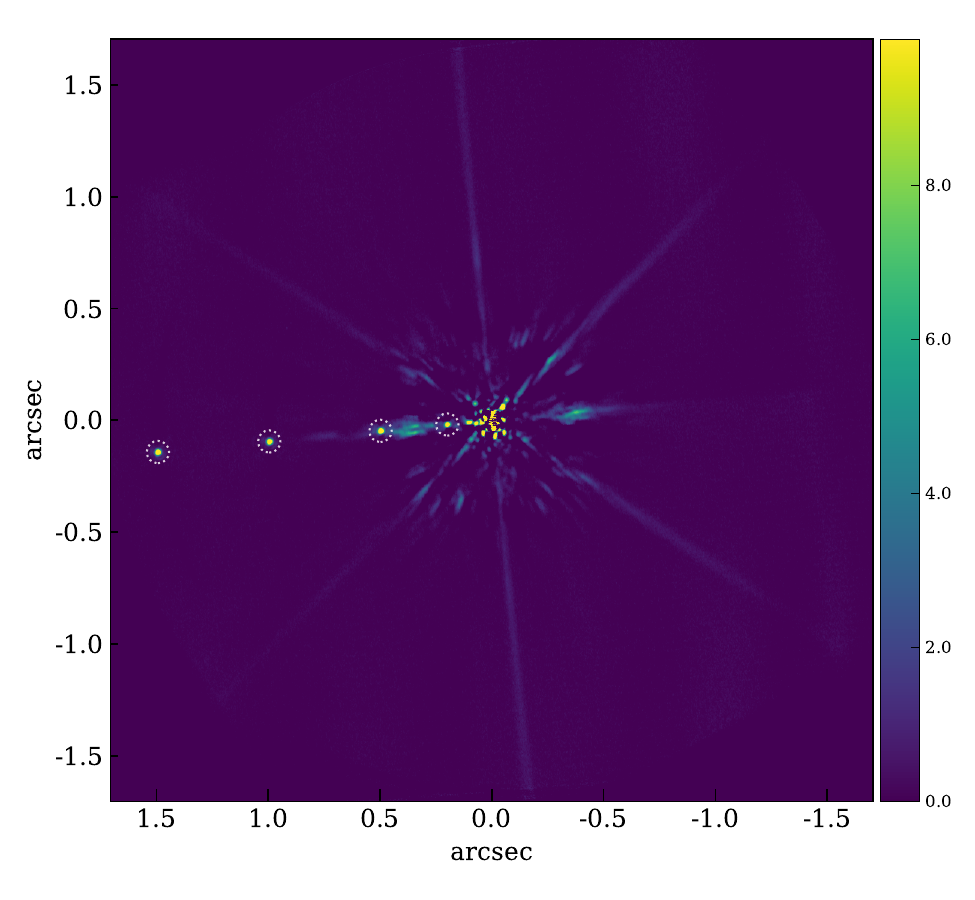}\includegraphics{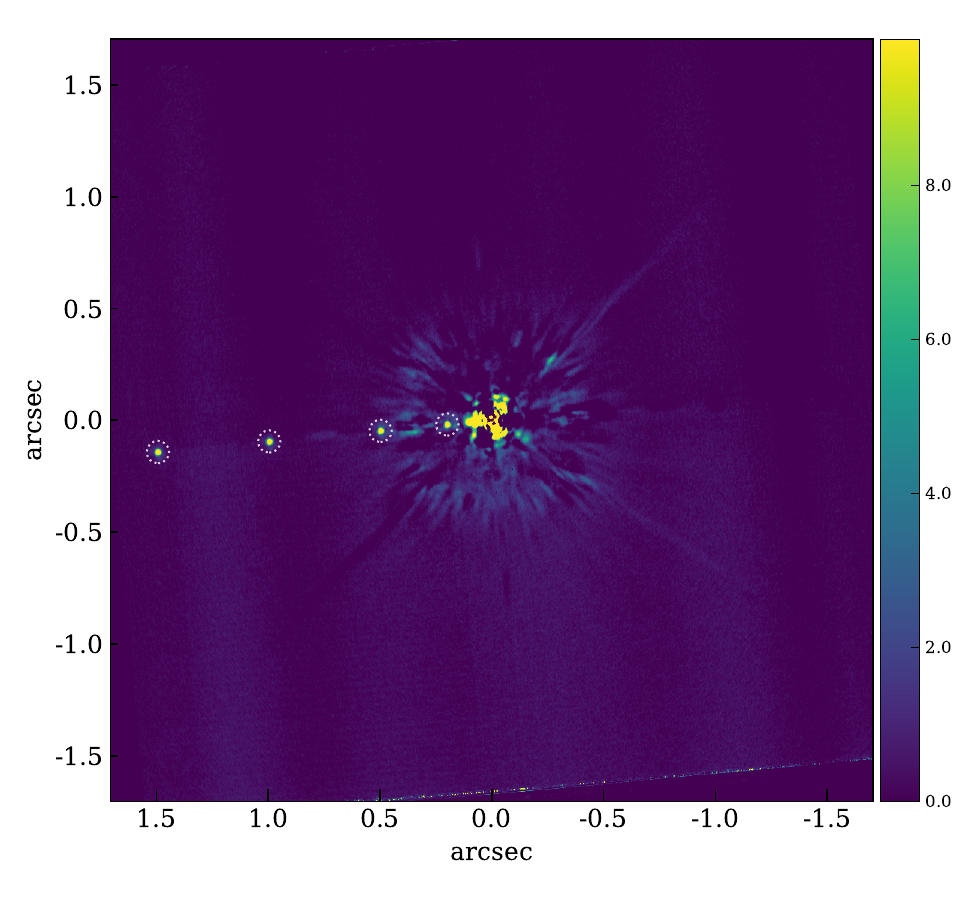}}
        \resizebox{\hsize}{!}{\includegraphics{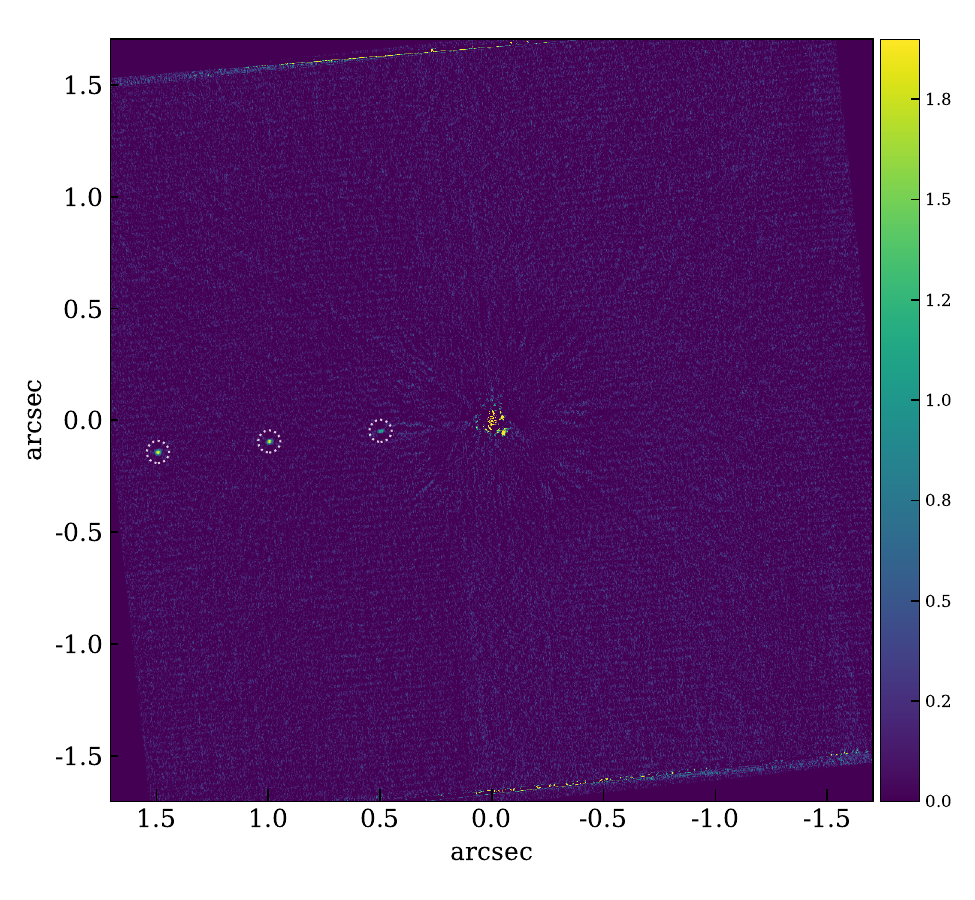}\includegraphics{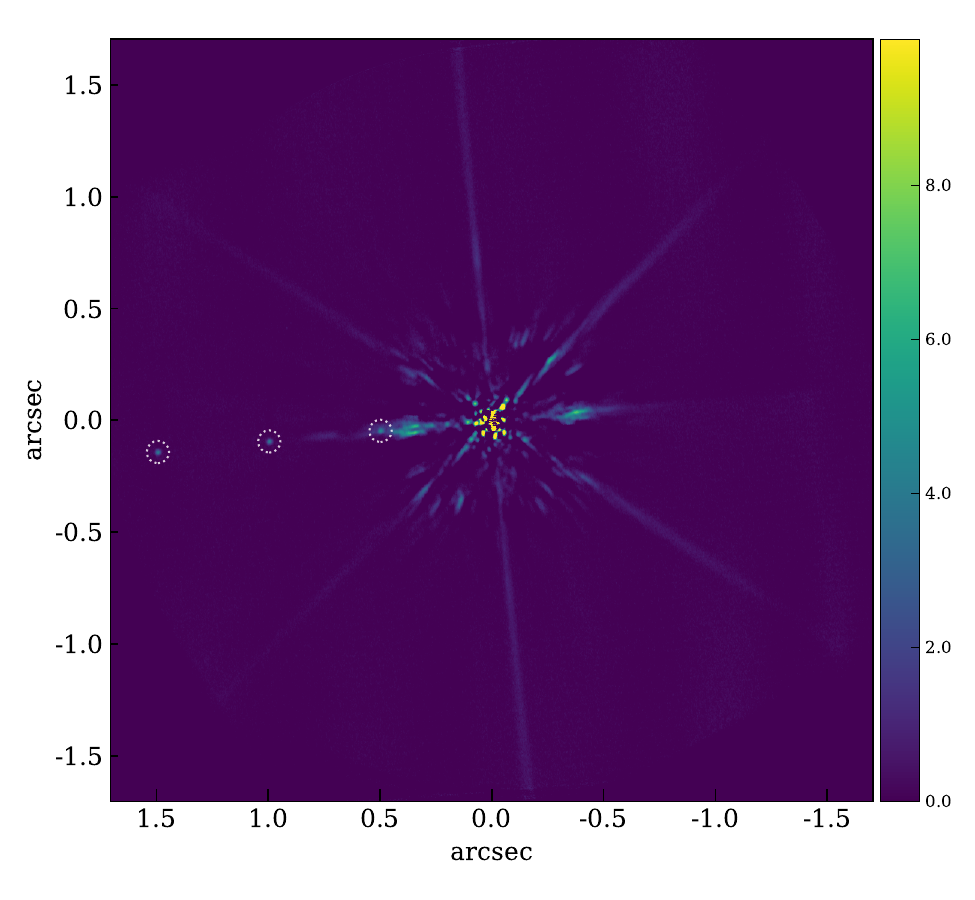}\includegraphics{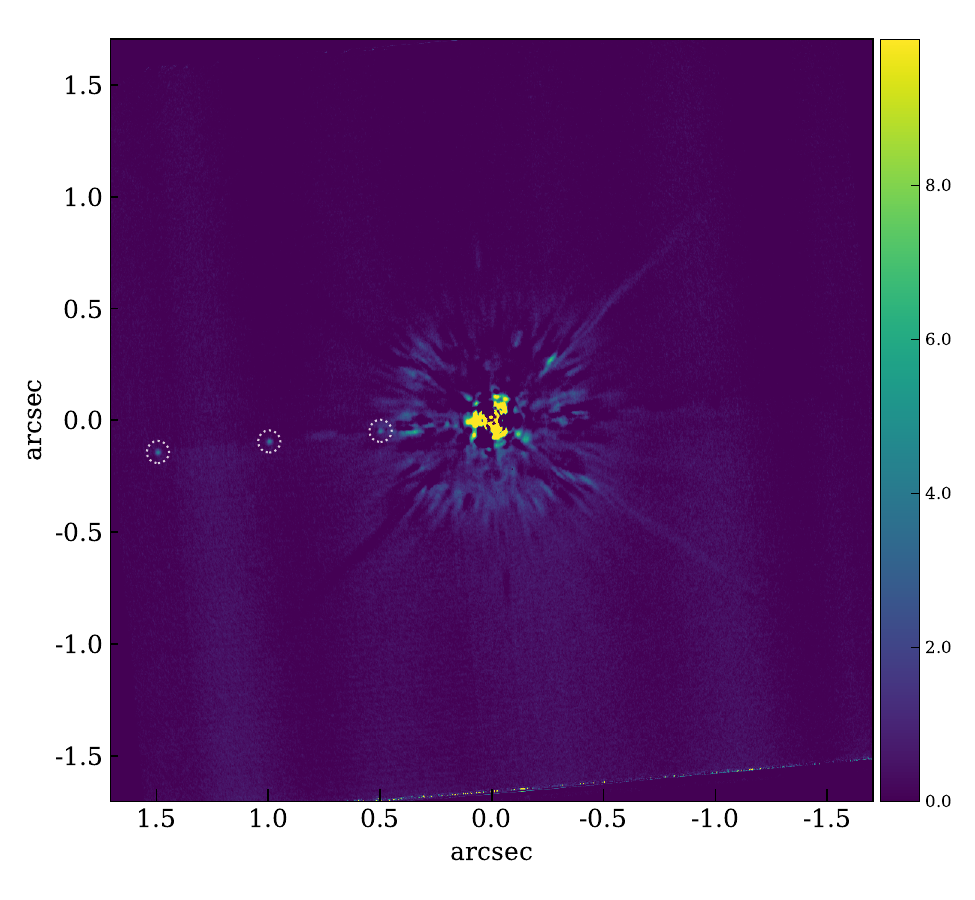}}
        \caption{\textit{Top:} Full-frame PCA-ADI results for the three ADI approaches ($\#1, \#2, \#3$) with injected companions of flux ratio of 0.1\,\%. \textit{Bottom:} Same as above but with flux ratio of 0.02\,\%. The companions with S/N $> 5$ are circled.}
        \label{fig__pca_maps}
 \end{figure*}
 
 Our observations were carried out in pupil tracking mode, which we expected would allow us to apply the ADI technique. However, the observing sequences have too short an exposure time, so  the variation of the parallactic angle (PA) in the cube ends up too small, possibly leading to self-subtracted companions (especially the closest ones). We therefore chose to test three different approaches of PCA-based ADI techniques to detect the companions. The first one (\#1) consisted of only taking the first and last 12 frames of the data cubes. For the second approach (\#2), we rotated the frame, $i$, of the cube by an additional angle of $0.5^\circ$, so that the PA of the frame $i$ has $+i*0.5^\circ$ (first frame having $i = 0$). We refer to this as a fake ADI. The last approach (\#3) is also a fake ADI but combined with a RDI-like method, namely, we rotated the second half of the cube by $+180^\circ$ and took the first half as a reference. Each approach uses one principal component because we noticed that larger values did not necessarily provide a better detection and  tended, rather, to remove the closest companions. In addition, this is a pre-analysis whose goal is to identify a possible companion and obtain its approximate location.
 
 To test which approach would be better in detecting companions, we took one of our SPHERE data cubes for which there is no visible bright companions (FF~Aql in band $I'$) and we injected fake companions at 0.1\arcsec, 0.2\arcsec, 0.5\arcsec, 1.0\arcsec, \ and 1.5\arcsec\ from the central star and at an angle of $185.4^\circ$. It corresponds to the mean parallactic angle of the cube and where we see the brightest speckle pattern within 0.5\arcsec\ (most pessimistic case for sensitivity). All the injected companions had the same flux ratio, $f_\mathrm{comp}/f_\mathrm{\star} = f = 0.1$\,\%, in an aperture where the  the full width at half maximum (FWHM) of the PSF was estimated by taking the median of the original data cube (i.e. no companions injected). Thus, we obtained an aperture of $1\times$FWHM.
  We used the functionalities included in the \vip\ package for that purpose. The data cube quality is very good, with a FWHM of 22.6\,mas, close to the theoretical diffraction limit of $\sim 20$\,mas in $V$. We display the image with injected companions in Fig.~\ref{fig__cube_injected_companions}.
 
 For each approach, we estimated the estimated the signal-to-noise ratio (S/N) in a resolution element at the expected positions of the companions. We used the \texttt{snr} module of the \vip\ package which implements the approach described in \citet{Mawet_2014_09_0} where a penalty term is applied to the S/N to account for the small sample statistics at small separations from the central star. Each companion location was tested against the background resolution elements at the same radial distance from the centre. We defined a detection threshold of S/N~$> 5$ above the background level\footnote{S/N is not the Gaussian significance, i.e. not the confidence level of a Gaussian n$\sigma$. S/N can be converted to n$\sigma$, however, values are similar for S/N $\lesssim 8$, while computer precision limits the conversion to higher S/N values.} (chosen as being the median background of the image plus $5\sigma$). Finally, we repeated  the entire process, this time with higher contrast companions of $f = 0.02$\,\%. We display the PCA-ADI results in Fig.~\ref{fig__pca_maps} for both flux ratios and the detection levels $n\sigma$ of each companion for all approaches are listed in Table~\ref{table__snr_pca_test}.
 
 As expected, the closest companion is not significantly detected (i.e. $\sigma < 5$) for any approaches, as well as the second companion at higher contrast. Between all approaches, \#2 offers the best detection level at any separations. Approach \#3 seems to be better than \#1 for small separations; however, the residual noise is larger in the central part. To perform a cross-check for possible detections, we  chose to search for companions using the three approaches across our full dataset. When a companion was identified, we applied a PCA on a single annulus with the true parallactic angles (width of $3\times$FWHM) encompassing the companion candidate. We optimised the S/N at its position by searching for an optimal number of PCs. The S/N and significance level for all detected companions are listed in Table~\ref{table__snr_pca_annular}.
 
\begin{figure*}[!h]
        \centering
        \resizebox{\hsize}{!}{\includegraphics{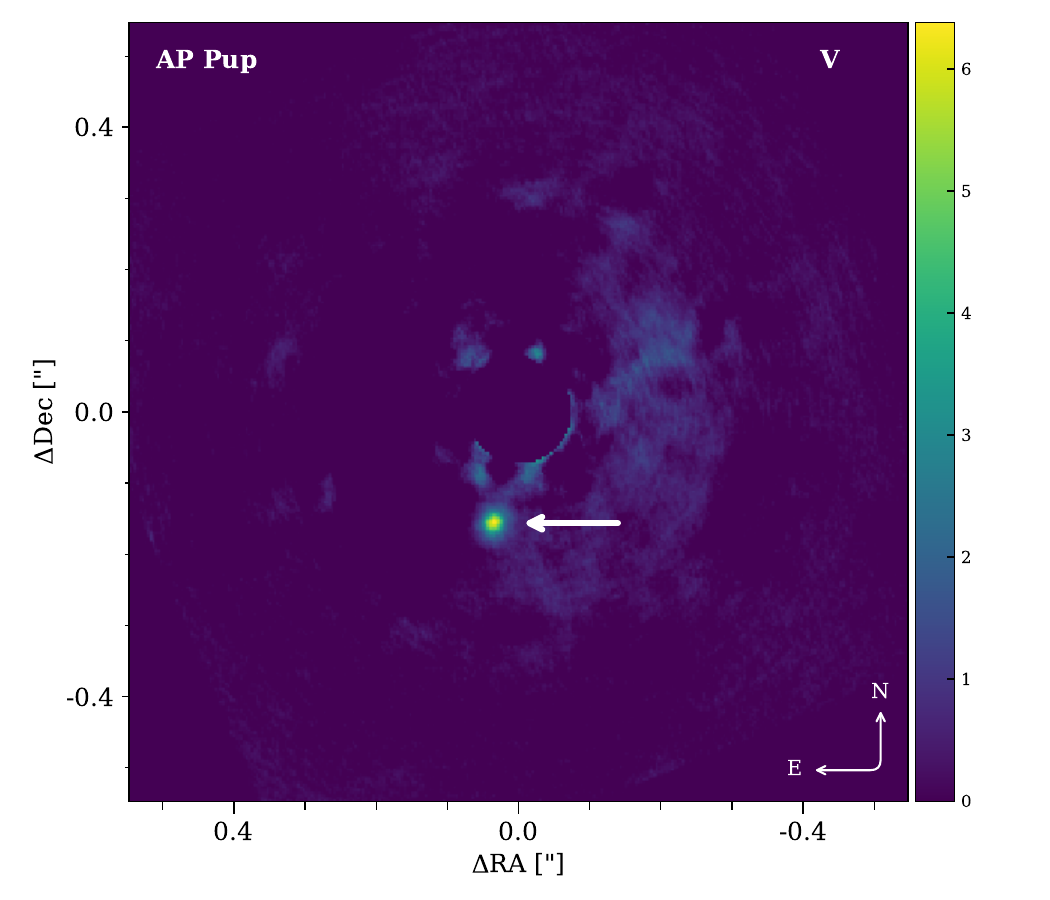}\includegraphics{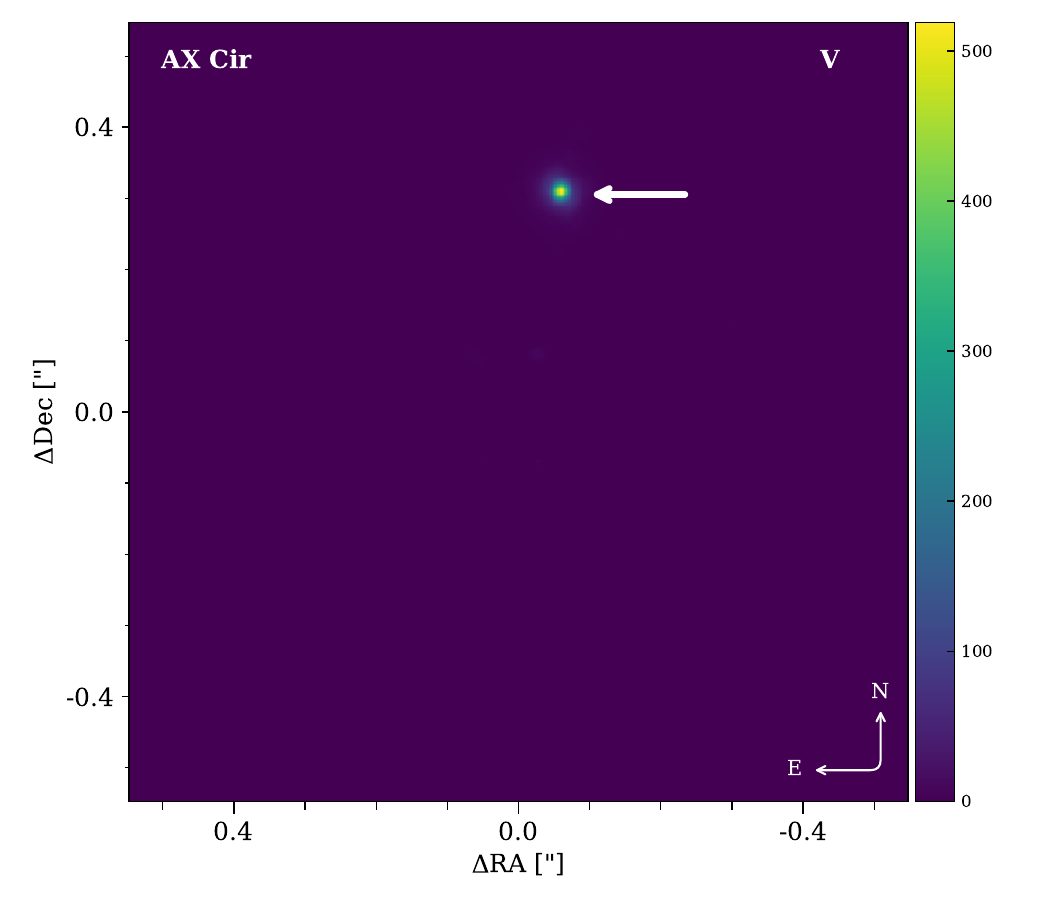}\includegraphics{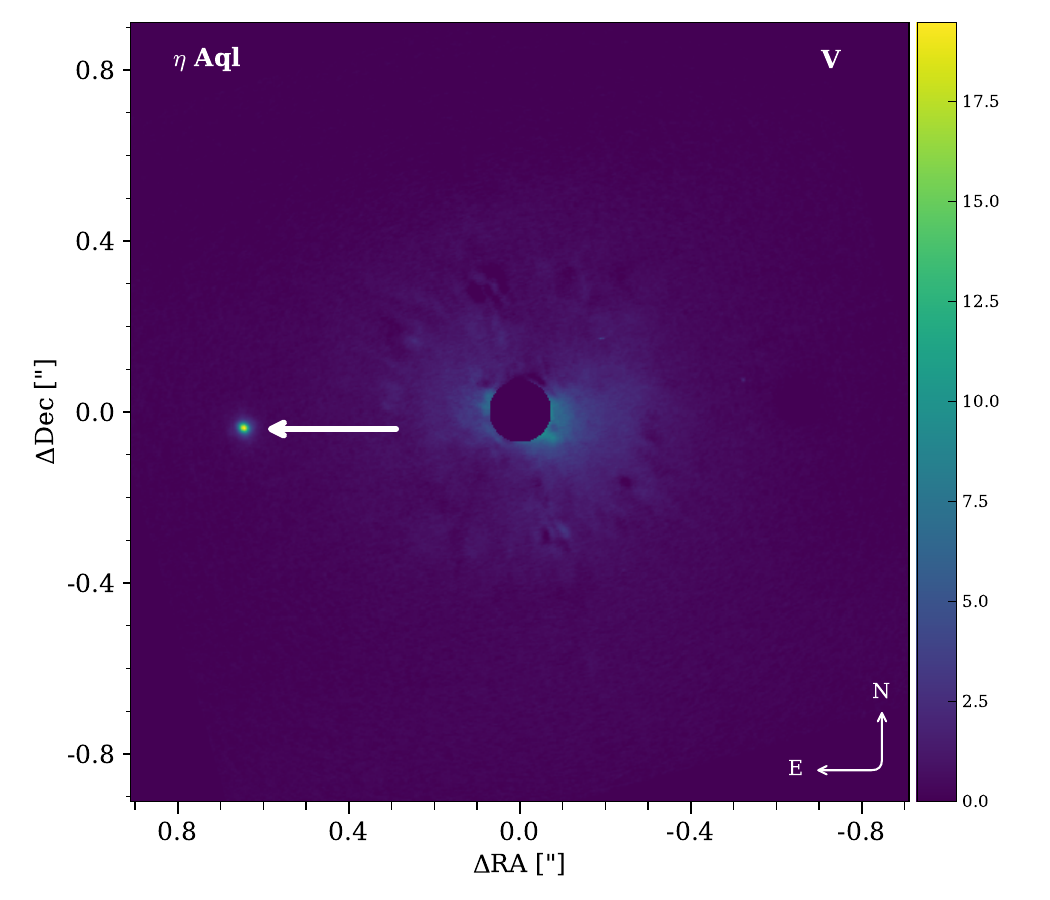}}
        \resizebox{\hsize}{!}{\includegraphics{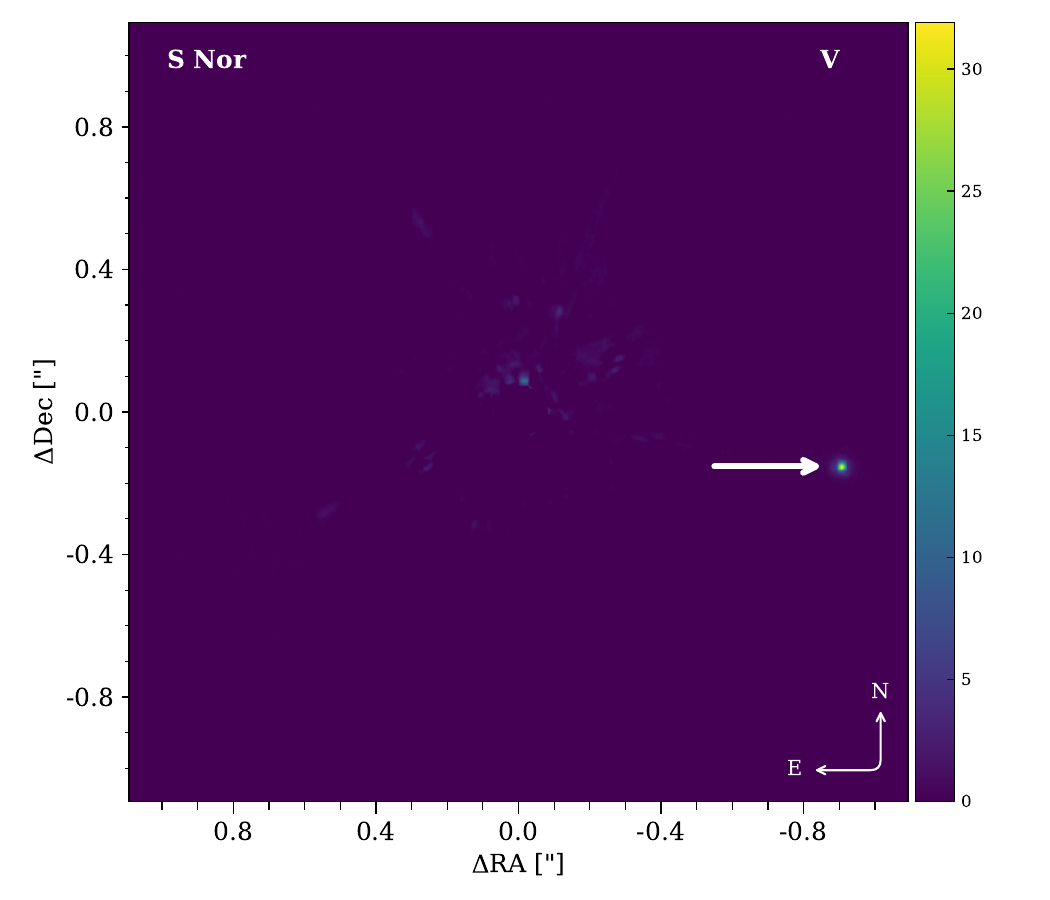}\includegraphics{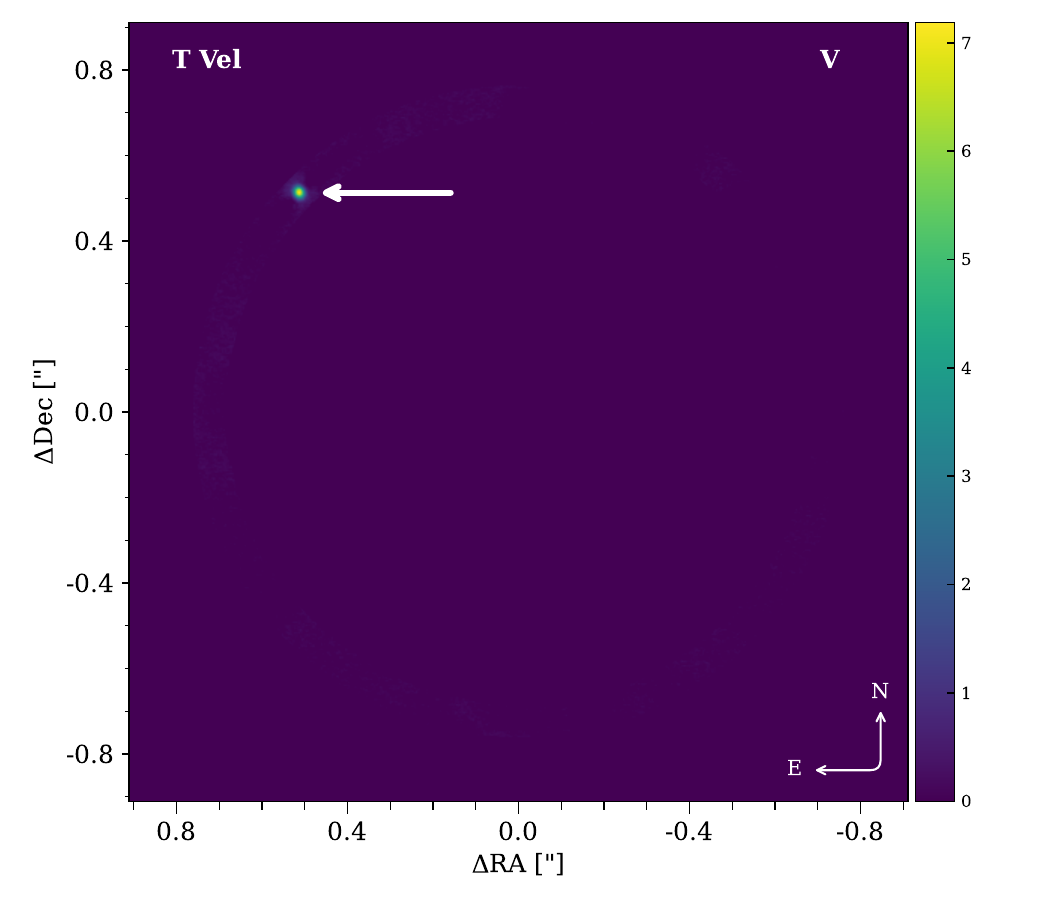}\includegraphics{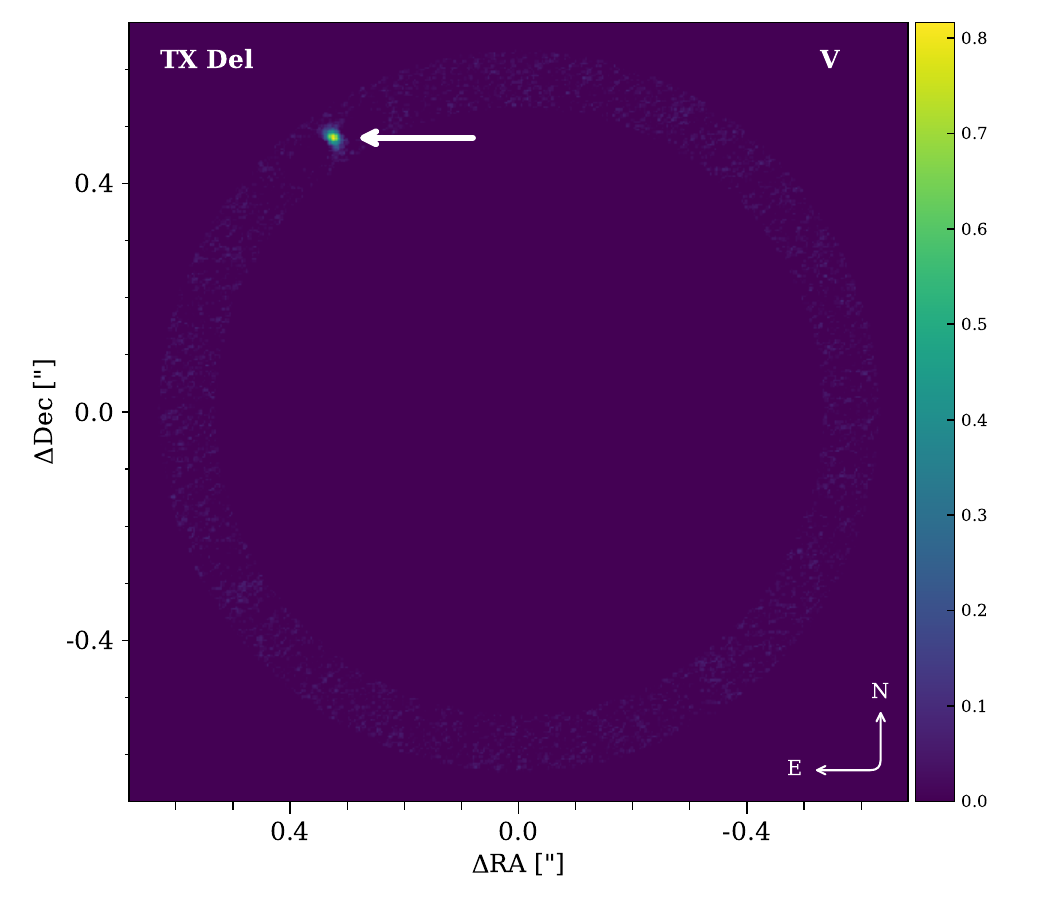}}
        \resizebox{0.6666\hsize}{!}{\includegraphics{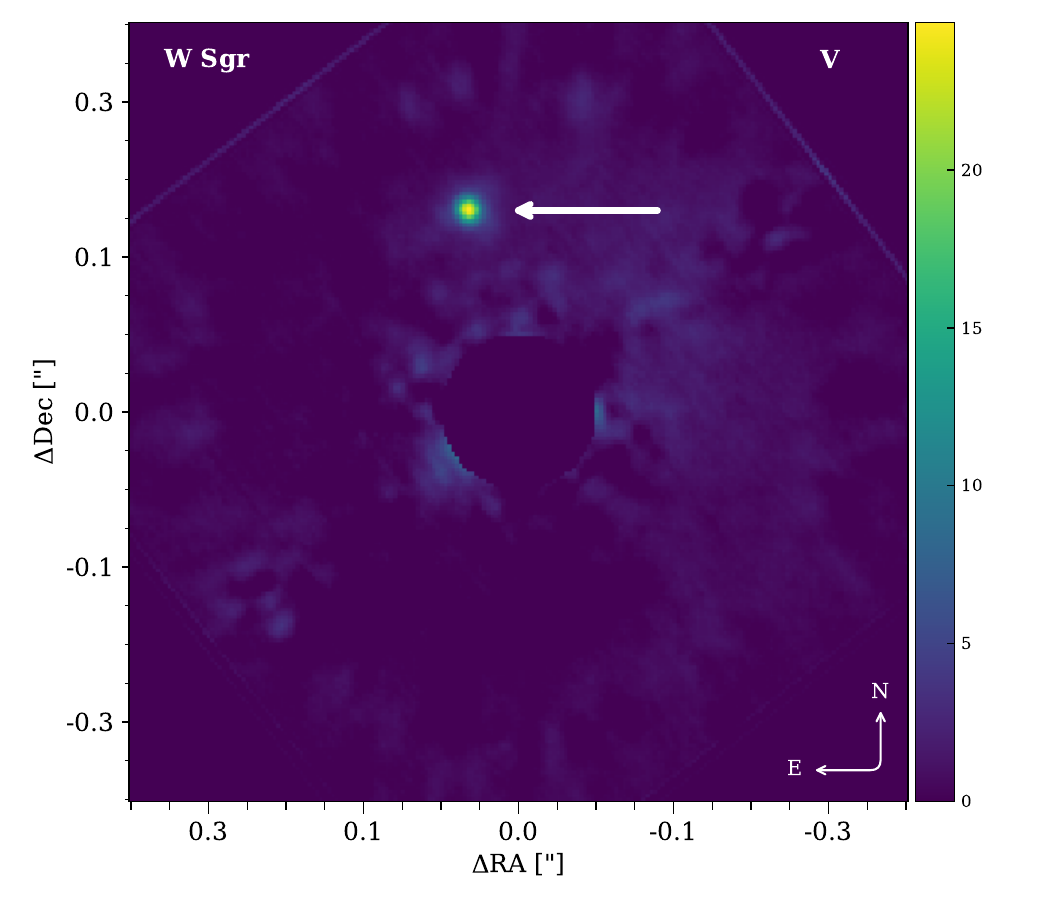}\includegraphics{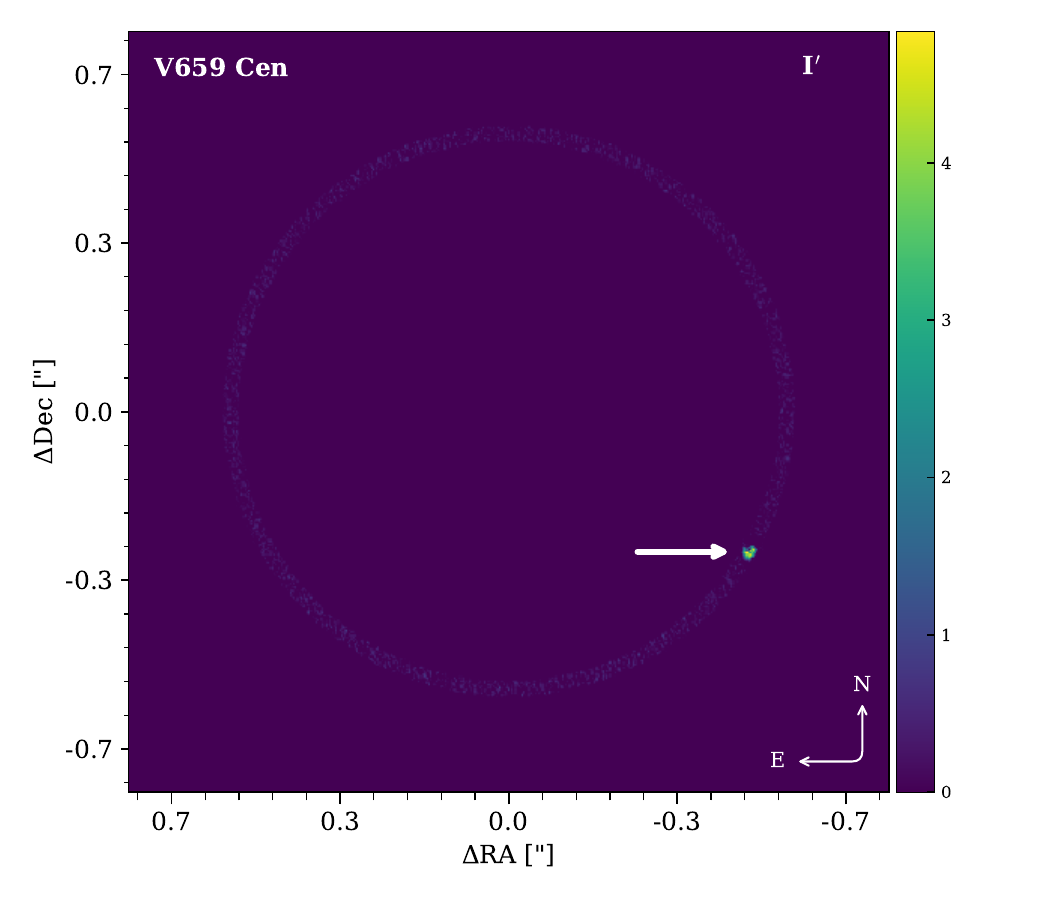}}
        \caption{PCA analysis of our Cepheids with detected companion in $V$, except for V659~Cen, which is in $I^\prime$ (see text).}
        \label{fig__detected_companions}
 \end{figure*}

We detected companions with $S/N > 5$ for only 8 Cepheids (i.e. 17\,\% of our sample). The PCA-ADI frames with the detected companion are displayed in Fig.~\ref{fig__detected_companions}. The fluxes and astrometry are determined in the next section. When no companions were detected, we estimated the detection limits according to the procedure explained in the following section.
 
        \begin{table}[!ht]
        \centering
        \caption{Detection level ($n\sigma$) of the injected companions for our three PCA-ADI approaches and two different flux ratios.}
        \begin{tabular}{c|ccc|ccc}
                \hline
                \hline
                & \multicolumn{3}{c|}{S/N} & \multicolumn{3}{c}{S/N} \\
                & \multicolumn{3}{c|}{$f = 0.1$\,\%} & \multicolumn{3}{c}{$f = 0.02$\,\%} \\
                &   $\#1$   &   $\#2$   &   $\#3$  &   $\#1$   &   $\#2$   &   $\#3$ \\
                \hline
                \#1     (0.1'') & 1.2     & 3.9     & 3.2 & 0.2   & 2.2  & 2.1 \\
                 \#2    (0.2'') & 7.7   & 8.2  & 6.2 &  3.7   & 4.3  & 2.9 \\
                \#3 (0.5'')     & $> 8.2$  & $> 8.2$  & $> 8.2$ & $> 8.2$ & 8.1  & 5.1 \\
                \#4     (1.0'') & $> 8.2$  & $> 8.2$  & $> 8.2$ & $> 8.2$ & $> 8.2$ & $> 8.2$ \\
                \#5     (1.5'') & $> 8.2$  & $> 8.2$  & $> 8.2$ & $> 8.2$ & $> 8.2$ & $> 8.2$ \\
                \hline
        \end{tabular}
        \label{table__snr_pca_test}
 \end{table}
 
        \begin{table}[!ht]
        \centering
        \caption{S/N of the detected companions around our Cepheids in all filters with the annular PCA analysis.}
        \begin{tabular}{c|ccc}
                \hline
                \hline
                & \multicolumn{3}{c}{S/N} \\
                &   $V$   &   $R^\prime$   &     $I^\prime$   \\
                \hline
                $\eta$~Aql       &  39.5   &  55.3  & 16.1  \\
                V659-Cen        &  --         &  33.5  & 44.1  \\
                AX~Cir                          &  18.9      &  21.8  & 16.2    \\
                TX~Del                   &  28.7    &  26.1   & 23.6  \\
                S~Nor                    &  35.3   &  28.9   & 18.6  \\
                AP~Pup                     &  11.9    & 14.4  & 25.0   \\
                W~Sgr               &  79.0 &  45.9 & 15.2  \\
                T~Vel                    &  48.7    &  33.1   & 37.1  \\
                \hline
        \end{tabular}
        \label{table__snr_pca_annular}
 \end{table}
 
 \subsection{Position and flux estimate of detected companions}
 
 We applied the method available in \vip\ using the negative fake companion technique \citep[NEGFC;][]{Lagrange_2010_07_0,Marois_2010_07_0} for the determination of the flux and position of the companion, coupled with Monte Carlo methods to estimate uncertainties. Briefly, the NEGFC technique consists of injecting a negative fake companion into the raw data cube with flux and position close to the expected values, performing a PCA on an annulus passing through the expected companion position and extracting the flux on a circular aperture centred at the position of the companion (chosen as three times the FWHM of the PSF). In our study, these steps were  iteratively repeated to minimise the total flux within the circular aperture, with the companion parameters defined by the \texttt{emcee} affine invariant algorithm \citep{Foreman-Mackey_2013_03_0} at each iteration. The astrometric positions are reported in Table~\ref{table__positions_fluxes}, where we took the mean value between the three filters and the corresponding standard deviation as uncertainty. However, due to the limited field rotation in our images, self-subtraction may lead to an underestimation of the flux. To obtain a more accurate flux measurement, we performed the  aperture photometry for both the Cepheids and their companions using the median image of the de-rotated cube. The resulting flux ratios between the companions and the Cepheids are also presented in Table~\ref{table__positions_fluxes}. Finally, position angles were corrected for the ZIMPOL true north offset of $-2^\circ$ \citep{Schmid_2017_06_0,Schmid_2018_11_0,Engler_2018_10_0}.
 
 \begin{table*}[!ht]
        \centering
        \caption{Position and flux ratios of the detected companions.}
        \begin{tabular}{c|ccccc|c}
                \hline
                \hline
                &    $\rho$  &  $\theta$   &   $f_V$  &  $f_{R^\prime}$    &   $f_{I^\prime}$ &  Sp.~Type\\
                &        (mas)         &  ($^\circ$)       &  (\%)  &      (\%)  &  (\%) & \\
                \hline
                $\eta$~Aql &     $639.4\pm0.3$      &  $91.57\pm0.09$   &   $0.42\pm0.06$    &     $0.41\pm0.01$    &    $0.32\pm0.01$  & A5V-F3V  \\
                V659~Cen  &     $597.3\pm1.1$      &  $238.3\pm0.4$   &   --    &     --    &    --   & --  \\
                AX~Cir        &      $309.9\pm0.3$   &    $368.6\pm0.1$      &   $7.2\pm0.1$      &    $6.12\pm0.02$   &     $4.58\pm0.02$  & B6V  \\
                TX~Del            &    $573.8\pm0.9$     &   $32.2\pm0.4$    &   $0.007\pm0.002$  &  $0.011\pm0.002$   &  $0.015\pm0.001$ & M0V-M1V  \\
                S~Nor              &    $907.3\pm0.9$     &   $258.3\pm0.4$   &   $0.275\pm0.007$      &   $0.247\pm0.004$   &  $0.236\pm0.004$ & A3V-A4V \\
                AP~Pup      &     $159.3\pm0.2$      &    $165.0\pm0.4$ &   $0.048\pm0.003$    &    $0.053\pm0.009$   &  $0.073\pm0.012$   & F9V-G2V  \\
                W~Sgr            &    $192.1\pm1.0$      &  $12.3\pm0.5$     &   $0.284\pm0.013$       &  $0.262\pm0.007$  &   $0.241\pm0.007$ & F0V-F2V \\
                T~Vel                   &    $716.5\pm0.8$     &   $43.2\pm0.4$     &    $0.022\pm0.003$ &   $0.025\pm0.005$    &   $0.027\pm0.004$ &    K2-K3V   \\
                \hline
        \end{tabular}
        \label{table__positions_fluxes}
        \tablefoot{$\rho$  amd  $\theta$ are the relative separation of the companion w.r.t. the Cepheid and its position angle, while $f_\mathrm{X}$ denotes the measured flux ratio at band $X$. Sp.~Type is our derived spectral type based on the flux ratio (see Sect.~\ref{subsection__detected_companions}).
        }
 \end{table*}
 
 \subsection{Detection limits}
 \label{section__detection_limits}
 
\begin{figure*}[!h]
        \centering
        \resizebox{\hsize}{!}{\includegraphics{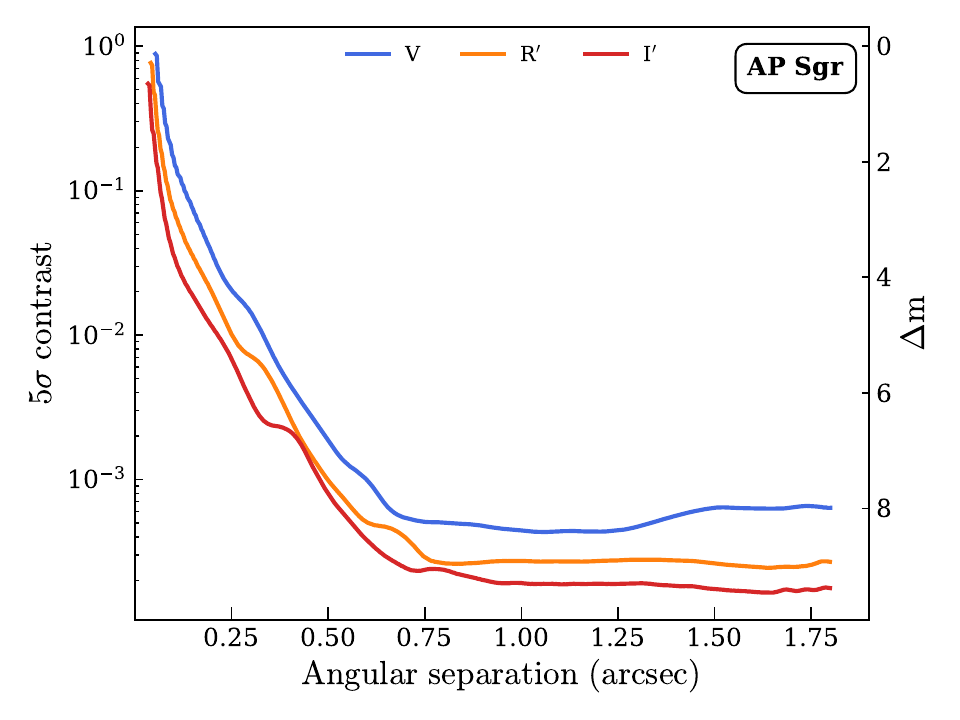}\includegraphics{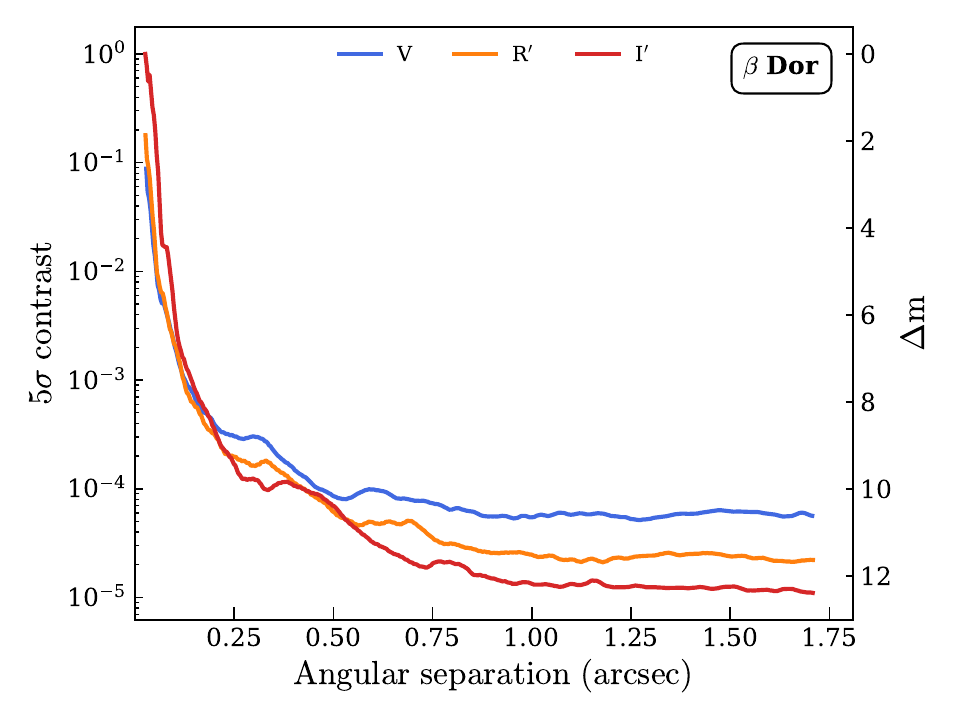}\includegraphics{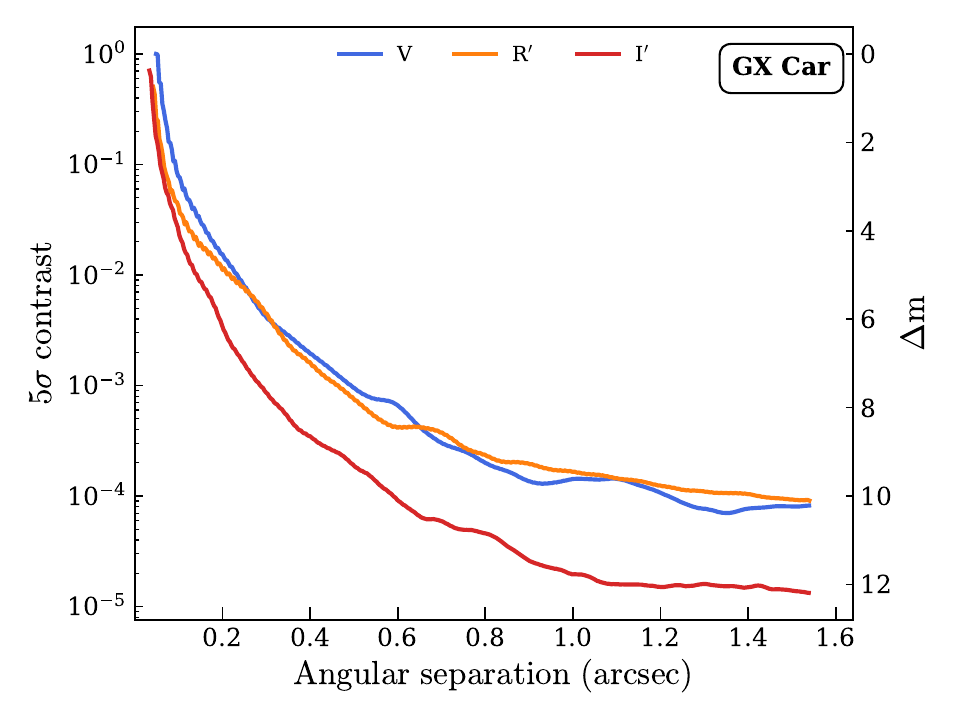}}
        \resizebox{\hsize}{!}{\includegraphics{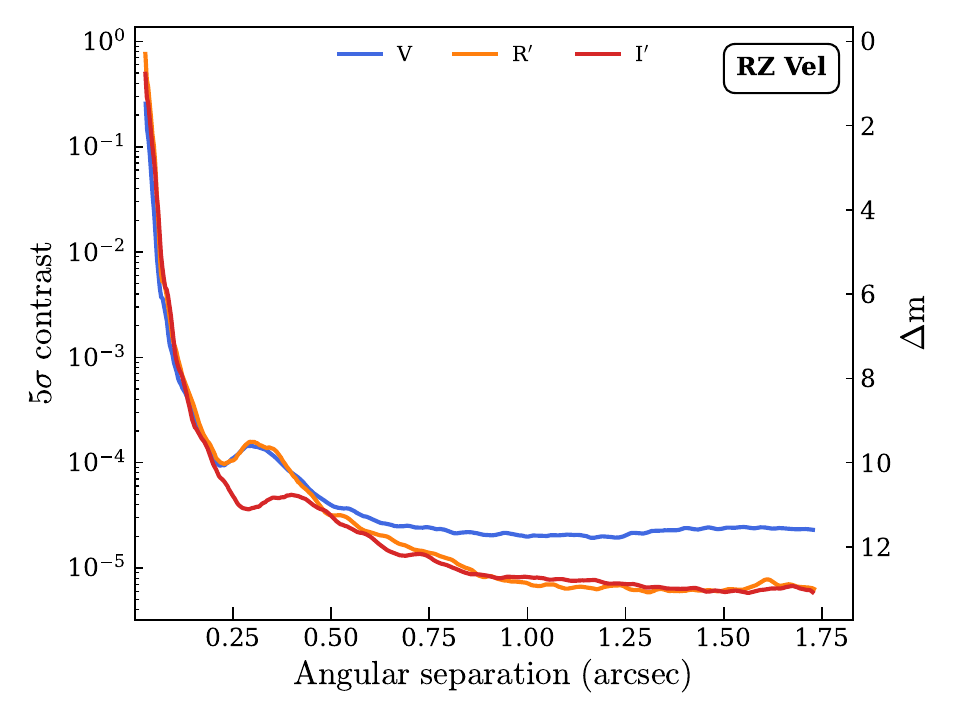}\includegraphics{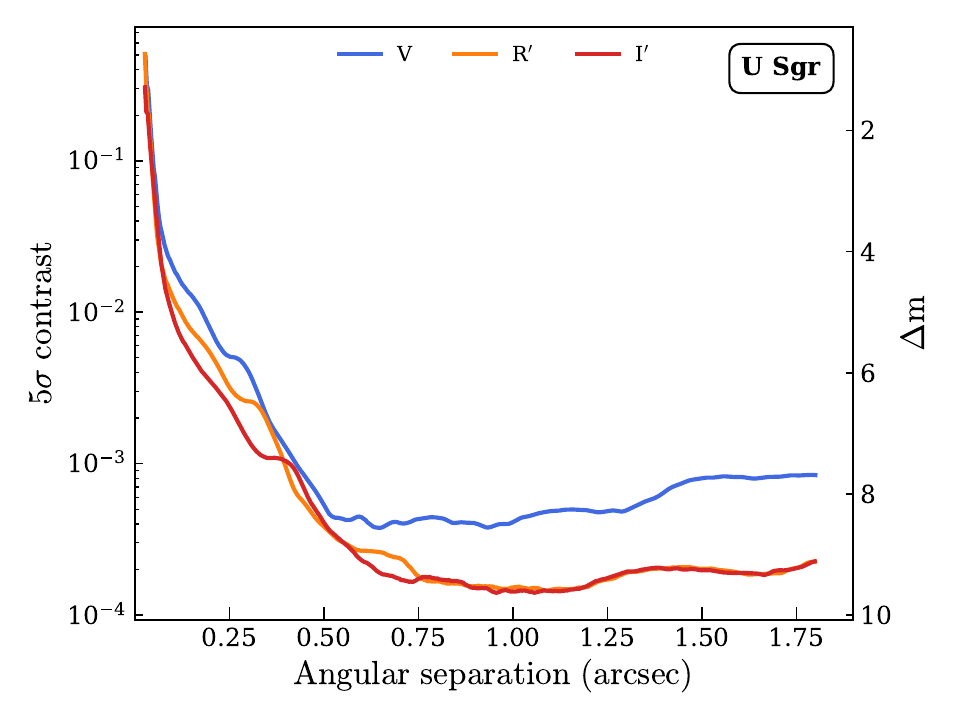}\includegraphics{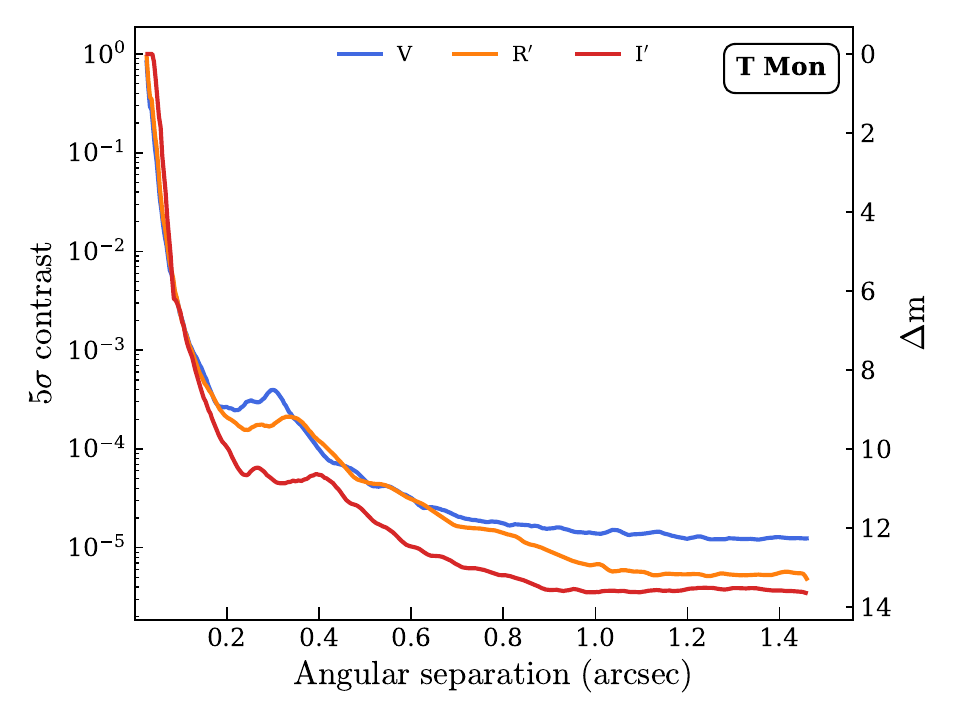}}
        \caption{$5\sigma$ contrast curves of six Cepheids of our sample with no wide companion detected.}
        \label{fig__contrast_curve}
 \end{figure*}

 For Cepheids where no companions were detected, we employed the \vip\ \texttt{contrast\_curve} function to produce contrast curves by injecting synthetic companions, assessing noise and throughput following PCA-based PSF subtraction, and determining the $5\sigma$ detectable flux ratio as a function of separation. A Student-t correction, as outlined by \citet{Mawet_2014_09_0}, was applied to account for small-sample statistics. In Fig.~\ref{fig__contrast_curve}, we display the contrast curve for eight Cepheids.
 
 \section{Discussion}
 \label{section__discussion}

In our sample, 41 out of 46 Cepheids (89\,\%) are confirmed binary or triple systems \citep{Szabados_2003_03_0}, with one additional Cepheid, BQ Ser, considered a suspected binary \citep{Gorynya_1996_01_0}. This aligns with the works from \citet{Kervella_2019_03_0}, where we estimated a binary fraction exceeding 80\,\% by analysing proper motion anomalies using combined \textsc{Hipparcos} and \textit{Gaia} DR2 positional data. However, in this study, we only identify a companion for eight of these Cepheids, indicating that the majority of them are likely spectroscopic binaries.

Three Cepheids have a visual companion, according to \citet{Szabados_2003_03_0}: V339~Cen, AX~Cir and RY~Sco. In the case of V339~Cen, no evidence of a companion is found in the existing literature, whether through spectroscopy or astrometry \citep{Kervella_2019_03_2}. Furthermore, our observations similarly show no visual component within 1.7\arcsec. AX Cir forms a triple system, exhibiting both a spectroscopic companion  \citep[e.g.][]{Petterson_2004_05_0,Gallenne_2014_01_0} and a visual companion \citep{Evans_2020_12_0}, with the latter's wider component having been confirmed by our SPHERE observations. For RY~Sco, its visual companion, positioned at $\sim2.2$\arcsec\ from the Cepheid, falls outside our field of view (FoV).

Additional candidate visual companions for V496~Aql, V350~Sgr, BB~Sgr, and RV~Sco, which are not mentioned in \citet{Szabados_2003_03_0}, were identified by \citet{Evans_2020_12_0}. However, these are situated beyond 2\arcsec\ from the Cepheid, placing them outside our FoV. \citet{Evans_2020_12_0} also reported visual companions within 2\arcsec\ for S~Nor, $\eta$~Aql, and AX~Cir, which we confirm in this study. We also report the new discovery of a wide companion associated with the Cepheids AP~Pup, T~Vel, and TX~Del.

The probability that any of the detected companions is a chance alignment with a field star is expected be low given the small FoV of our observations ($\sim 3.5\arcsec$). To quantify this, we queried the \textit{Gaia} DR3 catalogue for sources brighter than $G = 18$\,mag within a radius of 50\arcsec\ around each Cepheid hosting a detected companion and used the resulting counts to estimate the local surface density of field stars. Assuming Poisson statistics, we converted this density into the probability of finding at least one field star within a radius of 1\arcsec \ (i.e. the region encompassing all of our detected companions). For $\eta$~Aql, for example, we found 12 stars, corresponding to a density of 0.0015\,arcsec$^{-2}$, which implies a probability $< 0.5$\,\% that the detected companion is an unrelated field star. For the remaining Cepheids, the corresponding probabilities are $< 2$\,\%, with the exception of W~Sgr, for which we obtain $\sim 6$\,\%. However, when evaluated at the measured separation of the companion (0.2\arcsec), the probability that the detected companion is a field star physically unrelated to the Cepheid decreases to $\sim 0.2$\,\%.
 
 \subsection{Detected companions}
  \label{subsection__detected_companions}
  
In this section, we explore the companions we  identified and provide additional details regarding their specific characteristics. Unfortunately, performing absolute photometry was not possible due to the lack of photometric zero-point data for our filter bandpasses, which would only allow  for the measurement of flux ratios between the Cepheid and its companion. However, the Cepheid's magnitude at a specific phase can be estimated using the \cite{Berdnikov_2008_04_0} light curves, available in standard Johnson-Cousins filters. Although the transmission profiles of these filters differ slightly from those of SPHERE, they are sufficiently similar for practical use. By integrating stellar atmosphere models over the filter bandpasses across various temperature ranges, we determined a maximum magnitude difference of 0.13 mag at the lowest temperature (4500 K), which decreases with increasing temperature. Consequently, to estimate the companion's magnitude, we utilised the measured flux ratios combined with the expected Cepheid magnitude at the corresponding pulsation phase\footnote{estimated from a spline interpolation.}, as derived from the Berdnikov light curves, adopting an uncertainty of 0.1 mag.

To estimate the de-reddened magnitudes, we adopted the reddening law of \citet{Fouque_2007_12_0}, i.e. $R_\mathrm{V} = 3.23, A_\mathrm{V} = R_\mathrm{V} E(B-V), A_\mathrm{R_\mathrm{C}} = 0.645 A_\mathrm{V}$ and $A_\mathrm{I_\mathrm{C}} = 0.608 A_\mathrm{V}$. All uncertainties were propagated til the final results. The phases at our observing epoch were calculated using the ephemeris from \cite{Csornyei_2022_04_0}, and propagating the rate of pulsation change for $\eta$~Aql, AX~Cir, AP~Pup, W~Sgr, and T~Vel. For TX~Del, we used the ephemeris from \textit{Gaia} DR3 and the rate of period change from \citet{Percy_2000_01_0}, and from \citep{Trahin_2021_12_5} for S~Nor. 

\paragraph{$\eta$~Aql:} The wide companion of this Cepheid was already reported, and our observations of the companion’s position align with those reported by  \citet{Evans_2020_12_0} and \citet{Gallenne_2014_07_0}. The companion exhibited a positional shift of 4.1$^\circ$ between the initial and final epochs. Assuming a circular, face-on orbit, this translates to an angular velocity of approximately $0.546^\circ\,yr^{-1}$, suggesting a minimum orbital period of about 659\,yr.

From the light curves, the Cepheid magnitude at the pulsation phase 0.61 is $V = 3.9\pm0.1$\,mag, $R^\prime = 3.4\pm0.1$\,mag and $I^\prime = 3.0\pm0.1$\,mag (prime indices are omitted for simplicity and clarity). With a colour excess $E(B-V) = 0.16\pm0.02$ \citep{Trahin_2021_12_5} and our flux ratios, we calculated the companion's de-reddened magnitudes $V_0 = 9.4\pm0.2$\,mag, $R^\prime_0 = 9.1\pm0.1$\,mag and $I^\prime_0 = 9.0\pm0.1$\,mag. To compute the absolute magnitudes, an accurate distance to this bright star is required. Distance estimates from \textit{Gaia} ($270\pm14$\,pc\footnote{All parallaxes were corrected from the parallax offset \citep{Lindegren_2021_05_7}}), \citet[][$276\pm55$\,pc]{Kervella_2004_03_0} or \citet[][$252\pm27$\,pc]{Gieren_2018_12_0} are mutually consistent, providing a reliable accuracy. We adopted a mean distance of $d = 266\pm10$\,pc calculated from these sources. This results in absolute magnitudes of $M_\mathrm{V} = 2.2\pm0.2$\,mag, $M_\mathrm{R^\prime} = 1.9\pm0.1$\,mag, and $M_\mathrm{I^\prime} = 1.8\pm0.1$\,mag. Assuming the companion is a main sequence star and using the spectral type calibration from \citet{Pecaut_2013_09_0}, it would be classified between a A5V and a F3V star with a mass within $1.4-1.9\,M_\odot$. This is in agreement with our previous works using VLT/NACO \citep{Gallenne_2014_07_0}. This seems to also confirm the existence of a closer hotter companion, detected in an IUE spectrum, with a spectral type of B9.8V \citep{Evans_1991_05_0}. We will discuss this specific case in another publication combining different observing techniques (Gallenne et al., in prep.).

\paragraph{V659~Cen:} This star is member of a triple system comprising a spectroscopic binary and a wider component. \citet{Evans_2020_12_0} reported an approximate separation of 0.6\arcsec\ and a position angle of $\sim235^\circ$ for the wide companion using HST/WFC3 observations. Unfortunately, our SPHERE observations were hampered by poor observing conditions and sub-optimal adaptive optics corrections, preventing detection of the wide component in the $V$ band due to insufficient flux. However, in the $R^\prime$ and $I^\prime$ filters, we obtained sufficient flux to accurately measure the companion's astrometric positions; however it was not high enough for accurate photometry. Our measurement is in very good agreement with \citet{Evans_2020_12_0} and with the most recent position we determined in \citet[][$\rho \sim 0.590\arcsec$ and $PA = 237.8^\circ$]{Evans_2025_10_0} from the HST/STIS acquisition exposure. Assuming a circular orbit with low inclination, we estimated a minimum orbital period of approximately 749\,years based on the angular velocity between 2011 (HST) and 2018 (SPHERE).

\paragraph{AX~Cir:} This complex system comprises at least three components. The spectroscopic component was spatially resolved through long-baseline interferometry \citep{Gallenne_2014_01_0}, yielding a semi-major axis of approximately 27\,mas and an orbital inclination of about $\sim 55^\circ$ (Gallenne et al., in prep.). The $H$-band interferometric flux ratio indicates a B9V star. A preliminary analysis of the companion’s velocity from the STIS spectrum suggests it may itself be a binary. Additionally, a wide companion at approximately 0.3\arcsec\ was detected via speckle interferometry \citep[][$\rho = 0.313\arcsec$ and $PA = 346.9^\circ$]{Tokovinin_2019_07_0} and HST imaging \citep[][$\rho = 0.3\arcsec$ and $PA = 332.4^\circ$]{Evans_2020_12_0}. This component is the hottest one in the system and corresponds to the B6V star previously detected from IUE spectrum  \citep{Evans_1994_11_0}. Our SPHERE astrometry aligns well with these prior measurements and confirms orbital motion. The observed angular velocity of $5.1^\circ~yr^{-1}$ between the HST observations (2011) and ours (2018) suggests a minimum orbital period of approximately 70\,years.

Using the Cepheid light curves, we derived companion magnitudes of $V = 9.0\pm0.1$\,mag, $R^\prime = 8.6\pm0.1$\,mag and $I^\prime = 8.5\pm0.1$\,mag. Adopting a colour excess of $E(B-V) = 0.31\pm0.01$ \citep{Trahin_2019_11_0} and a \textit{Gaia} distance $d = 505\pm63$\,pc, we calculated absolute magnitudes of $M_\mathrm{V} = -0.6\pm0.3$\,mag, $M_\mathrm{R^\prime} = -0.6\pm0.3$\,mag and $M_\mathrm{I^\prime} = -0.7\pm0.3$\,mag. These values are consistent with the B6V spectral type reported by \citet{Evans_1994_11_0}.

 \paragraph{TX~Del:} This star is classified as a type II Cepheid \citep{Harris_1981_05_1,Balog_1995_01_0,Laney_1995_01_0}  and has been reported to be a spectroscopic binary \citep{Harris_1989_09_0}. The large  re-normalised unit weight error (RUWE) parameter of \textit{Gaia} \citep{Gaia-Collaboration_2018_08_0} and the proper motion anomaly \citep{Kervella_2019_03_2} confirm its binary nature.  \citet{Harris_1989_09_0} published the orbital elements, with a period of 133 days. Adopting the \textit{Gaia} distance of 1102\,pc, a mass of $3.5\,M_\odot$ for the Cepheid \citep{Kervella_2019_03_2} and assuming at most an equal mass for the companion, the maximum angular separation would be $\sim 0.9$\,mas. This is well below the spatial resolution of SPHERE.
 
 Here, we  report the astrometric detection of a third companion at an angular separation of 574\,mas. Assuming the same mass for the companion as the Cepheid, this would provide from the Kepler law an approximate orbital period of 6000 years. From the light curves of Berdnikov and our measured flux ratios, we calculated apparent magnitudes of $V = 19.4\pm0.3$\,mag, $R^\prime = 18.3\pm0.2$\,mag, and $I^\prime = 18.2\pm0.2$\,mag. Adopting a colour excess of $E(B-V) = 0.06\pm0.02$ \citep{Lallement_2018_08_4} and the \textit{Gaia} distance, we calculated the absolute magnitudes of $M_\mathrm{V} = 9.0\pm0.3$\,mag, $M_\mathrm{R^\prime} = 7.9\pm0.2$\,mag and $M_\mathrm{I^\prime} = 7.9\pm0.2$\,mag. This would correspond to a M0-M1V companion.
 
\paragraph{S~Nor:} This Cepheid is known to have a confirmed wide, resolved companion and there is a possible close-in spectroscopic companion. \citet{Szabados_1989_01_0} reported evidence of an inner companion, but orbital motion remains largely undetected. Recent high-precision radial velocity measurements \citep{Shetye_2024_10_0,Gallenne_2019_02_0} show no significant variations in systemic velocity, casting doubt on the presence of this close-in companion. Additionally, interferometric observations by \citet{Gallenne_2019_02_0} detected no companion brighter than a B7V star within 50 mas. The wide companion, located at $\sim 0.9\arcsec$, has been detected both spectrally and spatially. \citet{Evans_1992_04_0} identified a B9.5V companion through an IUE spectrum, consistent with the photometric detection reported by \citet{Evans_2020_12_0}. We calculated a very small angular motion of $\sim 0.12^\circ$ between our 2018 measurement and the HST (2011), suggesting a minimum orbital period of about 2951\,years. 

We estimated apparent magnitudes of $V = 12.5\pm0.1$\,mag, $R^\prime = 12.2\pm0.1$\,mag and $I^\prime = 11.8\pm0.1$\,mag. Adopting $E(B-V) = 0.29\pm0.02$ \citep{Trahin_2021_12_5} and the \textit{Gaia} distance of $910\pm18$\,pc, we computed absolute magnitudes of $M_\mathrm{V} = 1.8\pm0.1$\,mag, $M_\mathrm{R^\prime} = 1.8\pm0.1$\,mag and $M_\mathrm{I^\prime} = 1.4\pm0.2$\,mag. This would correspond to a A3V star, which is slightly colder than the B9.5V estimate from \citet{Evans_1992_04_0}. A more accurate determination of the spectral type will require multi-wavelength observations.

\paragraph{AP~Pup:} This Cepheid is considered a potential spectroscopic binary based on orbital motion evidence noted by \citet{Szabados_1989_01_0}. The IUE spectrum of \citet{Evans_1992_04_0} provided an upper limit spectral type of A3V. The imaging survey by \citet{Evans_2020_12_0} failed to detect the companion, likely due to limited spatial resolution. However, \citet{Kervella_2019_03_0}  identified a proper motion anomaly, indicating a close companion with a maximum separation of about 300\,mas and an orbital period up to 3100\,years. We confirm the detection of this companion. We derived apparent magnitudes of $V = 15.9\pm0.1$\,mag, $R^\prime = 15.3\pm0.2$\,mag and $I^\prime = 14.5\pm0.2$\,mag. Using the \textit{Gaia} distance of $1083\pm23$\,pc and $E(B-V) = 0.355\pm0.02$ \citep{Lallement_2018_08_4}, we estimated absolute magnitudes $M_\mathrm{V} = 4.6\pm0.2$\,mag, $M_\mathrm{R^\prime} = 4.4\pm0.2$\,mag and $M_\mathrm{I^\prime} = 3.6\pm0.2$\,mag. This would correspond to a spectral type in the range F9V-G2V.

 \paragraph{W~Sgr:} Initially proposed as a spectroscopic binary by \citet{Jacobsen_1974_08_0}, this Cepheid was later confirmed by \citet{Babel_1989_06_0}, who determined an orbital period of 1780 days and a high eccentricity of 0.52. \citet{Petterson_2004_05_0} refined these parameters using more precise RV data, establishing a period of 1580 days and an eccentricity of 0.4. \citet{Benedict_2007_04_0} detected the companion's astrometric motion using the HST Fine Guidance Sensor, deriving full orbital elements, including a semi-major axis of 12.9\,mas. \citet{Evans_2009_03_0} set a companion spectral type upper limit of F5V from HST/STIS observations. In \citet{Gallenne_2019_02_0}, we failed to detect the companion within 50 mas of the Cepheid using long-baseline interferometry, as the data were only sensitive to companions earlier than B8V. The companion's proximity prevents detection with VLT/SPHERE.
 
 \citet{Morgan_1978_06_0} identified a wider companion, initially mistaken for the spectroscopic companion, but its position at 0.116\arcsec\ was inconsistent with the orbital period reported by \citet{Babel_1989_06_0}. This companion aligns with a hotter star detected by \citet{Bohm-Vitense_1985_09_1} using IUE spectra, with an estimated temperature of 9400\,K, corresponding approximatively to an A1V spectral type. HST/STIS observations by \citet{Evans_2009_03_0} confirmed this wide companion as the hotter component, with a spectral type of A0V, located at a separation of $\rho = 0.1645\pm0.0006$ and $PA = 210.0\pm0.5^\circ$. This is the companion detected in our observations, which suggests (according to the angular motion) a minimum period of about 36\,yr. Finally, \citet{Evans_2016_05_0} also identified a possible third companion, located at 6.3\arcsec.
  
 We calculated the apparent magnitudes for the second companion detected by SPHERE of $V = 11.4\pm0.1$\,mag, $R^\prime = 11.0\pm0.1$\,mag and $I^\prime= 10.7\pm0.1$\,mag. Using the distance of $428\pm11$\,pc, based on the mean and standard deviation between \citet{Benedict_2007_04_0} and \textit{Gaia} parallaxes, along with $E(B-V) = 0.15\pm0.02$ \citep{Trahin_2021_12_5}, we estimated absolute magnitudes of $M_\mathrm{V} = 2.8\pm0.1$\,mag, $M_\mathrm{R^\prime} = 2.6\pm0.1$\,mag, and $M_\mathrm{I^\prime} = 2.2\pm0.1$\,mag, consistent with an F1V main sequence star. This is somewhat cooler than the A0-A1V classifications inferred from IUE and STIS spectroscopy. Given the different wavelength ranges and methods applied, small uncertainties in reddening, flux calibration, and contamination by the Cepheid can easily translate into shifts across several sub-classes, so we regard our estimate as broadly consistent with an early-type companion. We note also that this companion was not detected by \citet{Tokovinin_2020_07_0} from speckle interferometry, who set a detection limit of $\Delta I\sim 2.9$\,mag. This is well below the $\sim 6$\,mag expected to detect an A0V star. This also raises doubts about the detection  about 45 years ago by \citet{Morgan_1978_06_0} with the same technique. 
 
 \paragraph{T~Vel:}
 
 This short-period Cepheid has long been suspected of binarity based on slow systemic velocity shifts seen across multi-epoch radial-velocity campaigns. Early works have already flagged it as a possible spectroscopic binary with a red companion \citep{Gieren_1985_08_0}. No widely separated, hot UV-bright companion has been confirmed, pointing to a relatively close, faint companion instead. In the \textit{Gaia} era,  \cite{Kervella_2019_03_0} reported no significant proper motion anomaly at the \textit{Gaia} DR2 epoch.
 
 We report the detection of a relatively faint companion located at $\sim 717$\,mas from the Cepheid. We calculated its apparent magnitude of $V = 17.4\pm0.2$\,mag and $I^\prime = 16.0\pm0.2$\,mag. \cite{Berdnikov_2008_04_0}  does not contain photometric measurements in $R_\mathrm{C}$, we therefore interpolated the $V$ and $I_\mathrm{C}$ band light curve at 700\,nm to estimate $R^\prime = 16.6\pm0.2$\,mag. Using the \textit{Gaia} distance of $1064\pm18$\,pc and $E(B-V) = 0.29\pm0.02$ \citep{Trahin_2019_11_0}, we estimated the absolute magnitudes as $M_\mathrm{V} = 6.3\pm0.2$\,mag, $M_\mathrm{R^\prime} = 5.9\pm0.2$\,mag, and $M_\mathrm{I^\prime} = 5.3\pm0.2$\,mag. This would correspond to a K2-K3V companion.

 \subsection{Detection limits}

 For the remaining 39 Cepheids without a detected companion, we derived $5\sigma$ contrast limits for any unseen component, as described in Sect.~\ref{section__detection_limits}. Examples of these contrast curves are shown in Fig.~\ref{fig__contrast_curve} and the limits at various separations are given in Table~\ref{table__detection_limits}, together with the corresponding upper limits on the spectral type of an undetected main sequence companion. For the conversion, we also used the $V$-band light curves from \citet{Berdnikov_2008_04_0} and de-reddened the magnitudes with colour excesses from the literature \citep[e.g.][]{Csornyei_2022_04_0,Trahin_2021_12_5,Groenewegen_2020_03_0,Trahin_2019_11_0,Turner_2016_10_0,Fernie_1995_01_0}. The upper limits on the companion spectral types were then obtained from the corrected \textit{Gaia} parallaxes and the spectral-type calibration of \citet{Pecaut_2013_09_0}, as in the previous sections.
 
 We achieved maximum contrasts of approximately 10, 11, and 12\,mag at separations of 0.25\arcsec, 0.50\arcsec, and $> 1$\arcsec, respectively, while the mean contrasts over the whole sample are 6.1, 8.6, and 9.2 mag at these separations. For some Cepheids, at separations larger than 0.5\arcsec, we can exclude the presence of companions as late as K-type main sequence stars.
 
  \begin{table*}[!ht]
        \centering
        \caption{$5\sigma$ contrasts and upper limits spectral type for undetected main sequence companions.}
        \begin{tabular}{c|cc|cc|cc||c|cc|cc|cc}
                \hline
                \hline
                        Star    & \multicolumn{2}{c|}{$0.25\arcsec$} & \multicolumn{2}{c|}{$0.50\arcsec$} & \multicolumn{2}{c||}{$> 1\arcsec$ } &         Star    & \multicolumn{2}{c|}{$0.25\arcsec$} & \multicolumn{2}{c|}{$0.50\arcsec$} & \multicolumn{2}{c}{$> 1\arcsec$ } \\
                                                &  $\Delta V$ &  SpT & $\Delta V$ &  SpT & $\Delta V$ &  SpT  & &  $\Delta V$ &  SpT & $\Delta V$ &  SpT & $\Delta V$ &  SpT \\
                 \hline
FF~Aql & 5.5 & A3V & 7.5 & F6V & 8.5 & G2V & RV~Sco & 5.8 & F0V & 8.2 & G4V & 7.9 & G1V \\
FN~Aql & 5.5 & A3V & 7.5 & F6V & 8.5 & G2V & RY~Sco & 6.2 & A3V & 8.7 & F8V & 8.8 & F9V \\
KL~Aql & 2.4 & B1.5V & 4.4 & B7V & 7.1 & A9V & V482~Sco & 6.5 & F2V & 9.4 & K1V & 10.0 & K3V \\  
V496~Aql & 6.1 & F0V & 9.3 & K0V & 9.6 & K1V & V636~Sco & 5.9 & A6V & 8.7 & G4V & 9.6 & K1V \\  
V916~Aql & 3.8 & B3V & 6.1 & A0V & 6.1 & A1V & RU~Sct & 5.8 & A0V & 8.6 & F7V & 9.5 & G2V \\  
V1344~Aql & 6.4 & A9V & 9.6 & K0V & 9.9 & K1V & Y~Sct & 5.8 & A3V & 8.6 & G0V & 8.7 & G1V \\  
V340~Ara & 8.5 & F5V & 10.1 & G5V & 10.8 & G9V & S~Sge & 5.5 & A3V & 7.9 & F9V & 8.0 & F9V \\ 
GX~Car & 5.3 & F1V & 7.5 & G6V & 10.0 & K6V & BQ~Ser & 8.9 & G9V & 10.0 & K3V & 10.9 & K6V \\ 
U~Car & 8.9 & F5V & 10.2 & G1V & 10.7 & G7V & AP~Sgr & 4.2 & A2V & 6.8 & F8V & 8.2 & G8V \\
UW~Car & 6.3 & F1V & 9.0 & G9V & 10.3 & K4V & BB~Sgr & 5.3 & A3V & 7.9 & F9.5V & 7.7 & F8V \\   
V339~Cen & 7.5 & F5V & 10.0 & K0V & 10.5 & K3V & U~Sgr & 5.7 & A7V & 8.2 & G2V & 8.0 & G0V \\   
RZ~CMa & 5.1 & A7V & 8.3 & G8V & 8.6 & K0V & V350~Sgr & 6.3 & A9V & 9.2 & G9V & 9.9 & K2V \\ 
$\beta$~Dor & 8.8 & G3V & 10.2 & K2V & 10.6 & K3V & WZ~Sgr & 5.7 & B9V & 8.6 & F4V & 8.9 & F5V \\   
T~Mon & 8.8 & F2V & 10.8 & G5V & 12.2 & K3V & X~Sgr & 5.5 & A1V & 8.2 & F7V & 8.4 & F8V \\   
TX~Mon & 4.7 & B9.5V & 7.0 & F2V & 7.0 & F2V & Y~Sgr & 5.8 & A3V & 8.3 & F9.5V & 8.5 & G0V \\  
V465~Mon & 6.2 & A3V & 8.8 & G0V & 10.4 & K1V & YZ~Sgr & 6.1 & A3V & 9.2 & G3V & 9.9 & G9V \\  
BF~Oph & 5.5 & A8V & 8.1 & G4V & 7.6 & F9.5V & LR~Tra & 5.1 & A5V & 7.9 & G2V & 8.0 & G4V \\
Y~Oph & 6.1 & B9.5V & 9.2 & F7V & 9.3 & F8V & R~Tra & 6.2 & F3V & 9.0 & K1V & 8.9 & K0V \\ 
X~Pup & 4.6 & B3V & 7.1 & A2V & 8.5 & F1V & RZ~Vel & 9.9 & G7V & 11.0 & K3V & 11.6 & K4V \\   
V~Vel & 6.1 & A9V & 9.0 & G8V & 9.8 & K2V && & & & & & \\
                \hline
        \end{tabular}
        \label{table__detection_limits}
 \end{table*}
 
 \section{Conclusion}
 \label{section__conclusion}
 
 We present the first VLT/SPHERE survey of bright Galactic Cepheids, aimed at resolving faint visual companions on spatial scales between $\sim 0.02$\arcsec and 3.5\arcsec. Using ZIMPOL imaging in the $V, R^\prime$, and $I^\prime$ bands, we obtained high-quality point-spread functions with a spatial resolution down to $\sim 20$\,mas in $V$. Our observing strategy, combined with a PCA-based post-processing analysis, allowed us to measure accurate relative astrometry and flux ratios for detected companions and to derive stringent $5\sigma$ contrast curves for systems with no detections.
 
 We detected visual companions around eight Cepheids ($\eta$~Aql, V659~Cen, AX~Cir, TX~Del, S~Nor, AP~Pup, W~Sgr, and T~Vel), corresponding to about 17\,\% of our sample. For $\eta$~Aql, AX~Cir, S~Nor, W~Sgr, and V659~Cen we confirmed previously known wide companions and refined their astrometric positions, providing new constraints on their long orbital periods. For AP~Pup, T~Vel, and TX~Del we report the discovery of new wide companions with spectral types ranging from late F to mid-K on the main sequence. These systems highlight the wide range of configurations in which Cepheids are found.
 
 For the remaining 39 Cepheids, we derived $5\sigma$ detection limits as a function of separation, reaching typical maximum contrasts of $\sim 10$, 11, and 12\,mag at 0.25\arcsec, 0.50\arcsec, and $> 1$\arcsec, respectively, with mean contrasts of 6.1, 8.6, and 9.2\,mag across the sample. In several cases, we have been able to exclude, at separations beyond 0.5\arcsec, the presence of main sequence companions as late as K dwarfs. These limits demonstrate that most unresolved companions inferred from radial velocities and proper-motion anomalies must lie either at smaller separations than those probed here or at contrasts below our detection thresholds. Our SPHERE survey therefore provides a crucial high-contrast complement to spectroscopy, interferometry, and \textit{Gaia} astrometry, helping to map the full multiplicity architecture of Galactic Cepheids. Finally, the improved sensitivity and contrast expected from the SPHERE+ upgrade \citep{Boccaletti_2020_03_0} will enable us to extend this type of survey to fainter Galactic Cepheids, further refining the census of their companions and multiplicity properties.

 
 \begin{acknowledgements}
               This work is based on observations collected at the European Southern Observatory under ESO programme 0101.A-0214. 
               AG acknowledges the support of the Agencia Nacional de Investigaci\'on Cient\'ifica y Desarrollo (ANID) through the FONDECYT Regular grant 1241073 and ANID-ALMA fund No. ASTRO20-0059. 
               PK, GP and WG acknowledge funding from the European Research Council (ERC) under the European Union’s Horizon 2020 research and innovation program (project UniverScale, grant agreement 951549). GP acknowledges financial support from  the Polish Ministry of Science and Higher Education with the grant agreement 2024/WK/02. GP and PK acknowledge support from the Polish-French Marie Skłodowska-Curie and Pierre Curie Science Prize awarded by the Foundation for Polish Science. 
               B.P. acknowledges funding from the Polish National Science Center grant SONATA BIS 2020/38/E/ST9/00486. 
               This work has made use of data from the European Space Agency (ESA) mission {\it Gaia} (\url{https://www.cosmos.esa.int/gaia}), processed by the {\it Gaia} Data Processing and Analysis Consortium (DPAC; \url{https://www.cosmos.esa.int/web/gaia/dpac/consortium}). Funding for the DPAC has been provided by national institutions, in particular the institutions participating in the {\it Gaia} Multilateral Agreement. 
               This research made use of services provided by the Jean-Marie Mariotti Center (Aspro and SearchCal). 
               The SIMBAD database, and NASA's Astrophysics Data System Bibliographic Services were used in the preparation of this paper. 
               This work has made use of the High Contrast Data Centre, jointly operated by OSUG/IPAG (Grenoble), PYTHEAS/LAM/CeSAM (Marseille), OCA/Lagrange (Nice), Observatoire de Paris/LESIA (Paris), and Observatoire de Lyon/CRAL, and supported by a grant from Labex OSUG@2020 (Investissements d’avenir – ANR10 LABX56). 
               This research has made use of the Spanish Virtual Observatory (\url{https://svo.cab.inta-csic.es}) project funded by MCIN/AEI/10.13039/501100011033/ through grant PID2020-112949GB-I00.
 \end{acknowledgements}
 
 
 \bibliographystyle{aa}   
\bibliography{bibliographie}

@article{Evans_2022_10_0,
	adsurl = {https://ui.adsabs.harvard.edu/abs/2022ApJ...938..153E},
	archiveprefix = {arXiv},
	author = {{Evans}, Nancy Remage and {Engle}, Scott and {Pillitteri}, Ignazio and {Guinan}, Edward and {G{\"u}nther}, H. Moritz and {Wolk}, Scott and {Neilson}, Hilding and {Marengo}, Massimo and {Matthews}, Lynn D. and {Moschou}, Sofia and {Drake}, Jeremy J. and {Winston}, Elaine M. and {Moe}, Maxwell and {Kervella}, Pierre and {Breuval}, Louise},
	doi = {10.3847/1538-4357/ac6fdf},
	eid = {153},
	eprint = {2205.07967},
	journal = {\apj},
	month = oct,
	number = {2},
	pages = {153},
	primaryclass = {astro-ph.SR},
	title = {{X-Rays in Cepheids: Identifying Low-mass Companions of Intermediate-mass Stars}},
	volume = {938},
	year = 2022}

@article{Percy_2000_01_0,
	adsurl = {https://ui.adsabs.harvard.edu/abs/2000JAVSO..29...14P},
	author = {{Percy}, John R. and {Hoss}, Jonathan X.},
	journal = {JAAVSO},
	month = jan,
	number = {1},
	pages = {14-18},
	title = {{Period Changes in Population II Cepheids: TX Del and W Vir}},
	volume = {29},
	year = 2000}

@article{Hocde_2024_09_0,
	adsurl = {https://ui.adsabs.harvard.edu/abs/2024A&A...689A.224H},
	archiveprefix = {arXiv},
	author = {{Hocd{\'e}}, V. and {Moskalik}, P. and {Gorynya}, N.~A. and {Smolec}, R. and {Rathour}, R. Singh and {Zi{\'o}{\l}kowska}, O.},
	doi = {10.1051/0004-6361/202347798},
	eid = {A224},
	eprint = {2312.11407},
	journal = {\aap},
	month = sep,
	pages = {A224},
	primaryclass = {astro-ph.SR},
	title = {{Precise Fourier parameters of Cepheid radial velocity curves: Towards refining the Hertzsprung progression models}},
	volume = {689},
	year = 2024}

@article{Boccaletti_2020_03_0,
	adsurl = {https://ui.adsabs.harvard.edu/abs/2020arXiv200305714B},
	archiveprefix = {arXiv},
	author = {{Boccaletti}, A. and {Chauvin}, G. and {Mouillet}, D. and {Absil}, O. and {Allard}, F. and {Antoniucci}, S. and {Augereau}, J. -C. and {Barge}, P. and {Baruffolo}, A. and {Baudino}, J. -L. and {Baudoz}, P. and {Beaulieu}, M. and {Benisty}, M. and {Beuzit}, J. -L. and {Bianco}, A. and {Biller}, B. and {Bonavita}, B. and {Bonnefoy}, M. and {Bos}, S. and {Bouret}, J. -C. and {Brandner}, W. and {Buchschache}, N. and {Carry}, B. and {Cantalloube}, F. and {Cascone}, E. and {Carlotti}, A. and {Charnay}, B. and {Chiavassa}, A. and {Choquet}, E. and {Clenet}, Y. and {Crida}, A. and {De Boer}, J. and {De Caprio}, V. and {Desidera}, S. and {Desert}, J. -M. and {Delisle}, J. -B. and {Delorme}, P. and {Dohlen}, K. and {Doelman}, D. and {Dominik}, C. and {Orazi}, V. D and {Dougados}, C. and {Doute}, S. and {Fedele}, D. and {Feldt}, M. and {Ferreira}, F. and {Fontanive}, C. and {Fusco}, T. and {Galicher}, R. and {Garufi}, A. and {Gendron}, E. and {Ghedina}, A. and {Ginski}, C. and {Gonzalez}, J. -F. and {Gratadour}, D. and {Gratton}, R. and {Guillot}, T. and {Haffert}, S. and {Hagelberg}, J. and {Henning}, T. and {Huby}, E. and {Janson}, M. and {Kamp}, I. and {Keller}, C. and {Kenworthy}, M. and {Kervella}, P. and {Kral}, Q. and {Kuhn}, J. and {Lagadec}, E. and {Laibe}, G. and {Langlois}, M. and {Lagrange}, A. -M. and {Launhardt}, R. and {Leboulleux}, L. and {Le Coroller}, H. and {Li Causi}, G. and {Loupias}, M. and {Maire}, A.~L. and {Marleau}, G. and {Martinache}, F. and {Martinez}, P. and {Mary}, D. and {Mattioli}, M. and {Mazoyer}, J. and {Meheut}, H. and {Menard}, F. and {Mesa}, D. and {Meunier}, N. and {Miguel}, Y. and {Milli}, J. and {Min}, M. and {Molliere}, P. and {Mordasini}, C. and {Moretto}, G. and {Mugnier}, L. and {Muro Arena}, G. and {Nardetto}, N. and {Diaye}, M. N and {Nesvadba}, N. and {Pedichini}, F. and {Pinilla}, P. and {Por}, E. and {Potier}, A. and {Quanz}, S. and {Rameau}, J. and {Roelfsema}, R. and {Rouan}, D. and {Rigliaco}, E. and {Salasnich}, B. and {Samland}, M. and {Sauvage}, J. -F. and {Schmid}, H. -M. and {Segransan}, D. and {Snellen}, I. and {Snik}, F. and {Soulez}, F. and {Stadler}, E. and {Stam}, D. and {Tallon}, M. and {Thebault}, P. and {Thiebaut}, E. and {Tschudi}, C. and {Udry}, S. and {van Holstein}, R. and {Vernazza}, P. and {Vidal}, F. and {Vigan}, A. and {Waters}, R. and {Wildi}, F. and {Willson}, M. and {Zanutta}, A. and {Zavagno}, A. and {Zurlo}, A.},
	doi = {10.48550/arXiv.2003.05714},
	eid = {arXiv:2003.05714},
	eprint = {2003.05714},
	journal = {arXiv e-prints},
	month = mar,
	pages = {arXiv:2003.05714},
	primaryclass = {astro-ph.EP},
	title = {{SPHERE+: Imaging young Jupiters down to the snowline}},
	year = 2020}

@article{Schmid_2017_06_0,
	adsurl = {https://ui.adsabs.harvard.edu/abs/2017A&A...602A..53S},
	archiveprefix = {arXiv},
	author = {{Schmid}, H.~M. and {Bazzon}, A. and {Milli}, J. and {Roelfsema}, R. and {Engler}, N. and {Mouillet}, D. and {Lagadec}, E. and {Sissa}, E. and {Sauvage}, J.-F. and {Ginski}, C. and {Baruffolo}, A. and {Beuzit}, J.~L. and {Boccaletti}, A. and {Bohn}, A.~J. and {Claudi}, R. and {Costille}, A. and {Desidera}, S. and {Dohlen}, K. and {Dominik}, C. and {Feldt}, M. and {Fusco}, T. and {Gisler}, D. and {Girard}, J.~H. and {Gratton}, R. and {Henning}, T. and {Hubin}, N. and {Joos}, F. and {Kasper}, M. and {Langlois}, M. and {Pavlov}, A. and {Pragt}, J. and {Puget}, P. and {Quanz}, S.~P. and {Salasnich}, B. and {Siebenmorgen}, R. and {Stute}, M. and {Suarez}, M. and {Szul{\'a}gyi}, J. and {Thalmann}, C. and {Turatto}, M. and {Udry}, S. and {Vigan}, A. and {Wildi}, F.},
	doi = {10.1051/0004-6361/201629416},
	eid = {A53},
	eprint = {1703.05624},
	journal = {\aap},
	month = jun,
	pages = {A53},
	primaryclass = {astro-ph.SR},
	title = {{SPHERE/ZIMPOL observations of the symbiotic system R Aquarii. I. Imaging of the stellar binary and the innermost jet clouds}},
	volume = {602},
	year = 2017}

@article{Engler_2018_10_0,
	adsurl = {https://ui.adsabs.harvard.edu/abs/2018A&A...618A.151E},
	archiveprefix = {arXiv},
	author = {{Engler}, N. and {Schmid}, H.~M. and {Quanz}, S.~P. and {Avenhaus}, H. and {Bazzon}, A.},
	doi = {10.1051/0004-6361/201832674},
	eid = {A151},
	eprint = {1808.04373},
	journal = {\aap},
	month = oct,
	pages = {A151},
	primaryclass = {astro-ph.EP},
	title = {{Detection of scattered light from the hot dust in HD 172555}},
	volume = {618},
	year = 2018}

@article{Kervella_2022_01_0,
	adsurl = {https://ui.adsabs.harvard.edu/abs/2022A&A...657A...7K},
	archiveprefix = {arXiv},
	author = {{Kervella}, Pierre and {Arenou}, Fr{\'e}d{\'e}ric and {Th{\'e}venin}, Fr{\'e}d{\'e}ric},
	doi = {10.1051/0004-6361/202142146},
	eid = {A7},
	eprint = {2109.10912},
	journal = {\aap},
	month = jan,
	pages = {A7},
	primaryclass = {astro-ph.SR},
	title = {{Stellar and substellar companions from Gaia EDR3. Proper-motion anomaly and resolved common proper-motion pairs}},
	volume = {657},
	year = 2022}

@article{Csornyei_2022_04_0,
	adsurl = {https://ui.adsabs.harvard.edu/abs/2022MNRAS.511.2125C},
	archiveprefix = {arXiv},
	author = {{Cs{\"o}rnyei}, G. and {Szabados}, L. and {Moln{\'a}r}, L. and {Cseh}, B. and {Egei}, N. and {Kalup}, Cs and {Kecskem{\'e}thy}, V. and {K{\"o}nyves-T{\'o}th}, R. and {S{\'a}rneczky}, K. and {Szak{\'a}ts}, R.},
	doi = {10.1093/mnras/stac115},
	eprint = {2201.04748},
	journal = {\mnras},
	month = apr,
	number = {2},
	pages = {2125-2146},
	primaryclass = {astro-ph.SR},
	title = {{Study of changes in the pulsation period of 148 Galactic Cepheid variables}},
	volume = {511},
	year = 2022}

@article{Evans_2025_10_0,
	adsurl = {https://ui.adsabs.harvard.edu/abs/2025AJ....170..242E},
	archiveprefix = {arXiv},
	author = {{Evans}, Nancy Remage and {Proffitt}, Charles and {Kervella}, Pierre and {Kuraszkiewicz}, Joanna and {G{\"u}nther}, H. Moritz and {Anderson}, Richard I. and {Gallenne}, Alexandre and {M{\'e}rand}, Antoine and {Trahin}, Boris and {Viviani}, Giordano and {Shetye}, Shreeya},
	doi = {10.3847/1538-3881/ae03bf},
	eid = {242},
	eprint = {2509.15053},
	journal = {\aj},
	month = oct,
	number = {4},
	pages = {242},
	primaryclass = {astro-ph.SR},
	title = {{V659 Cen: System Members Updated}},
	volume = {170},
	year = 2025}

@article{Tokovinin_2020_07_0,
	adsurl = {https://ui.adsabs.harvard.edu/abs/2020AJ....160....7T},
	archiveprefix = {arXiv},
	author = {{Tokovinin}, Andrei and {Mason}, Brian D. and {Mendez}, Rene A. and {Costa}, Edgardo and {Horch}, Elliott P.},
	doi = {10.3847/1538-3881/ab91c1},
	eid = {7},
	eprint = {2005.05305},
	journal = {\aj},
	month = jul,
	number = {1},
	pages = {7},
	primaryclass = {astro-ph.SR},
	title = {{Speckle Interferometry at SOAR in 2019}},
	volume = {160},
	year = 2020}

@article{Jacobsen_1974_08_0,
	adsurl = {https://ui.adsabs.harvard.edu/abs/1974ApJ...191..691J},
	author = {{Jacobsen}, T.~S.},
	doi = {10.1086/153010},
	journal = {\apj},
	month = aug,
	pages = {691-697},
	title = {{New high-dispersion radial-velocity curves of W Sagittarii.}},
	volume = {191},
	year = 1974}

@article{Shetye_2024_10_0,
	adsurl = {https://ui.adsabs.harvard.edu/abs/2024A&A...690A.284S},
	archiveprefix = {arXiv},
	author = {{Shetye}, Shreeya S. and {Viviani}, Giordano and {Anderson}, Richard I. and {Mowlavi}, Nami and {Eyer}, Laurent and {Evans}, Nancy R. and {Szabados}, L{\'a}szl{\'o}},
	doi = {10.1051/0004-6361/202450185},
	eid = {A284},
	eprint = {2405.19840},
	journal = {\aap},
	month = oct,
	pages = {A284},
	primaryclass = {astro-ph.SR},
	title = {{VELOcities of CEpheids (VELOCE): II. Systematic search for spectroscopic binary cepheids}},
	volume = {690},
	year = 2024}

@article{Harris_1989_09_0,
	adsurl = {https://ui.adsabs.harvard.edu/abs/1989AJ.....98..981H},
	author = {{Harris}, Hugh C. and {Welch}, Douglas L.},
	doi = {10.1086/115190},
	journal = {\aj},
	month = sep,
	pages = {981},
	title = {{The Binary Type II Cepheids IX CAS and TX Del}},
	volume = {98},
	year = 1989}

@article{Harris_1981_05_1,
	adsurl = {https://ui.adsabs.harvard.edu/abs/1981AJ.....86..719H},
	author = {{Harris}, H.~C.},
	doi = {10.1086/112937},
	journal = {\aj},
	month = may,
	pages = {719-729},
	title = {{Photometric abundances of type II Cepheid variables.}},
	volume = {86},
	year = 1981}

@inproceedings{Laney_1995_01_0,
	adsurl = {https://ui.adsabs.harvard.edu/abs/1995ASPC...83..367L},
	author = {{Laney}, C.~D.},
	booktitle = {IAU Colloq. 155: Astrophysical Applications of Stellar Pulsation},
	editor = {{Stobie}, R.~S. and {Whitelock}, P.~A.},
	month = jan,
	pages = {367},
	series = {Astronomical Society of the Pacific Conference Series},
	title = {{Type II? Cepheid Radii and TX Del}},
	volume = {83},
	year = 1995}

@article{Balog_1995_01_0,
	adsurl = {https://ui.adsabs.harvard.edu/abs/1995IBVS.4150....1B},
	author = {{Balog}, Z. and {Vinko}, J.},
	journal = {Information Bulletin on Variable Stars},
	month = jan,
	pages = {1},
	title = {{Is TX Delphini a Population I (classical) Cepheid?}},
	volume = {4150},
	year = 1995}

@article{Tokovinin_2019_07_0,
	adsurl = {https://ui.adsabs.harvard.edu/abs/2019AJ....158...48T},
	archiveprefix = {arXiv},
	author = {{Tokovinin}, Andrei and {Mason}, Brian D. and {Mendez}, Rene A. and {Horch}, Elliott P. and {Brice{\~n}o}, Cesar},
	doi = {10.3847/1538-3881/ab24e4},
	eid = {48},
	eprint = {1905.10436},
	journal = {\aj},
	month = jul,
	number = {1},
	pages = {48},
	primaryclass = {astro-ph.SR},
	title = {{Speckle Interferometry at SOAR in 2018}},
	volume = {158},
	year = 2019}

@article{Kervella_2019_03_2,
	adsurl = {https://ui.adsabs.harvard.edu/abs/2019A&A...623A..72K},
	archiveprefix = {arXiv},
	author = {{Kervella}, Pierre and {Arenou}, Fr{\'e}d{\'e}ric and {Mignard}, Fran{\c{c}}ois and {Th{\'e}venin}, Fr{\'e}d{\'e}ric},
	doi = {10.1051/0004-6361/201834371},
	eid = {A72},
	eprint = {1811.08902},
	journal = {\aap},
	month = mar,
	pages = {A72},
	primaryclass = {astro-ph.SR},
	title = {{Stellar and substellar companions of nearby stars from Gaia DR2. Binarity from proper motion anomaly}},
	volume = {623},
	year = 2019}

@article{Lallement_2018_08_4,
	adsurl = {https://ui.adsabs.harvard.edu/abs/2018A&A...616A.132L},
	archiveprefix = {arXiv},
	author = {{Lallement}, R. and {Capitanio}, L. and {Ruiz-Dern}, L. and {Danielski}, C. and {Babusiaux}, C. and {Vergely}, L. and {Elyajouri}, M. and {Arenou}, F. and {Leclerc}, N.},
	doi = {10.1051/0004-6361/201832832},
	eid = {A132},
	eprint = {1804.06060},
	journal = {\aap},
	month = aug,
	pages = {A132},
	primaryclass = {astro-ph.GA},
	title = {{Three-dimensional maps of interstellar dust in the Local Arm: using Gaia, 2MASS, and APOGEE-DR14}},
	volume = {616},
	year = 2018}

@article{Lindegren_2021_05_7,
	adsurl = {https://ui.adsabs.harvard.edu/abs/2021A&A...649A...4L},
	archiveprefix = {arXiv},
	author = {{Lindegren}, L. and {Bastian}, U. and {Biermann}, M. and {Bombrun}, A. and {de Torres}, A. and {Gerlach}, E. and {Geyer}, R. and {Hern{\'a}ndez}, J. and {Hilger}, T. and {Hobbs}, D. and {Klioner}, S.~A. and {Lammers}, U. and {McMillan}, P.~J. and {Ramos-Lerate}, M. and {Steidelm{\"u}ller}, H. and {Stephenson}, C.~A. and {van Leeuwen}, F.},
	doi = {10.1051/0004-6361/202039653},
	eid = {A4},
	eprint = {2012.01742},
	journal = {\aap},
	month = may,
	pages = {A4},
	primaryclass = {astro-ph.IM},
	title = {{Gaia Early Data Release 3. Parallax bias versus magnitude, colour, and position}},
	volume = {649},
	year = 2021}

@article{Trahin_2021_12_5,
	adsurl = {https://ui.adsabs.harvard.edu/abs/2021A&A...656A.102T},
	archiveprefix = {arXiv},
	author = {{Trahin}, B. and {Breuval}, L. and {Kervella}, P. and {M{\'e}rand}, A. and {Nardetto}, N. and {Gallenne}, A. and {Hocd{\'e}}, V. and {Gieren}, W.},
	doi = {10.1051/0004-6361/202141680},
	eid = {A102},
	eprint = {2111.09125},
	journal = {\aap},
	month = dec,
	pages = {A102},
	primaryclass = {astro-ph.SR},
	title = {{Inspecting the Cepheid parallax of pulsation using Gaia EDR3 parallaxes. Projection factor and period-luminosity and period-radius relations}},
	volume = {656},
	year = 2021}

@article{Gomez-Gonzalez_2016_05_1,
	adsurl = {https://ui.adsabs.harvard.edu/abs/2016A&A...589A..54G},
	archiveprefix = {arXiv},
	author = {{Gomez Gonzalez}, C.~A. and {Absil}, O. and {Absil}, P. -A. and {Van Droogenbroeck}, M. and {Mawet}, D. and {Surdej}, J.},
	doi = {10.1051/0004-6361/201527387},
	eid = {A54},
	eprint = {1602.08381},
	journal = {\aap},
	month = may,
	pages = {A54},
	primaryclass = {astro-ph.IM},
	title = {{Low-rank plus sparse decomposition for exoplanet detection in direct-imaging ADI sequences. The LLSG algorithm}},
	volume = {589},
	year = 2016}

@article{Lee_1999_10_5,
	adsurl = {https://ui.adsabs.harvard.edu/abs/1999Natur.401..788L},
	author = {{Lee}, Daniel D. and {Seung}, H. Sebastian},
	doi = {10.1038/44565},
	journal = {\nat},
	month = oct,
	number = {6755},
	pages = {788-791},
	title = {{Learning the parts of objects by non-negative matrix factorization}},
	volume = {401},
	year = 1999}

@article{Lafreniere_2007_05_2,
	adsurl = {https://ui.adsabs.harvard.edu/abs/2007ApJ...660..770L},
	archiveprefix = {arXiv},
	author = {{Lafreni{\`e}re}, David and {Marois}, Christian and {Doyon}, Ren{\'e} and {Nadeau}, Daniel and {Artigau}, {\'E}tienne},
	doi = {10.1086/513180},
	eprint = {astro-ph/0702697},
	journal = {\apj},
	month = may,
	number = {1},
	pages = {770-780},
	primaryclass = {astro-ph},
	title = {{A New Algorithm for Point-Spread Function Subtraction in High-Contrast Imaging: A Demonstration with Angular Differential Imaging}},
	volume = {660},
	year = 2007}

@inproceedings{Gerard_2016_07_2,
	adsurl = {https://ui.adsabs.harvard.edu/abs/2016SPIE.9909E..58G},
	archiveprefix = {arXiv},
	author = {{Gerard}, Benjamin L. and {Marois}, Christian},
	booktitle = {Adaptive Optics Systems V},
	doi = {10.1117/12.2231905},
	editor = {{Marchetti}, Enrico and {Close}, Laird M. and {V{\'e}ran}, Jean-Pierre},
	eid = {990958},
	eprint = {1609.08692},
	month = jul,
	pages = {990958},
	primaryclass = {astro-ph.IM},
	series = {Society of Photo-Optical Instrumentation Engineers (SPIE) Conference Series},
	title = {{Planet detection down to a few {\ensuremath{\lambda}}/D: an RSDI/TLOCI approach to PSF subtraction}},
	volume = {9909},
	year = 2016}

@article{Soummer_2011_11_2,
	adsurl = {https://ui.adsabs.harvard.edu/abs/2011ApJ...741...55S},
	archiveprefix = {arXiv},
	author = {{Soummer}, R{\'e}mi and {Hagan}, J. Brendan and {Pueyo}, Laurent and {Thormann}, Adrien and {Rajan}, Abhijith and {Marois}, Christian},
	doi = {10.1088/0004-637X/741/1/55},
	eid = {55},
	eprint = {1110.1382},
	journal = {\apj},
	month = nov,
	number = {1},
	pages = {55},
	primaryclass = {astro-ph.EP},
	title = {{Orbital Motion of HR 8799 b, c, d Using Hubble Space Telescope Data from 1998: Constraints on Inclination, Eccentricity, and Stability}},
	volume = {741},
	year = 2011}

@article{Lafreniere_2009_04_6,
	adsurl = {https://ui.adsabs.harvard.edu/abs/2009ApJ...694L.148L},
	archiveprefix = {arXiv},
	author = {{Lafreni{\`e}re}, David and {Marois}, Christian and {Doyon}, Ren{\'e} and {Barman}, Travis},
	doi = {10.1088/0004-637X/694/2/L148},
	eprint = {0902.3247},
	journal = {\apjl},
	month = apr,
	number = {2},
	pages = {L148-L152},
	primaryclass = {astro-ph.EP},
	title = {{HST/NICMOS Detection of HR 8799 b in 1998}},
	volume = {694},
	year = 2009}

@article{Marois_2006_04_9,
	adsurl = {https://ui.adsabs.harvard.edu/abs/2006ApJ...641..556M},
	archiveprefix = {arXiv},
	author = {{Marois}, Christian and {Lafreni{\`e}re}, David and {Doyon}, Ren{\'e} and {Macintosh}, Bruce and {Nadeau}, Daniel},
	doi = {10.1086/500401},
	eprint = {astro-ph/0512335},
	journal = {\apj},
	month = apr,
	number = {1},
	pages = {556-564},
	primaryclass = {astro-ph},
	title = {{Angular Differential Imaging: A Powerful High-Contrast Imaging Technique}},
	volume = {641},
	year = 2006}

@article{Kouwenhoven_2005_01_9,
	adsurl = {https://ui.adsabs.harvard.edu/abs/2005A&A...430..137K},
	archiveprefix = {arXiv},
	author = {{Kouwenhoven}, M.~B.~N. and {Brown}, A.~G.~A. and {Zinnecker}, H. and {Kaper}, L. and {Portegies Zwart}, S.~F.},
	doi = {10.1051/0004-6361:20048124},
	eprint = {astro-ph/0410106},
	journal = {\aap},
	month = jan,
	pages = {137-154},
	primaryclass = {astro-ph},
	title = {{The primordial binary population. I. A near-infrared adaptive optics search for close visual companions to A star members of Scorpius OB2}},
	volume = {430},
	year = 2005}

@article{Chini_2012_08_3,
	adsurl = {https://ui.adsabs.harvard.edu/abs/2012MNRAS.424.1925C},
	archiveprefix = {arXiv},
	author = {{Chini}, R. and {Hoffmeister}, V.~H. and {Nasseri}, A. and {Stahl}, O. and {Zinnecker}, H.},
	doi = {10.1111/j.1365-2966.2012.21317.x},
	eprint = {1205.5238},
	journal = {\mnras},
	month = aug,
	number = {3},
	pages = {1925-1929},
	primaryclass = {astro-ph.SR},
	title = {{A spectroscopic survey on the multiplicity of high-mass stars}},
	volume = {424},
	year = 2012}

@article{Duchene_2013_08_4,
	adsurl = {https://ui.adsabs.harvard.edu/abs/2013ARA&A..51..269D},
	archiveprefix = {arXiv},
	author = {{Duch{\^e}ne}, Gaspard and {Kraus}, Adam},
	doi = {10.1146/annurev-astro-081710-102602},
	eprint = {1303.3028},
	journal = {\araa},
	month = aug,
	number = {1},
	pages = {269-310},
	primaryclass = {astro-ph.SR},
	title = {{Stellar Multiplicity}},
	volume = {51},
	year = 2013}

@article{Mawet_2014_09_0,
	adsurl = {https://ui.adsabs.harvard.edu/abs/2014ApJ...792...97M},
	archiveprefix = {arXiv},
	author = {{Mawet}, D. and {Milli}, J. and {Wahhaj}, Z. and {Pelat}, D. and {Absil}, O. and {Delacroix}, C. and {Boccaletti}, A. and {Kasper}, M. and {Kenworthy}, M. and {Marois}, C. and {Mennesson}, B. and {Pueyo}, L.},
	doi = {10.1088/0004-637X/792/2/97},
	eid = {97},
	eprint = {1407.2247},
	journal = {\apj},
	month = sep,
	number = {2},
	pages = {97},
	primaryclass = {astro-ph.IM},
	title = {{Fundamental Limitations of High Contrast Imaging Set by Small Sample Statistics}},
	volume = {792},
	year = 2014}

@article{Lagrange_2010_07_0,
	adsurl = {https://ui.adsabs.harvard.edu/abs/2010Sci...329...57L},
	archiveprefix = {arXiv},
	author = {{Lagrange}, A. -M. and {Bonnefoy}, M. and {Chauvin}, G. and {Apai}, D. and {Ehrenreich}, D. and {Boccaletti}, A. and {Gratadour}, D. and {Rouan}, D. and {Mouillet}, D. and {Lacour}, S. and {Kasper}, M.},
	doi = {10.1126/science.1187187},
	eprint = {1006.3314},
	journal = {Science},
	month = jul,
	number = {5987},
	pages = {57},
	primaryclass = {astro-ph.EP},
	title = {{A Giant Planet Imaged in the Disk of the Young Star {\ensuremath{\beta}} Pictoris}},
	volume = {329},
	year = 2010}

@inproceedings{Marois_2010_07_0,
	adsurl = {https://ui.adsabs.harvard.edu/abs/2010SPIE.7736E..1JM},
	author = {{Marois}, Christian and {Macintosh}, Bruce and {V{\'e}ran}, Jean-Pierre},
	booktitle = {Adaptive Optics Systems II},
	doi = {10.1117/12.857225},
	editor = {{Ellerbroek}, Brent L. and {Hart}, Michael and {Hubin}, Norbert and {Wizinowich}, Peter L.},
	eid = {77361J},
	month = jul,
	pages = {77361J},
	series = {Society of Photo-Optical Instrumentation Engineers (SPIE) Conference Series},
	title = {{Exoplanet imaging with LOCI processing: photometry and astrometry with the new SOSIE pipeline}},
	volume = {7736},
	year = 2010}

@article{Gomez-Gonzalez_2017_07_0,
	adsurl = {https://ui.adsabs.harvard.edu/abs/2017AJ....154....7G},
	archiveprefix = {arXiv},
	author = {{Gomez Gonzalez}, Carlos Alberto and {Wertz}, Olivier and {Absil}, Olivier and {Christiaens}, Valentin and {Defr{\`e}re}, Denis and {Mawet}, Dimitri and {Milli}, Julien and {Absil}, Pierre-Antoine and {Van Droogenbroeck}, Marc and {Cantalloube}, Faustine and {Hinz}, Philip M. and {Skemer}, Andrew J. and {Karlsson}, Mikael and {Surdej}, Jean},
	doi = {10.3847/1538-3881/aa73d7},
	eid = {7},
	eprint = {1705.06184},
	journal = {\aj},
	month = jul,
	number = {1},
	pages = {7},
	primaryclass = {astro-ph.IM},
	title = {{VIP: Vortex Image Processing Package for High-contrast Direct Imaging}},
	volume = {154},
	year = 2017}

@article{Amara_2012_12_0,
	adsurl = {https://ui.adsabs.harvard.edu/abs/2012MNRAS.427..948A},
	archiveprefix = {arXiv},
	author = {{Amara}, Adam and {Quanz}, Sascha P.},
	doi = {10.1111/j.1365-2966.2012.21918.x},
	eprint = {1207.6637},
	journal = {\mnras},
	month = dec,
	number = {2},
	pages = {948-955},
	primaryclass = {astro-ph.IM},
	title = {{PYNPOINT: an image processing package for finding exoplanets}},
	volume = {427},
	year = 2012}

@inproceedings{Delorme_2017_12_0,
	adsurl = {https://ui.adsabs.harvard.edu/abs/2017sf2a.conf..347D},
	archiveprefix = {arXiv},
	author = {{Delorme}, P. and {Meunier}, N. and {Albert}, D. and {Lagadec}, E. and {Le Coroller}, H. and {Galicher}, R. and {Mouillet}, D. and {Boccaletti}, A. and {Mesa}, D. and {Meunier}, J. -C. and {Beuzit}, J. -L. and {Lagrange}, A. -M. and {Chauvin}, G. and {Sapone}, A. and {Langlois}, M. and {Maire}, A. -L. and {Montarg{\`e}s}, M. and {Gratton}, R. and {Vigan}, A. and {Surace}, C.},
	booktitle = {SF2A-2017: Proceedings of the Annual meeting of the French Society of Astronomy and Astrophysics},
	editor = {{Reyl{\'e}}, C. and {Di Matteo}, P. and {Herpin}, F. and {Lagadec}, E. and {Lan{\c{c}}on}, A. and {Meliani}, Z. and {Royer}, F.},
	eprint = {1712.06948},
	month = dec,
	pages = {Di},
	primaryclass = {astro-ph.IM},
	title = {{The SPHERE Data Center: a reference for high contrast imaging processing}},
	year = 2017}

@article{Schmid_2018_11_0,
	adsurl = {https://ui.adsabs.harvard.edu/abs/2018A&A...619A...9S},
	archiveprefix = {arXiv},
	author = {{Schmid}, H.~M. and {Bazzon}, A. and {Roelfsema}, R. and {Mouillet}, D. and {Milli}, J. and {Menard}, F. and {Gisler}, D. and {Hunziker}, S. and {Pragt}, J. and {Dominik}, C. and {Boccaletti}, A. and {Ginski}, C. and {Abe}, L. and {Antoniucci}, S. and {Avenhaus}, H. and {Baruffolo}, A. and {Baudoz}, P. and {Beuzit}, J.~L. and {Carbillet}, M. and {Chauvin}, G. and {Claudi}, R. and {Costille}, A. and {Daban}, J. -B. and {de Haan}, M. and {Desidera}, S. and {Dohlen}, K. and {Downing}, M. and {Elswijk}, E. and {Engler}, N. and {Feldt}, M. and {Fusco}, T. and {Girard}, J.~H. and {Gratton}, R. and {Hanenburg}, H. and {Henning}, Th. and {Hubin}, N. and {Joos}, F. and {Kasper}, M. and {Keller}, C.~U. and {Langlois}, M. and {Lagadec}, E. and {Martinez}, P. and {Mulder}, E. and {Pavlov}, A. and {Podio}, L. and {Puget}, P. and {Quanz}, S.~P. and {Rigal}, F. and {Salasnich}, B. and {Sauvage}, J. -F. and {Schuil}, M. and {Siebenmorgen}, R. and {Sissa}, E. and {Snik}, F. and {Suarez}, M. and {Thalmann}, Ch. and {Turatto}, M. and {Udry}, S. and {van Duin}, A. and {van Holstein}, R.~G. and {Vigan}, A. and {Wildi}, F.},
	doi = {10.1051/0004-6361/201833620},
	eid = {A9},
	eprint = {1808.05008},
	journal = {\aap},
	month = nov,
	pages = {A9},
	primaryclass = {astro-ph.IM},
	title = {{SPHERE/ZIMPOL high resolution polarimetric imager. I. System overview, PSF parameters, coronagraphy, and polarimetry}},
	volume = {619},
	year = 2018}

@article{Evans_2020_12_0,
	adsurl = {https://ui.adsabs.harvard.edu/abs/2020ApJ...905...81E},
	archiveprefix = {arXiv},
	author = {{Evans}, Nancy Remage and {G{\"u}nther}, H. Moritz and {Bond}, Howard E. and {Schaefer}, Gail H. and {Mason}, Brian D. and {Karovska}, Margarita and {Tingle}, Evan and {Wolk}, Scott and {Engle}, Scott and {Guinan}, Edward and {Pillitteri}, Ignazio and {Proffitt}, Charles and {Kervella}, Pierre and {Gallenne}, Alexandre and {Anderson}, Richard I. and {Moe}, Maxwell},
	doi = {10.3847/1538-4357/abc1f1},
	eid = {81},
	eprint = {2010.07996},
	journal = {\apj},
	month = dec,
	number = {1},
	pages = {81},
	primaryclass = {astro-ph.SR},
	title = {{Hubble Space Telescope Snapshot Survey for Resolved Companions of Galactic Cepheids: Final Results}},
	volume = {905},
	year = 2020}

@phdthesis{Trahin_2019_11_0,
	author = {{Trahin}, B.},
	journal = {Universit\'e PS:L},
	month = {Septembre},
	school = {Universit\'e PSL (https://hal.archives-ouvertes.fr/tel-02372923)},
	title = {{\'Etalonnage de l'\'echelle des distances dans l'\`ere Gaia: Les \'etoiles pulsantes RR Lyrae et C\'eph\'eides}},
	year = {2019}}

@article{Beuzit_2019_11_0,
	adsurl = {https://ui.adsabs.harvard.edu/abs/2019A&A...631A.155B},
	archiveprefix = {arXiv},
	author = {{Beuzit}, J. -L. and {Vigan}, A. and {Mouillet}, D. and {Dohlen}, K. and {Gratton}, R. and {Boccaletti}, A. and {Sauvage}, J. -F. and {Schmid}, H.~M. and {Langlois}, M. and {Petit}, C. and {Baruffolo}, A. and {Feldt}, M. and {Milli}, J. and {Wahhaj}, Z. and {Abe}, L. and {Anselmi}, U. and {Antichi}, J. and {Barette}, R. and {Baudrand}, J. and {Baudoz}, P. and {Bazzon}, A. and {Bernardi}, P. and {Blanchard}, P. and {Brast}, R. and {Bruno}, P. and {Buey}, T. and {Carbillet}, M. and {Carle}, M. and {Cascone}, E. and {Chapron}, F. and {Charton}, J. and {Chauvin}, G. and {Claudi}, R. and {Costille}, A. and {De Caprio}, V. and {de Boer}, J. and {Delboulb{\'e}}, A. and {Desidera}, S. and {Dominik}, C. and {Downing}, M. and {Dupuis}, O. and {Fabron}, C. and {Fantinel}, D. and {Farisato}, G. and {Feautrier}, P. and {Fedrigo}, E. and {Fusco}, T. and {Gigan}, P. and {Ginski}, C. and {Girard}, J. and {Giro}, E. and {Gisler}, D. and {Gluck}, L. and {Gry}, C. and {Henning}, T. and {Hubin}, N. and {Hugot}, E. and {Incorvaia}, S. and {Jaquet}, M. and {Kasper}, M. and {Lagadec}, E. and {Lagrange}, A. -M. and {Le Coroller}, H. and {Le Mignant}, D. and {Le Ruyet}, B. and {Lessio}, G. and {Lizon}, J. -L. and {Llored}, M. and {Lundin}, L. and {Madec}, F. and {Magnard}, Y. and {Marteaud}, M. and {Martinez}, P. and {Maurel}, D. and {M{\'e}nard}, F. and {Mesa}, D. and {M{\"o}ller-Nilsson}, O. and {Moulin}, T. and {Moutou}, C. and {Orign{\'e}}, A. and {Parisot}, J. and {Pavlov}, A. and {Perret}, D. and {Pragt}, J. and {Puget}, P. and {Rabou}, P. and {Ramos}, J. and {Reess}, J. -M. and {Rigal}, F. and {Rochat}, S. and {Roelfsema}, R. and {Rousset}, G. and {Roux}, A. and {Saisse}, M. and {Salasnich}, B. and {Santambrogio}, E. and {Scuderi}, S. and {Segransan}, D. and {Sevin}, A. and {Siebenmorgen}, R. and {Soenke}, C. and {Stadler}, E. and {Suarez}, M. and {Tiph{\`e}ne}, D. and {Turatto}, M. and {Udry}, S. and {Vakili}, F. and {Waters}, L.~B.~F.~M. and {Weber}, L. and {Wildi}, F. and {Zins}, G. and {Zurlo}, A.},
	doi = {10.1051/0004-6361/201935251},
	eid = {A155},
	eprint = {1902.04080},
	journal = {\aap},
	month = nov,
	pages = {A155},
	primaryclass = {astro-ph.IM},
	title = {{SPHERE: the exoplanet imager for the Very Large Telescope}},
	volume = {631},
	year = 2019}

@article{Groenewegen_2020_03_0,
	adsurl = {https://ui.adsabs.harvard.edu/abs/2020A&A...635A..33G},
	archiveprefix = {arXiv},
	author = {{Groenewegen}, M.~A.~T.},
	doi = {10.1051/0004-6361/201937060},
	eid = {A33},
	eprint = {2002.02186},
	journal = {\aap},
	month = mar,
	pages = {A33},
	primaryclass = {astro-ph.SR},
	title = {{Analysing the spectral energy distributions of Galactic classical Cepheids}},
	volume = {635},
	year = 2020}

@article{Gieren_1985_08_0,
	adsurl = {https://ui.adsabs.harvard.edu/abs/1985ApJ...295..507G},
	author = {{Gieren}, W.~P.},
	doi = {10.1086/163395},
	journal = {\apj},
	month = {Aug},
	pages = {507-520},
	title = {{A search for more Cepheid binaries.}},
	volume = {295},
	year = {1985}}

@article{Evans_1988_03_0,
	adsurl = {https://ui.adsabs.harvard.edu/abs/1988ApJS...66..343E},
	author = {{Evans}, Nancy R.},
	doi = {10.1086/191260},
	journal = {\apjs},
	month = {Mar},
	pages = {343},
	title = {{The Orbit of the Classical Cepheid SU Cygni}},
	volume = {66},
	year = {1988}}

@article{Kervella_2019_03_0,
	adsurl = {http://adsabs.harvard.edu/abs/2019A%26A...623A.116K},
	archiveprefix = {arXiv},
	arxivurl = {http://arxiv.org/abs/1903.03632},
	author = {{Kervella}, P. and {Gallenne}, A. and {Remage Evans}, N. and {Szabados}, L. and {Arenou}, F. and {M{\'e}rand}, A. and {Proto}, Y. and {Karczmarek}, P. and {Nardetto}, N. and {Gieren}, W. and {Pietrzynski}, G.},
	doi = {10.1051/0004-6361/201834210},
	eid = {A116},
	eprint = {1903.03632},
	journal = {\aap},
	month = mar,
	pages = {A116},
	primaryclass = {astro-ph.SR},
	title = {{Multiplicity of Galactic Cepheids and RR Lyrae stars from Gaia DR2. I. Binarity from proper motion anomaly}},
	volume = 623,
	year = 2019}

@article{Kervella_2019_03_1,
	adsurl = {http://adsabs.harvard.edu/abs/2019A%26A...623A.117K},
	author = {{Kervella}, P. and {Gallenne}, A. and {Evans}, N.~R. and {Szabados}, L. and {Arenou}, F. and {M{\'e}rand}, A. and {Nardetto}, N. and {Gieren}, W. and {Pietrzynski}, G.},
	doi = {10.1051/0004-6361/201834211},
	eid = {A117},
	journal = {\aap},
	month = mar,
	pages = {A117},
	title = {{Multiplicity of Galactic Cepheids and RR Lyrae stars from Gaia DR2. II. Resolved common proper motion pairs}},
	volume = 623,
	year = 2019}

@article{Gallenne_2019_02_0,
	adsurl = {http://adsabs.harvard.edu/abs/2019A%26A...622A.164G},
	archiveprefix = {arXiv},
	arxivurl = {http://arxiv.org/abs/1812.09989},
	author = {{Gallenne}, A. and {Kervella}, P. and {Borgniet}, S. and {M{\'e}rand}, A. and {Pietrzy{\'n}ski}, G. and {Gieren}, W. and {Monnier}, J.~D. and {Schaefer}, G.~H. and {Evans}, N.~R. and {Anderson}, R.~I. and {Baron}, F. and {Roettenbacher}, R.~M. and {Karczmarek}, P.},
	doi = {10.1051/0004-6361/201834614},
	eid = {A164},
	eprint = {1812.09989},
	journal = {\aap},
	month = feb,
	pages = {A164},
	primaryclass = {astro-ph.SR},
	title = {{Multiplicity of Galactic Cepheids from long-baseline interferometry. IV. New detected companions from MIRC and PIONIER observations}},
	volume = 622,
	year = 2019}

@article{Gieren_2018_12_0,
	adsurl = {http://adsabs.harvard.edu/abs/2018A%26A...620A..99G},
	archiveprefix = {arXiv},
	arxivurl = {http://arxiv.org/abs/1809.04073},
	author = {{Gieren}, W. and {Storm}, J. and {Konorski}, P. and {G{\'o}rski}, M. and {Pilecki}, B. and {Thompson}, I. and {Pietrzy{\'n}ski}, G. and {Graczyk}, D. and {Barnes}, T.~G. and {Fouqu{\'e}}, P. and {Nardetto}, N. and {Gallenne}, A. and {Karczmarek}, P. and {Suchomska}, K. and {Wielg{\'o}rski}, P. and {Taormina}, M. and {Zgirski}, B.},
	doi = {10.1051/0004-6361/201833263},
	eid = {A99},
	eprint = {1809.04073},
	journal = {\aap},
	month = dec,
	pages = {A99},
	primaryclass = {astro-ph.SR},
	title = {{The effect of metallicity on Cepheid period-luminosity relations from a Baade-Wesselink analysis of Cepheids in the Milky Way and Magellanic Clouds{\star}}},
	volume = 620,
	year = 2018}

@article{Gallenne_2018_11_0,
	adsurl = {http://adsabs.harvard.edu/abs/2018ApJ...867..121G},
	archiveprefix = {arXiv},
	arxivurl = {http://arxiv.org/abs/1809.07486},
	author = {{Gallenne}, A. and {Kervella}, P. and {Evans}, N.~R. and {Proffitt}, C.~R. and {Monnier}, J.~D. and {M{\'e}rand}, A. and {Nelan}, E. and {Winston}, E. and {Pietrzy{\'n}ski}, G. and {Schaefer}, G. and {Gieren}, W. and {Anderson}, R.~I. and {Borgniet}, S. and {Kraus}, S. and {Roettenbacher}, R.~M. and {Baron}, F. and {Pilecki}, B. and {Taormina}, M. and {Graczyk}, D. and {Mowlavi}, N. and {Eyer}, L.},
	doi = {10.3847/1538-4357/aae373},
	eid = {121},
	eprint = {1809.07486},
	journal = {\apj},
	month = nov,
	pages = {121},
	primaryclass = {astro-ph.SR},
	title = {{A Geometrical 1\% Distance to the Short-period Binary Cepheid V1334 Cygni}},
	volume = 867,
	year = 2018}

@article{Evans_2018_10_0,
	adsurl = {http://adsabs.harvard.edu/abs/2018ApJ...866...30E},
	archiveprefix = {arXiv},
	arxivurl = {http://arxiv.org/abs/1808.10472},
	author = {{Evans}, N.~R. and {Proffitt}, C. and {Carpenter}, K.~G. and {Winston}, E.~M. and {Kober}, G.~V. and {G{\"u}nther}, H.~M. and {Gorynya}, N. and {Rastorguev}, A. and {Inno}, L.},
	doi = {10.3847/1538-4357/aade03},
	eid = {30},
	eprint = {1808.10472},
	journal = {\apj},
	month = oct,
	pages = {30},
	primaryclass = {astro-ph.SR},
	title = {{The Mass of the Cepheid V350 Sgr}},
	volume = 866,
	year = 2018}

@article{Pecaut_2013_09_0,
	adsurl = {http://adsabs.harvard.edu/abs/2013ApJS..208....9P},
	archiveprefix = {arXiv},
	arxivurl = {http://arxiv.org/abs/1307.2657},
	author = {{Pecaut}, M.~J. and {Mamajek}, E.~E.},
	doi = {10.1088/0067-0049/208/1/9},
	eid = {9},
	eprint = {1307.2657},
	journal = {\apjs},
	month = sep,
	pages = {9},
	primaryclass = {astro-ph.SR},
	title = {{Intrinsic Colors, Temperatures, and Bolometric Corrections of Pre-main-sequence Stars}},
	volume = 208,
	year = 2013}

@article{Evans_2018_08_1,
	adsurl = {http://adsabs.harvard.edu/abs/2018ApJ...863..187E},
	archiveprefix = {arXiv},
	arxivurl = {http://arxiv.org/abs/1807.06115},
	author = {{Evans}, N.~R. and {Karovska}, M. and {Bond}, H.~E. and {Schaefer}, G.~H. and {Sahu}, K.~C. and {Mack}, J. and {Nelan}, E.~P. and {Gallenne}, A. and {Tingle}, E.~D.},
	doi = {10.3847/1538-4357/aad410},
	eid = {187},
	eprint = {1807.06115},
	journal = {\apj},
	month = aug,
	pages = {187},
	primaryclass = {astro-ph.SR},
	title = {{The Orbit of the Close Companion of Polaris: Hubble Space Telescope Imaging, 2007 to 2014}},
	volume = 863,
	year = 2018}

@article{Gaia-Collaboration_2018_08_0,
	adsurl = {http://adsabs.harvard.edu/abs/2018A%26A...616A...1G},
	archiveprefix = {arXiv},
	arxivurl = {http://arxiv.org/abs/1804.09365},
	author = {{Gaia Collaboration} and {Brown}, A.~G.~A. and {Vallenari}, A. and {Prusti}, T. and {de Bruijne}, J.~H.~J. and {Babusiaux}, C. and {Bailer-Jones}, C.~A.~L. and {Biermann}, M. and {Evans}, D.~W. and {Eyer}, L. and et al.},
	doi = {10.1051/0004-6361/201833051},
	eid = {A1},
	eprint = {1804.09365},
	journal = {\aap},
	month = aug,
	pages = {A1},
	title = {{Gaia Data Release 2. Summary of the contents and survey properties}},
	volume = 616,
	year = 2018}

@article{Pilecki_2018_07_0,
	adsurl = {http://adsabs.harvard.edu/abs/2018ApJ...862...43P},
	archiveprefix = {arXiv},
	arxivurl = {http://arxiv.org/abs/1806.01391},
	author = {{Pilecki}, B. and {Gieren}, W. and {Pietrzy{\'n}ski}, G. and {Thompson}, I.~B. and {Smolec}, R. and {Graczyk}, D. and {Taormina}, M. and {Udalski}, A. and {Storm}, J. and {Nardetto}, N. and {Gallenne}, A. and {Kervella}, P. and {Soszy{\'n}ski}, I. and {G{\'o}rski}, M. and {Wielg{\'o}rski}, P. and {Suchomska}, K. and {Karczmarek}, P. and {Zgirski}, B.},
	doi = {10.3847/1538-4357/aacb32},
	eid = {43},
	eprint = {1806.01391},
	journal = {\apj},
	month = jul,
	pages = {43},
	primaryclass = {astro-ph.SR},
	title = {{The Araucaria Project: High-precision Cepheid Astrophysics from the Analysis of Variables in Double-lined Eclipsing Binaries}},
	volume = 862,
	year = 2018}

@article{Bohm-Vitense_1985_09_1,
	adsurl = {http://adsabs.harvard.edu/abs/1985ApJ...296..169B},
	author = {{Bohm-Vitense}, E.},
	doi = {10.1086/163432},
	journal = {\apj},
	month = sep,
	pages = {169-174},
	title = {{Cepheid distances from blue main-sequence companions}},
	volume = 296,
	year = 1985}

@article{Foreman-Mackey_2013_03_0,
	adsurl = {http://adsabs.harvard.edu/abs/2013PASP..125..306F},
	archiveprefix = {arXiv},
	arxivurl = {http://arxiv.org/abs/1202.3665},
	author = {{Foreman-Mackey}, D. and {Hogg}, D.~W. and {Lang}, D. and {Goodman}, J.},
	doi = {10.1086/670067},
	eprint = {1202.3665},
	journal = {\pasp},
	month = mar,
	pages = {306},
	primaryclass = {astro-ph.IM},
	title = {{emcee: The MCMC Hammer}},
	volume = 125,
	year = 2013}

@article{Evans_2011_09_0,
	adsurl = {http://adsabs.harvard.edu/abs/2011AJ....142...87E},
	author = {{Evans}, N.~R. and {Berdnikov}, L. and {Gorynya}, N. and {Rastorguev}, A. and {Eaton}, J.},
	doi = {10.1088/0004-6256/142/3/87},
	eid = {87},
	journal = {\aj},
	month = sep,
	pages = {87},
	title = {{The Orbit of the Cepheid V350 Sgr Revisited}},
	volume = 142,
	year = 2011}

@article{Turner_2016_10_0,
	adsurl = {http://adsabs.harvard.edu/abs/2016RMxAA..52..223T},
	arxivurl = {http://arxiv.org/abs/1603.02276},
	author = {{Turner}, D.~G.},
	journal = {\rmxaa},
	month = oct,
	pages = {223-239},
	title = {{The Scale of Reddening for Classical Cepheid Variables}},
	volume = 52,
	year = 2016}

@article{Anderson_2016_10_0,
	adsurl = {http://adsabs.harvard.edu/abs/2016ApJS..226...18A},
	archiveprefix = {arXiv},
	arxivurl = {http://arxiv.org/abs/1608.00556},
	author = {{Anderson}, R.~I. and {Casertano}, S. and {Riess}, A.~G. and {Melis}, C. and {Holl}, B. and {Semaan}, T. and {Papics}, P.~I. and {Blanco-Cuaresma}, S. and {Eyer}, L. and {Mowlavi}, N. and {Palaversa}, L. and {Roelens}, M.},
	doi = {10.3847/0067-0049/226/2/18},
	eid = {18},
	eprint = {1608.00556},
	journal = {\apjs},
	month = oct,
	pages = {18},
	primaryclass = {astro-ph.SR},
	title = {{Vetting Galactic Leavitt Law Calibrators Using Radial Velocities: On the Variability, Binarity, and Possible Parallax Error of 19 Long-period Cepheids}},
	volume = 226,
	year = 2016}

@article{Szabados_2012_11_1,
	adsurl = {http://adsabs.harvard.edu/abs/2012MNRAS.426.3148S},
	archiveprefix = {arXiv},
	arxivurl = {http://arxiv.org/abs/1210.5064},
	author = {{Szabados}, L. and {Neh{\'e}z}, D.},
	doi = {10.1111/j.1365-2966.2012.21872.x},
	eprint = {1210.5064},
	journal = {\mnras},
	month = nov,
	pages = {3148-3153},
	primaryclass = {astro-ph.SR},
	title = {{Binarity among Cepheids in the Magellanic Clouds}},
	volume = 426,
	year = 2012}

@article{Neilson_2016_06_0,
	adsurl = {http://adsabs.harvard.edu/abs/2016ApJ...824....1N},
	archiveprefix = {arXiv},
	arxivurl = {http://arxiv.org/abs/1604.03128},
	author = {{Neilson}, H.~R. and {Engle}, S.~G. and {Guinan}, E.~F. and {Bisol}, A.~C. and {Butterworth}, N.},
	doi = {10.3847/0004-637X/824/1/1},
	eid = {1},
	eprint = {1604.03128},
	journal = {\apj},
	month = jun,
	pages = {1},
	primaryclass = {astro-ph.SR},
	title = {{The Secret Lives of Cepheids: Evolution, Mass-Loss, and Ultraviolet Emission of the Long-period Classical Cepheid}},
	volume = 824,
	year = 2016}

@article{Evans_2016_05_0,
	adsurl = {http://adsabs.harvard.edu/abs/2016AJ....151..129E},
	archiveprefix = {arXiv},
	arxivurl = {http://arxiv.org/abs/1603.02224},
	author = {{Evans}, N.~R. and {Bond}, H.~E. and {Schaefer}, G.~H. and {Mason}, B.~D. and {Tingle}, E. and {Karovska}, M. and {Pillitteri}, I.},
	doi = {10.3847/0004-6256/151/5/129},
	eid = {129},
	eprint = {1603.02224},
	journal = {\aj},
	month = may,
	pages = {129},
	primaryclass = {astro-ph.SR},
	title = {{Hubble Space Telescope Snapshot Survey for Resolved Companions of Galactic Cepheids}},
	volume = 151,
	year = 2016}

@article{Evans_2005_08_0,
	adsurl = {http://adsabs.harvard.edu/abs/2005AJ....130..789E},
	arxivurl = {http://arxiv.org/abs/astro-ph/0504169},
	author = {{Evans}, N.~R. and {Carpenter}, K.~G. and {Robinson}, R. and {Kienzle}, F. and {Dekas}, A.~E.},
	doi = {10.1086/430458},
	eprint = {astro-ph/0504169},
	journal = {\aj},
	month = aug,
	pages = {789-793},
	title = {{High-Mass Triple Systems: The Classical Cepheid Y Carinae}},
	volume = 130,
	year = 2005}

@article{Evans_2015_07_0,
	adsurl = {http://adsabs.harvard.edu/abs/2015AJ....150...13E},
	archiveprefix = {arXiv},
	arxivurl = {http://arXiv.org/abs/1505.05823},
	author = {{Evans}, N.~R. and {Berdnikov}, L. and {Lauer}, J. and {Morgan}, D. and {Nichols}, J. and {G{\"u}nther}, H.~M. and {Gorynya}, N. and {Rastorguev}, A. and {Moskalik}, P.},
	doi = {10.1088/0004-6256/150/1/13},
	eid = {13},
	eprint = {1505.05823},
	journal = {\aj},
	month = jul,
	pages = {13},
	primaryclass = {astro-ph.SR},
	title = {{Binary Properties from Cepheid Radial Velocities (CRaV)}},
	volume = 150,
	year = 2015}

@article{Gallenne_2015_07_0,
	adsurl = {http://adsabs.harvard.edu/abs/2015A%26A...579A..68G},
	archiveprefix = {arXiv},
	author = {{Gallenne}, A. and {M{\'e}rand}, A. and {Kervella}, P. and {Monnier}, J.~D. and {Schaefer}, G.~H. and {Baron}, F. and {Breitfelder}, J. and {Le Bouquin}, J.~B. and {Roettenbacher}, R.~M. and {Gieren}, W. and {Pietrzy{\'n}ski}, G. and {McAlister}, H. and {ten Brummelaar}, T. and {Sturmann}, J. and {Sturmann}, L. and {Turner}, N. and {Ridgway}, S. and {Kraus}, S.},
	doi = {10.1051/0004-6361/201525917},
	eid = {A68},
	eprint = {1505.02715},
	journal = {\aap},
	month = jul,
	pages = {A68},
	primaryclass = {astro-ph.SR},
	title = {{Robust high-contrast companion detection from interferometric observations. The CANDID algorithm and an application to six binary Cepheids}},
	volume = 579,
	year = 2015}

@article{Pilecki_2015_06_0,
	adsurl = {http://adsabs.harvard.edu/abs/2015ApJ...806...29P},
	archiveprefix = {arXiv},
	arxivurl = {http://arXiv.org/abs/1504.04611},
	author = {{Pilecki}, B. and {Graczyk}, D. and {Gieren}, W. and {Pietrzy{\'n}ski}, G. and {Thompson}, I.~B. and {Smolec}, R. and {Udalski}, A. and {Soszy{\'n}ski}, I. and {Konorski}, P. and {Taormina}, M. and {Gallenne}, A. and {Minniti}, D. and {Catelan}, M.},
	doi = {10.1088/0004-637X/806/1/29},
	eid = {29},
	eprint = {1504.04611},
	journal = {\apj},
	month = jun,
	pages = {29},
	primaryclass = {astro-ph.SR},
	title = {{The Araucaria Project: the First-overtone Classical Cepheid in the Eclipsing System OGLE-LMC-CEP-2532}},
	volume = 806,
	year = 2015}

@article{Bohm-Vitense_1997_03_0,
	adsurl = {http://adsabs.harvard.edu/abs/1997ApJ...477..916B},
	author = {{B{\"o}hm-Vitense}, E. and {Remage Evans}, N. and {Carpenter}, K. and {Beck-Winchatz}, B. and {Robinson}, R.},
	journal = {\apj},
	month = mar,
	pages = {916-925},
	title = {{The Mass of the Classical Cepheid S Muscae}},
	volume = 477,
	year = 1997}

@article{Neilson_2015_02_0,
	adsurl = {http://adsabs.harvard.edu/abs/2015A%26A...574A...2N},
	archiveprefix = {arXiv},
	arxivurl = {http://arXiv.org/abs/1412.3468},
	author = {{Neilson}, H.~R. and {Schneider}, F.~R.~N. and {Izzard}, R.~G. and {Evans}, N.~R. and {Langer}, N.},
	doi = {10.1051/0004-6361/201424408},
	eid = {A2},
	eprint = {1412.3468},
	journal = {\aap},
	month = feb,
	pages = {A2},
	primaryclass = {astro-ph.SR},
	title = {{The occurrence of classical Cepheids in binary systems}},
	volume = 574,
	year = 2015}

@article{Gorynya_1996_01_0,
	adsurl = {http://adsabs.harvard.edu/abs/1996AstL...22...33G},
	author = {{Gorynya}, N.~A. and {Rastorguev}, A.~S. and {Samus}, N.~N.},
	journal = {Astronomy Letters},
	month = jan,
	pages = {33-38},
	title = {{Mean radial velocities and binarity of cepheids from the 1987-1995 measurements}},
	volume = 22,
	year = 1996}

@article{Evans_1995_05_0,
	adsurl = {http://adsabs.harvard.edu/abs/1995ApJ...445..393E},
	author = {{Evans}, N.~R.},
	doi = {10.1086/175704},
	journal = {\apj},
	month = may,
	pages = {393-405},
	title = {{The mass ratios of Cepheid binaries}},
	volume = 445,
	year = 1995}

@article{Gallenne_2014_07_0,
	adsurl = {http://adsabs.harvard.edu/abs/2014A%26A...567A..60G},
	archiveprefix = {arXiv},
	author = {{Gallenne}, A. and {Kervella}, P. and {M{\'e}rand}, A. and {Evans}, N.~R. and {Girard}, J.~H.~V. and {Gieren}, W. and {Pietrzy{\'n}ski}, G.},
	doi = {10.1051/0004-6361/201423872},
	eid = {A60},
	eprint = {1406.0493},
	journal = {\aap},
	month = jul,
	pages = {A60},
	primaryclass = {astro-ph.SR},
	title = {{Searching for visual companions of close Cepheids. VLT/NACO lucky imaging of Y Oph, FF Aql, X Sgr, W Sgr, and {$\eta$} Aql}},
	volume = 567,
	year = 2014}

@inproceedings{Gallenne_2013_02_0,
	adsurl = {http://adsabs.harvard.edu/abs/2013EAS....64..197G},
	archiveprefix = {arXiv},
	author = {{Gallenne}, A. and {Kervella}, P. and {M{\'e}rand}, A. and {Monnier}, J.~D. and {Breitfelder}, J. and {Pietrzy{\'n}ski}, G. and {Gieren}, W.},
	booktitle = {EAS Publications Series},
	doi = {10.1051/eas/1364027},
	editor = {{Pavlovski} K., {Tkachenko}, A. and {Torres}, G.},
	eprint = {1311.3207},
	month = feb,
	pages = {197-204},
	primaryclass = {astro-ph.SR},
	title = {{Binary Cepheids from optical interferometry}},
	volume = 64,
	year = 2013}

@article{Evans_2009_03_0,
	adsurl = {http://adsabs.harvard.edu/abs/2009AJ....137.3700E},
	archiveprefix = {arXiv},
	arxivurl = {http://arXiv.org/abs/0902.3281},
	author = {{Evans}, N.~R. and {Massa}, D. and {Proffitt}, C.},
	doi = {10.1088/0004-6256/137/3/3700},
	eprint = {0902.3281},
	journal = {\aj},
	month = mar,
	pages = {3700-3705},
	primaryclass = {astro-ph.SR},
	title = {{Massive Star Multiplicity: The Cepheid W Sgr}},
	volume = 137,
	year = 2009}

@article{Babel_1989_06_0,
	adsurl = {http://adsabs.harvard.edu/abs/1989A%26A...216..125B},
	author = {{Babel}, J. and {Burki}, G. and {Mayor}, M. and {Chmielewski}, Y. and {Waelkens}, C.},
	journal = {\aap},
	month = jun,
	pages = {125-134},
	title = {{W Sagittarii - Pulsation and orbit}},
	volume = 216,
	year = 1989}

@article{Morgan_1978_06_0,
	adscomment = {AAA ID. AAA021.118.023},
	adsurl = {http://adsabs.harvard.edu/abs/1978MNRAS.183..701M},
	author = {{Morgan}, B.~L. and {Beddoes}, D.~R. and {Scaddan}, R.~J. and {Dainty}, J.~C.},
	journal = {\mnras},
	month = jun,
	pages = {701-710},
	title = {{Observations of binary stars by speckle interferometry. I}},
	volume = 183,
	year = 1978}

@article{Gallenne_2014_01_0,
	adsurl = {http://adsabs.harvard.edu/abs/2014A%26A...561L...3G},
	archiveprefix = {arXiv},
	arxivurl = {http://arXiv.org/abs/1312.1950},
	author = {{Gallenne}, A. and {M{\'e}rand}, A. and {Kervella}, P. and {Breitfelder}, J. and {Le Bouquin}, J.-B. and {Monnier}, J.~D. and {Gieren}, W. and {Pilecki}, B. and {Pietrzy{\'n}ski}, G.},
	doi = {10.1051/0004-6361/201322883},
	eid = {L3},
	eprint = {1312.1950},
	journal = {\aap},
	month = jan,
	pages = {L3},
	primaryclass = {astro-ph.SR},
	title = {{Multiplicity of Galactic Cepheids from long-baseline interferometry. II. The Companion of AX Circini revealed with VLTI/PIONIER}},
	volume = 561,
	year = 2014}

@article{Evans_2013_10_0,
	adsurl = {http://adsabs.harvard.edu/abs/2013AJ....146...93R},
	archiveprefix = {arXiv},
	arxivurl = {http://arXiv.org/abs/1307.7123},
	author = {{Evans}, N.~R. and {Bond}, H.~E. and {Schaefer}, G.~H. and {Mason}, B.~D. and {Karovska}, M. and {Tingle}, E.},
	doi = {10.1088/0004-6256/146/4/93},
	eid = {93},
	eprint = {1307.7123},
	journal = {\aj},
	month = oct,
	pages = {93},
	primaryclass = {astro-ph.SR},
	title = {{Binary Cepheids: Separations and Mass Ratios in 5 M $_{&sun;}$ Binaries}},
	volume = 146,
	year = 2013}

@article{Szabados_2013_09_0,
	adsurl = {http://adsabs.harvard.edu/abs/2013MNRAS.434..870S},
	archiveprefix = {arXiv},
	arxivurl = {http://arXiv.org/abs/1308.1855},
	author = {{Szabados}, L. and {Anderson}, R.~I. and {Derekas}, A. and {Kiss}, L.~L. and {Szalai}, T. and {Sz{\'e}kely}, P. and {Christiansen}, J.~L.},
	doi = {10.1093/mnras/stt1079},
	eprint = {1308.1855},
	journal = {\mnras},
	month = sep,
	pages = {870-877},
	primaryclass = {astro-ph.SR},
	title = {{Discovery of the spectroscopic binary nature of three bright southern Cepheids}},
	volume = 434,
	year = 2013}

@article{Szabados_2003_03_0,
	adsurl = {http://adsabs.harvard.edu/abs/2003IBVS.5394....1S},
	author = {{Szabados}, L.},
	journal = {Information Bulletin on Variable Stars},
	month = mar,
	pages = {1},
	title = {{Database on Binaries among Galactic Classical Cepheids}},
	volume = 5394,
	year = 2003}

@article{Massa_2008_01_0,
	adsurl = {http://adsabs.harvard.edu/abs/2008MNRAS.383..139M},
	archiveprefix = {arXiv},
	arxivurl = {http://arXiv.org/abs/0704.1051},
	author = {{Massa}, D. and {Evans}, N.~R.},
	doi = {10.1111/j.1365-2966.2007.12520.x},
	eprint = {0704.1051},
	journal = {\mnras},
	month = jan,
	pages = {139-149},
	title = {{The angular separation of the components of the Cepheid AWPer}},
	volume = 383,
	year = 2008}

@article{Lloyd-Evans_1982_06_0,
	adsurl = {http://adsabs.harvard.edu/abs/1982MNRAS.199..925L},
	author = {{Lloyd Evans}, T.},
	journal = {\mnras},
	month = jun,
	pages = {925-941},
	title = {{Cepheid binaries. II - New southern examples}},
	volume = 199,
	year = 1982}

@article{Evans_1994_11_0,
	adsurl = {http://adsabs.harvard.edu/abs/1994ApJ...436..273E},
	author = {{Evans}, N.~R.},
	doi = {10.1086/174902},
	journal = {\apj},
	month = nov,
	pages = {273-285},
	title = {{Classical Cepheids with terminal age main-sequence companions: Constraints on evolution}},
	volume = 436,
	year = 1994}

@article{Neilson_2011_05_0,
	adsurl = {http://adsabs.harvard.edu/abs/2011A%26A...529L...9N},
	archiveprefix = {arXiv},
	author = {{Neilson}, H.~R. and {Cantiello}, M. and {Langer}, N.},
	doi = {10.1051/0004-6361/201116920},
	eid = {L9},
	eprint = {1104.1638},
	journal = {\aap},
	month = may,
	pages = {L9},
	primaryclass = {astro-ph.SR},
	title = {{The Cepheid mass discrepancy and pulsation-driven mass loss}},
	volume = 529,
	year = 2011}

@article{Evans_1992_04_0,
	adsurl = {http://adsabs.harvard.edu/abs/1992ApJ...389..657E},
	author = {{Evans}, N.~R.},
	doi = {10.1086/171238},
	journal = {\apj},
	month = apr,
	pages = {657-664},
	title = {{New calibrators for the Cepheid period-luminosity relation}},
	volume = 389,
	year = 1992}

@article{Welch_1987_07_0,
	adsurl = {http://adsabs.harvard.edu/abs/1987PASP...99..610W},
	author = {{Welch}, D.~L. and {Evans}, N.~R. and {Lyons}, R.~W. and {Harris}, H.~C. and {Barnes}, III, T.~G. and {Slovak}, M.~H. and {Moffett}, T.~J.},
	doi = {10.1086/132022},
	journal = {\pasp},
	month = jul,
	pages = {610-616},
	title = {{The orbit of the classical Cepheid U Aquilae}},
	volume = 99,
	year = 1987}

@article{Evans_1998_03_0,
	adsurl = {http://adsabs.harvard.edu/abs/1998ApJ...494..768E},
	author = {{Evans}, N.~R. and {Boehm-Vitense}, E. and {Carpenter}, K. and {Beck-Winchatz}, B. and {Robinson}, R.},
	doi = {10.1086/305242},
	journal = {\apj},
	month = mar,
	pages = {768},
	title = {{Classical Cepheid Masses: U Aquilae}},
	volume = 494,
	year = 1998}

@article{Evans_1992_01_0,
	adsurl = {http://cdsads.u-strasbg.fr/abs/1992ApJ...384..220E},
	author = {{Evans}, N.~R.},
	doi = {10.1086/170865},
	journal = {\apj},
	month = jan,
	pages = {220-233},
	title = {{A magnitude-limited survey of Cepheid companions in the ultraviolet}},
	volume = 384,
	year = 1992}

@article{Petterson_2004_05_0,
	adsurl = {http://adsabs.harvard.edu/abs/2004MNRAS.350...95P},
	author = {{Petterson}, O.~K.~L. and {Cottrell}, P.~L. and {Albrow}, M.~D.},
	doi = {10.1111/j.1365-2966.2004.07555.x},
	journal = {\mnras},
	month = may,
	pages = {95-112},
	title = {{The orbits of southern binary Cepheids}},
	volume = 350,
	year = 2004}

@article{Evans_2008_09_0,
	adsurl = {http://cdsads.u-strasbg.fr/abs/2008AJ....136.1137E},
	archiveprefix = {arXiv},
	author = {{Evans}, N.~R. and {Schaefer}, G.~H. and {Bond}, H.~E. and {Bono}, G. and {Karovska}, M. and {Nelan}, E. and {Sasselov}, D. and {Mason}, B.~D.},
	doi = {10.1088/0004-6256/136/3/1137},
	eprint = {0806.4904},
	journal = {\aj},
	month = sep,
	pages = {1137-1146},
	title = {{Direct Detection of the Close Companion of Polaris with the Hubble Space Telescope}},
	volume = 136,
	year = 2008}

@article{Evans_1991_05_0,
	adsurl = {http://cdsads.u-strasbg.fr/abs/1991ApJ...372..597E},
	author = {{Evans}, N.~R.},
	doi = {10.1086/170004},
	journal = {\apj},
	month = may,
	pages = {597-609},
	title = {{Classical Cepheid luminosities from binary companions}},
	volume = 372,
	year = 1991}

@article{Szabados_1991_01_0,
	adsurl = {http://adsabs.harvard.edu/abs/1991CoKon..96..123S},
	author = {{Szabados}, L.},
	journal = {Commun. of the Konkoly Observatory Hungary},
	month = jan,
	pages = {123-244},
	title = {{Northern Cepheids: Period Update and Duplicity Effects}},
	volume = 96,
	year = 1991}

@article{Berdnikov_2008_04_0,
	adsurl = {http://adsabs.harvard.edu/abs/2008yCat.2285....0B},
	author = {{Berdnikov}, L.~N.},
	journal = {VizieR Online Data Catalog: II/285, originally published in: Sternberg Astronomical Institute, Moscow},
	month = apr,
	title = {{Photoelectric observations of Cepheids in UBV(RI)c (Berdnikov, 2008)}},
	volume = 2285,
	year = 2008}

@article{Fernie_1995_01_0,
	adsurl = {http://adsabs.harvard.edu/abs/1995IBVS.4148....1F},
	author = {{Fernie}, J.~D. and {Evans}, N.~R. and {Beattie}, B. and {Seager}, S.},
	journal = {Inf. Bul. on Variable Stars},
	month = jan,
	pages = {1},
	title = {{A Database of Galactic Classical Cepheids}},
	volume = 4148,
	year = 1995}

@article{Benedict_2007_04_0,
	adsurl = {http://adsabs.harvard.edu/abs/2007AJ....133.1810B},
	author = {{Benedict}, G.~F. and {McArthur}, B.~E. and {Feast}, M.~W. and {Barnes}, T.~G. and {Harrison}, T.~E. and {Patterson}, R.~J. and {Menzies}, J.~W. and {Bean}, J.~L. and {Freedman}, W.~L.},
	doi = {10.1086/511980},
	eprint = {arXiv:astro-ph/0612465},
	journal = {\aj},
	month = apr,
	pages = {1810-1827},
	title = {{Hubble Space Telescope Fine Guidance Sensor Parallaxes of Galactic Cepheid Variable Stars: Period-Luminosity Relations}},
	volume = 133,
	year = 2007}

@article{Kervella_2004_03_0,
	adsurl = {http://adsabs.harvard.edu/abs/2004A%26A...416..941K},
	author = {{Kervella}, P. and {Nardetto}, N. and {Bersier}, D. and {Mourard}, D. and {Coud{\'e} du Foresto}, V.},
	doi = {10.1051/0004-6361:20031743},
	eprint = {arXiv:astro-ph/0311525},
	journal = {\aap},
	month = mar,
	pages = {941-953},
	title = {{Cepheid distances from infrared long-baseline interferometry. I. VINCI/VLTI observations of seven Galactic Cepheids}},
	volume = 416,
	year = 2004}

@article{Fouque_2007_12_0,
	adsurl = {http://adsabs.harvard.edu/abs/2007A%26A...476...73F},
	archiveprefix = {arXiv},
	author = {{Fouqu{\'e}}, P. and {Arriagada}, P. and {Storm}, J. and {Barnes}, T.~G. and {Nardetto}, N. and {M{\'e}rand}, A. and {Kervella}, P. and {Gieren}, W. and {Bersier}, D. and {Benedict}, G.~F. and {McArthur}, B.~E.},
	doi = {10.1051/0004-6361:20078187},
	eprint = {0709.3255},
	journal = {\aap},
	month = dec,
	pages = {73-81},
	title = {{A new calibration of Galactic Cepheid period-luminosity relations from B to K bands, and a comparison to LMC relations}},
	volume = 476,
	year = 2007}

@article{Szabados_1989_01_0,
	adsurl = {http://adsabs.harvard.edu/abs/1989CoKon..94....1S},
	author = {{Szabados}, L.},
	journal = {Commun. of the Konkoly Observatory Hungary},
	month = jan,
	pages = {1-85},
	title = {{Period changes of bright southern Cepheids}},
	volume = 94,
	year = 1989}

@article{Neilson_2008_09_0,
	adsurl = {http://adsabs.harvard.edu/abs/2008ApJ...684..569N},
	archiveprefix = {arXiv},
	author = {{Neilson}, H.~R. and {Lester}, J.~B.},
	doi = {10.1086/588650},
	eprint = {0803.4198},
	journal = {\apj},
	month = sep,
	pages = {569-587},
	title = {{On the Enhancement of Mass Loss in Cepheids Due to Radial Pulsation}},
	volume = 684,
	year = 2008}

@article{Keller_2008_04_0,
	adsurl = {http://adsabs.harvard.edu/abs/2008ApJ...677..483K},
	archiveprefix = {arXiv},
	author = {{Keller}, S.~C.},
	doi = {10.1086/529366},
	eprint = {0801.1342},
	journal = {\apj},
	month = apr,
	pages = {483-487},
	title = {{Cepheid Mass loss and the Pulsation-Evolutionary Mass Discrepancy}},
	volume = {677},
	year = {2008}}
 
 
 \begin{appendix} 
        
        \onecolumn
        \section{Log of our SPHERE observations}
        
                \begin{longtable}{ccccccccc}
                        \caption{\label{table__log}Log of the SPHERE/ZIMPOL observations.}\\
                        \hline\hline
                        Star    &  UT  &        MJD & Filter & NDF & DIT  & N$_\mathrm{DIT}$   & N$_\mathrm{exp}$   &   Seeing      \\
                                                &         &     (day)   &   &   & (s)      &              &               &       (\arcsec)                       \\
                        \hline
                        \object{$\eta$ Aql} & 2018-05-25 &  58263.392  &  V/I'  &  ND2 &  1  &  20  &  6  &  0.61 \\
                        $\eta$ Aql  & 2018-05-25  &  58263.390  &  V/R' & ND2 &  1  &  20  &  6  &  0.65                          \\
                        \object{FF Aql}   & 2018-04-20 &  58228.394  &  V/I' & ND2 &  5  &  20  &  3  &  0.55                \\
                        FF Aql   & 2018-04-20 &  58228.385  &  V/R' & ND2 &  5  &  20  &  6  &  0.65                              \\
                        \object{FN Aql}   & 2018-08-11 &  58341.224  &  V/I' & ND1 &  5  &  20  &  6  &  0.67                \\
                        FN Aql   & 2018-08-11 &  58341.217  &  V/R' & ND1 &  5  &  20  &  6  &  0.53                              \\
                        \object{KL Aql}   & 2018-08-09 &  58339.250  &  V/I' & -- &  30  &  10  &  3  &  0.71                \\
                        KL Aql  & 2018-08-09 &  58339.224  &  V/R' &  -- &  30  &  10  &  6  &  0.99                             \\
                        \object{V496 Aql}  & 2018-08-10 &  58340.257  &  V/I' & ND1 &  4  &  20  &  6  &  0.51   \\
                        V496 Aql   & 2018-08-10 &  58340.251  &  V/R' & ND1 &  4  &  20  &  6  &  0.75                      \\
                        \object{V916 Aql}  & 2018-07-11 &  58310.149  &  V/I' & ND1 &  60  &  8  &  3  &  0.48   \\
                        V916 Aql   & 2018-07-11 &  58310.132  &  V/R' & ND1 &  60  &  8  &  3  &  0.61                      \\
                        \object{V1344 Aql}   & 2018-08-07  &  58337.237  &  V/I' & ND1 &  4  &  20  &  6  &  0.51        \\
                        V1344 Aql   & 2018-08-07 &  58337.23  &  V/R' & ND1 &  4  &  20  &  6  &  0.50                      \\
                        \object{V340 Ara}   & 2018-08-17 &  58347.168  &  V/I' & ND1 &  40 &  8  &  3  &  0.58   \\
                        V340 Ara   & 2018-08-17 &  58347.142  &  V/R' & ND1 &  40  &  9  &  6  &  0.61                      \\
                        \object{GX Car}   & 2020-01-28  &  58876.302  &  V/I' & -- &  5  &  20  &  9  &  0.95            \\
                        GX Car   & 2020-01-28 &  58876.308  &  V/R' & --  &  5  &  20  &  9  &  0.84                             \\
                        \object{U Car}  & 2018-05-20  &  58258.147  &  V/I' & ND2 &  9  &  20  &  6  &  0.52                \\
                        U Car   & 2018-05-20 &  58258.134  &  V/R' & ND2 &  9  &  20  &  6  &  0.68                              \\
                        \object{UW Car}   & 2020-01-28 &  58876.29  &  V/I' & -- &  5  &  20  &  9  &  0.77                 \\
                        UW Car  & 2020-01-28 &  58876.278  &  V/R' & -- &  5  &  20  &  9  &  0.95                                \\
                        \object{V339 Cen}  & 2018-07-11 &  58310.118  &  V/I' & ND1 &  8  &  20  &  3  &  0.47   \\
                        V339 Cen   & 2018-07-11  &  58310.106  &  V/R' & ND1 &  8  &  20  &  6  &  0.38                          \\
                        \object{V659 Cen}   & 2018-07-03  &  58302.094  &  V/I' & ND1 &  2.5  &  20  &  3  &  1.49        \\
                        V659 Cen  &  2018-07-03 &  58302.090  &  V/R' & ND1 &  2.5  &  20  &  6  &  1.56                            \\
                        \object{AX Cir}   & 2018-08-17  &  58347.049  &  V/I' & ND2 &  8.7  &  20  &  3  &  0.51                 \\
                        AX Cir   & 2018-08-17 &  58347.036  &  V/R' & ND2 &  8.7  &  20  &  6  &  0.61                            \\
                        \object{RZ CMa}   & 2020-02-23 &  58902.144  &  V/I' & -- &  3  &  20  &  12  &  0.61                \\
                        RZ CMa  & 2020-02-23 &  58902.132  &  V/R' & -- &  8  &  10  &  12  &  1.16                               \\
                        \object{TX Del}  & 2018-07-14  &  58313.283  &  V/I' & -- &  3  &  100  &  3  &  0.49                \\
                        TX Del  & 2018-07-14 &  58313.261  &  V/R' & -- &  3  &  100  &  6  &  0.57                               \\
                        \object{$\beta$ Dor}  & 2020-01-17 &  58865.149  &  V/I' & ND2 &  3  &  20  &  9  &  0.57  \\
                        $\beta$ Dor  & 2020-01-17 &  58865.141  &  V/R' & ND2 &  3  &  20  &  9  &  0.78                          \\
                        \object{T Mon}  & 2020-02-15 &  58894.071  &  V/I' & ND1 &  2  &  20  &  15  &  0.46               \\
                        T Mon   & 2020-02-15 &  58894.062  &  V/R' & ND1 &  1  &  40  &  15  &  0.53                             \\
                        \object{TX Mon} & 2020-02-24  &  58903.144  &  V/I' & -- &  12  &  20  &  6  &  0.53                \\
                        TX Mon & 2020-02-24  &  58903.124  &  V/R' & -- &  12  &  20  &  6  &  0.68                               \\
                        \object{V465 Mon} & 2020-03-23  &  58931.084  &  V/I' & -- &  12  &  20  &  6  &  0.65   \\
                        V465 Mon  & 2020-03-23 &  58931.066  &  V/R' & -- &  12  &  20  &  6  &  0.65                     \\
                        \object{S Nor}   &  2018-08-17 &  58347.119  &  V/I' & ND1 &  1.2  &  20  &  3  &  0.54              \\
                        S Nor   & 2018-08-17 &  58347.116  &  V/R' & ND1 &  1.2  &  20  &  6  &  0.53                            \\
                        \object{BF Oph}  & 2018-07-15 &  58314.231  &  V/I' & ND1 &  1.7  &  20  &  3  &  0.63              \\
                        BF Oph  & 2018-07-15 &  58314.228  &  V/R' & ND1 &  1.7  &  20  &  6  &  0.55                            \\
                        \object{Y Oph}  & 2018-05-25 &  58263.376  &  V/I' & ND2 &  11 &  20  &  3  &  0.54                \\
                        Y Oph  & 2018-05-25 &  58263.360  &  V/R' & ND2 &  11  &  20  &  6  &  0.76                               \\
                        \object{AP Pup}  & 2020-03-17 &  58925.114  &  V/I' & ND1 &  8  &  20  &  6  &  0.94                \\
                        AP Pup  & 2020-03-17 &  58925.102  &  V/R' & ND1 &  8  &  20  &  6  &  1.0                               \\
                        \object{X Pup}  & 2020-03-17 &  58925.082  &  V/I' & -- &  2 &  60  &  9  &  1.28  \\
                        X Pup  & 2020-03-17 &  58925.064  &  V/R' & -- &  2  &  60  &  9  &  1.11  \\
                        \object{RY Sco}   & 2018-06-12 &  58281.274  &  V/I' & ND1 &  3  &  20  &  3  &  0.77  \\
                        RY Sco  & 2018-06-12 &  58281.269  &  V/R' & ND1 &  3  &  20  &  6  &  0.86  \\
                        \object{RV Sco}  & 2018-08-14 &  58344.097  &  V/I'  & ND1 &  1.3  &  20  &  3  &  1.06  \\
                        RV Sco  & 2018-08-14 &  58344.094  &  V/R' & ND1 &  1.3  &  20  &  6  &  1.15  \\
                        \object{V482 Sco}   & 2018-08-14 &  58344.152  &  V/I' & ND1 &  4.3  &  20  &  3  &  0.65  \\
                        V482 Sco  & 2018-08-14 &  58344.145  &  V/R' & ND1 &  4.3  &  20  &  6  &  0.69  \\
                        \object{V636 Sco}  & 2018-08-07  &  58337.223  &  V/I' & ND1 &  3  &  10  &  3  &  0.51  \\
                        V636 Sco  & 2018-08-07 &  58337.218  &  V/R' & ND1 &  1.5  &  20  &  6  &  0.51  \\
                        \object{RU Sct} & 2018-08-06  &  58336.251  &  V/I' & -- &  1  &  20  &  3  &  0.92  \\
                        RU Sct   & 2018-08-06 &  58336.234  &  V/R' & -- &  5  &  40  &  6  &  1.02  \\
                        \object{Y Sct}  & 2018-08-14 &  58344.194  &  V/I' & ND1 &  34  &  10  &  3  &  0.56  \\
                        Y Sct  & 2018-08-14 &  58344.169  &  V/R' & ND1 &  34  &  10  &  6  &  0.58  \\
                        \object{BQ Ser}  & 2018-08-14 &  58344.126  &  V/I' & ND1 &  30  &  10  &  3  &  0.73 \\
                        BQ Ser   & 2018-08-14 &  58344.104  &  V/R' & ND1 &  30  &  10  &  6  &  0.93  \\
                        \object{S Sge}  & 2018-05-25 &  58263.407  &  V/I' & ND2 &  5  &  20  &  3  &  0.73  \\
                        S Sge  & 2018-05-25 &  58263.399  &  V/R'  & ND2 &  5  &  20  &  6  &  0.54  \\
                        \object{AP Sgr}  & 2018-06-14 &  58283.321  &  V/I' & -- &  1.3  &  20  &  3  &  0.36  \\
                        AP Sgr  & 2018-06-14 &  58283.319  &  V/R' & -- &  1.3  &  20  &  6  &  0.46  \\
                        \object{BB Sgr}  & 2018-06-14 &  58283.367  &  V/I' & ND1 &  1.5  &  20  &  3  &  0.74  \\
                        BB Sgr  & 2018-06-14 &  58283.364  &  V/R' & ND1 &  1.5  &  20  &  6  &  0.77  \\
                        \object{U Sgr}  & 2018-06-14 &  58283.332  &  V/I' & ND1 &  1.4  &  20  &  3  &  0.38  \\
                        U Sgr  & 2018-06-14 &  58283.33  &  V/R' & ND1 &  1.4  &  20  &  6  &  0.40  \\
                        \object{V350 Sgr} & 2018-08-10  &  58340.241  &  V/I' & ND1 &  3  &  20  &  3  &  0.54  \\
                        V350 Sgr  & 2-19-08-10 &  58340.234  &  V/R' & ND1 &  3  &  20  &  6  &  0.54  \\
                        \object{W Sgr} &  2018-06-12 &  58281.288  &  V/I' & ND2 &  2.6  &  20  &  3  &  0.78  \\
                        W Sgr  & 2-18-06-12 &  58281.284  &  V/R' & ND2 &  2.6  &  20  &  6  &  0.69  \\
                        \object{WZ Sgr}  & 2018-08-11 &  58341.207  &  V/I' & ND1 &  3  &  20  &  3  &  0.75  \\
                        WZ Sgr   & 2018-08-11 &  58341.202  &  V/R' & ND1 &  3  &  20  &  6  &  0.88  \\
                        \object{X Sgr}  & 2018-07-17 &  58316.256  &  V/I' & ND2 &  2.2  &  20  &  3  &  0.90  \\
                        X Sgr   & 2018-07-17 &  58316.252  &  V/R' & ND2 &  2.2  &  20  &  6  &  1.29  \\
                        \object{Y Sgr}  & 2018-06-14 &  58283.352  &  V/I' & ND2 &  7.3  &  20  &  3  &  0.48  \\
                        Y Sgr  & 2018-06-14 &  58283.341  &  V/R' & ND2 &  7.3  &  20  &  6  &  0.37  \\
                        \object{YZ Sgr}  & 2018-08-07 &  58337.254  &  V/I' & ND2 &  4  &  10  &  3  &  0.45  \\
                        YZ Sgr &  2018-08-07 &  58337.25  &  V/R' & ND2 &  4 &  10  &  6  &  0.45  \\
                        \object{LR Tra}  & 2018-06-11 &  58280.155  &  V/I' & ND1 &  4  &  20  &  3  &  0.74  \\
                        LR Tra   & 2018-06-11 &  58280.148  &  V/R' & ND1 &  4  &  20  &  6  &  0.79  \\
                        \object{R Tra}  & 2018-07-11 &  58310.097  &  V/I' & ND1 &  1.1  &  20  &  3  &  0.37  \\
                        R Tra   & 2018-07-11 &  58310.094  &  V/R' & ND1 &  1.1  &  20  &  6  &  0.41  \\
                        \object{RZ Vel}  & 2020-01-18  &  58866.195  &  V/I' & ND1 &  4  &  30  &  9  &  0.30  \\
                        RZ Vel   & 2020-01-18 &  58866.179  &  V/R' & ND1 &  6  &  20  &  9  &  0.40  \\
                        \object{V Vel}   & 2020-01-19 &  58867.154  &  V/I' & ND1 &  8  &  30  &  6  &  0.44  \\
                        V Vel  & 2020-01-19 &  58867.146  &  V/R' & ND1 &  10  &  6  &  6  &  0.47  \\
                        \object{T Vel}  & 2020-01-18 &  58866.156  &  V/I' & ND1 &  12  &  20  &  6  &  0.46  \\
                        T Vel  & 2020-01-18 &  58866.138  &  V/R' & ND1 &  12  &  20  &  6  &  0.88  \\
                        \hline
                \end{longtable}
                \tablefoot{NDF: neutral density filter (attenuation factor of $\sim 12$ and $\sim 120$, respectively for ND1 and ND2). DIT: detector integration time. N$_\mathrm{exp}$: number of exposure.}
                
                \FloatBarrier 
                \twocolumn
                
 \end{appendix}

\end{document}